\documentclass[useAMS,usenatbib,usegraphicx]{mn2e}
\usepackage[english]{babel}
\usepackage[T1]{fontenc}
\usepackage{bm}
\usepackage{lmodern} %
\usepackage{color}
\usepackage[flushleft]{threeparttable}

\title[Faraday rotation studies of six AGN jets]
  {18-22 cm VLBA Faraday rotation studies of six AGN jets}
\author[J.C. Motter \& D.C. Gabuzda]
  {J.C.~Motter,$^{1,2}$\thanks{juliana.motter@usp.br}
  D.C.~Gabuzda,$^2$\thanks{d.gabuzda@ucc.ie} \\
  $^1$Instituto de Astronomia, Geof\'isica e Ci\^encias Atmosf\'ericas, Universidade de S\~ao Paulo, 05508-090, S\~ao Paulo, Brasil \\
  $^2$Physics Department, University College Cork, T12 K8AF, Cork, Ireland \\}
\date{Released 2002 Xxxxx XX}

\pagerange{\pageref{firstpage}--\pageref{lastpage}} \pubyear{2002}

\def\LaTeX{L\kern-.36em\raise.3ex\hbox{a}\kern-.15em
    T\kern-.1667em\lower.7ex\hbox{E}\kern-.125emX}

\begin{document}

\label{firstpage}

\maketitle

\begin{abstract}

The formation of relativistic jets in active galactic nuclei (AGN) is related to accretion on to their central supermassive black holes, and magnetic fields are believed to play a central role in launching, collimating and accelerating the jet streams from very compact regions out to kiloparsec or megaparsec scales. In the presence of helical or toroidal magnetic fields threading the AGN jets and their immediate vicinity, gradients in the observed Faraday rotation measures are expected due to the systematic change in the line-of-sight component of the magnetic field across the jet. We have analysed total intensity, linear polarization, fractional polarization and Faraday rotation maps based on Very Long Baseline Array data obtained at four wavelengths in the 18-22~cm range for six AGN (OJ~287, 3C~279, PKS~1510-089, 3C~345, BL~Lac and 3C~454.3). These observations typically probe projected distances out to tens of parsecs from the observed core, and are well suited for Faraday rotation studies due to the relatively long wavelengths used and the similarity of the structures measured at the different wavelengths. We have identified statistically significant, monotonic, transverse Faraday rotation gradients across the jets of four of these six sources, as well as a tentative transverse Faraday rotation gradient across the jet of OJ~287, providing evidence for the presence of toroidal magnetic fields, which may be one component of helical magnetic fields associated with these AGN jets.

\end{abstract}


\section{Introduction} \label{sec:intro}

The current paradigm for explaining the active galactic nuclei (AGN) phenomenon invokes the presence of a `central engine' constituted by a supermassive black hole surrounded by a hot accretion disc. In this scenario, energy is generated by the gravitational infall of matter, which is heated to high temperatures in the disc. The relativistic jets of AGN seem to originate in a region close to the black hole through some process that would involve the extraction of energy from the black hole spin \citep{1977MNRAS.179..433B} or from the accretion disc \citep{1982MNRAS.199..883B}. Both these processes require the presence of magnetic ($\bm{B}$) fields. Current studies have shown that magnetic fields play a crucial role in launching, collimating and directing these jets (e.g \citealt{2001Sci...291...84M}; \citealt{2009MNRAS.394L.126M}).

In the radio regime, AGN jets are detected due to the synchrotron radiation they emit, thus indicating the presence of relativistic electrons accelerated by local magnetic fields. Synchrotron radiation is intrinsically linearly polarized (up to about 75 per cent in optically thin regions with uniform $\bm{B}$ fields); therefore linear polarization observations, especially those carried out with high resolution using very long baseline interferometry (VLBI), are an important tool to provide information about the degree of order and the direction of the $\bm{B}$-field giving rise to the observed synchrotron emission. Multi-wavelength polarization observations also provide information about Faraday rotation occurring between the source and observer.

Faraday rotation of the plane of the linear polarization occurs when a linearly polarized electromagnetic wave travels through magnetized plasma (containing free electrons and magnetic field). The amount of rotation is given by the relation

\newcommand{\ud}{\mathrm{d}}
\begin{equation}
	\chi_{obs} - \chi_o = \frac{e^3\lambda^2}{8\pi^2\epsilon_0 m^2 c^3} \int n_e \bm{B}\cdot \ud \mathbf{l} \equiv \mathrm{RM}\lambda^2
	\label{eq_rm}
\end{equation}
where $\chi_{obs}$ and $\chi_o$ are the observed and unrotated (intrinsic) polarization angles, respectively, $-e$ and $m$ are the charge and mass of the particles causing the Faraday rotation, usually taken to be electrons, $c$ is the speed of light, $n_e$ is the electron density, $\bm{B}$ is the magnetic field in the region of Faraday rotation, $\ud\mathbf{l}$ is a path-length element along the line of sight, $\lambda$ is the observing wavelength and the integral is carried out along the line of sight from the source to the observer. RM is the rotation measure, which can be determined from multi-wavelength observations of the polarization angles. 

Faraday rotation measurements thus carry information about the line-of-sight magnetic field and the electron density in the region where the Faraday rotation is occurring, and $\chi_o$, which provides information about the intrinsic $\bm{B}$-field geometry associated with the source projected onto the plane of the sky. 

In general, many polarization structures are observed in AGN jets, but there is an overall trend for the jet polarization to be oriented close to parallel or perpendicular to the local jet direction (\citealt{2005AJ....130.1389L}, \citealt{2005MNRAS.360..869L}, \citealt{2015ASSL..414..117G}) and for the Faraday rotation to be stronger in the core regions (e.g., \citealt{2003ApJ...589..126Z}; \citeyear{2004ApJ...612..749Z}). When the first AGN to have a `spine+sheath'  magnetic field structure was identified by \citet{1999ApJ...518L..87A}, it was proposed by them that the `sheath'  (orthogonal polarization, longitudinal B field near the jet edges) could be caused by the interaction of the jet with the surrounding medium, and that the `spine'  (longitudinal polarization, orthogonal B field near the central axis of the jet) could be a consequence of transverse shocks propagating along the jet. However, since more AGN jets have now been found to have this structure, it has also been argued that the `spine+sheath' composition could be evidence for the presence of helical magnetic fields threading these jets (e.g., \citealt{2005MNRAS.356..859P}, \citealt{2005MNRAS.360..869L}, \citeauthor*{2015aMNRAS.450.2441G} 2015a), with the azimuthal component dominating near their central axis and the longitudinal component becoming dominant near the jet edges.

\citet{blandfordr.d.1993} pointed out that, in the presence of a helical magnetic field, transverse RM gradients should be observed across the jet due to the systematic change in the associated line-of-sight component of the $\bm{B}$-field. Many works have reported the detection of such transverse gradients on parsec scales, starting with the one published by \citet{2002PASJ...54L..39A} for the jet of the quasar 3C~273 and later confirmed by \citet{2005ApJ...626L..73Z}, and with the most recent studies being those of \citet{2012AJ....144..105H}, \citet{2013MNRAS.431..695M}, and \citeauthor{2014aMNRAS.438L...1G} (2\citeyear{2014aMNRAS.438L...1G}, 2\citeyear{2014bMNRAS.444..172G}, 2\citeyear{2015aMNRAS.450.2441G}). The detection of these transverse gradients only provides evidence for the existence of an azimuthal field component, and demonstrating that this is one component of a helical $\bm{B}$ field requires the detection of a longitudinal field component as well. Therefore it is important to search for other signs of the presence of a helical magnetic field geometry, such as asymmetries in the transverse polarization structure across the jet \citep*{2013MNRAS.430.1504M}.

The main goal of this work is to carry out total intensity, linear polarization and Faraday rotation studies for six AGN through the analysis of data obtained at four wavelengths in the range 18-22~cm with the Very Long Baseline Array (VLBA). These six sources were chosen for this study because they have been investigated by the radio astronomy group at the University of S\~ao Paulo for a number of years in the context of precessional models for the parsec-scale jets and multi-waveband analyses (\citealt{1998ApJ...496..172A}, \citealt{2000A&A...355..915A}, \citealt{2004ApJ...602..625C}, \citealt{2011ATel.3799....1B} and \citealt{2013MNRAS.428..280C}). This wavelength range is ideally suited for probing projected distances out to tens of parsecs from the observed core, thus providing a link between the VLBA milliarcsecond-scale structures observed at centimetre wavelengths and the arcsecond scales seen by the Very Large Array (VLA). In our analysis, we have identified statistically significant, monotonic, transverse Faraday rotation gradients across the jets of four out of these six sources, indicating the presence of a toroidal magnetic field component, which, in turn, may be associated with helical magnetic fields threading these AGN jets. The polarization structures are consistent with those observed on smaller scales in the 2 cm MOJAVE maps (Monitoring of Jets in AGN with VLBA Experiments; \citealt{2005AJ....130.1389L}).

This work is structured as follows. In Section \ref{obs}, we present a brief description of the observational data used in our analysis. In Section \ref{res}, we present the results obtained for each source. In Section \ref{sec:dis}, we discuss the reliability of the transverse RM gradients detected and other evidence for helical magnetic fields threading the AGN jets studied in this work. Finally, in Section \ref{sec:conc}, we summarize our main findings.

\begin{table*} 
 \scriptsize 
 \caption{Observation dates and source properties. \label{obj_prop}
 } 
 \begin{threeparttable}
 \begin{center} 
 \begin{tabular}{@{}cccccccc}
  \hline
  Source & J2000 & Observation & Redshift & Reference & pc/mas\tnote{a} & Integrated & Reference \\
  name & name & date &  & for the redshift &  & RM (rad m$^{-2}$) & for integrated RM \\
  \hline
  OJ 287 & J0854+2006 & 2010 February 2 & 0.306 & \citet*{1989AAS...80..103S} & 4.48 & 31${\pm}$3 & \citet{1988PhDT........24R} \\
  
  3C 279 & J1256-0547 & 2010 March 7 & 0.536 & \citet{1996ApJS..104...37M} & 6.31 & 27${\pm}$2 & \citet{1988PhDT........24R} \\
  
  PKS 1510-089 & J1512-0905 & 2010 March 7 & 0.360 & \citet*{1990PASP..102.1235T} & 5.00 & -6.8${\pm}$0.8 & \citet{2009ApJ...702.1230T} \\
  
  3C 345 & J1642+3948 & 2010 March 7 & 0.593 & \citet{1996ApJS..104...37M} & 6.63 & 18.0${\pm}$0.3 & \citet{2009ApJ...702.1230T} \\
   
  BL Lac & J2202+4216 & 2010 August 23 & 0.0686 & \citet{1995ApJ...452L...5V} & 1.29 & -205${\pm}$5 & \citet{1988PhDT........24R} \\
  
  3C 454.3 & J2253+1608 & 2010 August 23 & 0.859 & \citet{1991MNRAS.250..414J} & 7.70 & -60.5${\pm}$0.2 & \citet{2009ApJ...702.1230T} \\
   
  \hline
 \end{tabular}
 \begin{tablenotes}
  \item[a] Cosmological parameters: H$_0$ = 71 km\ s$^{-1}$\ Mpc$^{-1}$; $\Omega_{\Lambda}$ = 0.73; $\Omega_M$ = 0.27.
 \end{tablenotes}
 \end{center}
 \end{threeparttable}
\end{table*} 

\section{Observations and Data Analysis} \label{obs}

The observations analysed for this study were made with the National Radio Astronomy Observatory (NRAO) VLBA in 2010, and include polarimetric observations of the 135 AGN of the MOJAVE-I sample (\citealt{2005AJ....130.1389L}) simultaneously at 1358, 1430, 1493 and 1665 MHz (22.1, 21.0, 20.1, 18.0 cm, respectively; \citealt{2011arXiv1101.5942C}). The objects were observed in nine 24-h sessions at a total aggregate bit rate of 128 Mbits~s$^{-1}$. They were observed in scans of about 3.5 min spread out in time when the source was visible by most or all of the array, and the total observing time per source was about 35-45 min. 

In this work, six well known AGN from the sample were studied. The source names, redshifts, pc~mas$^{-1}$ values, integrated (taken to be predominantly Galactic) Faraday rotations and observing dates are shown in Table \ref{obj_prop}. The redshifts and pc/mas values were taken from the MOJAVE project website (http://www.physics.purdue.edu/MOJAVE/, \citeauthor{2009aAJ....137.3718L} 2009a), and correspond to the cosmological parameters H$_0$ = 71 km\ s$^{-1}$\ Mpc$^{-1}$, $\Omega_{\Lambda}$ = 0.73 and $\Omega_{\mathrm{M}}$ = 0.27. References to the integrated RMs are indicated in the last column of Table \ref{obj_prop}.

The preliminary amplitude, phase, polarization (D-term) and electric vector position angle (EVPA) calibrations were done in the NRAO Astronomical Imaging Processing System (AIPS) using standard techniques. For the phase calibration, the Los Alamos station was used as the reference antenna. The calibration of the EVPAs was done by comparing the VLA and total VLBI polarization measurements of the compact, polarized source J0006-0623 obtained at nearly simultaneous epochs and rotating the EVPAs for the total VLBI polarization to agree with those for the VLA polarization. The EVPA corrections applied to the sources at each frequency and each epoch are shown in Table \ref{calib}. More details about the calibration procedures can be found in \citet{2011arXiv1101.5942C} and \citet{fionahealy}.

We used the calibrated visibility data to iteratively construct maps of the total intensity \textit{I} and of the Stokes parameters \textit{Q} and \textit{U} using the AIPS task `IMAGR' with robust weighting (parameter ROBUST = 0). The final maps were then convolved with the lowest-frequency beam to match their resolutions. Polarization angle maps were constructed from the Stokes \textit{Q} and \textit{U} maps using the task `COMB'. Before constructing the Faraday RM maps using the AIPS task `RM', we removed the effect of the integrated RM to provide a better picture of the Faraday rotation occurring in the vicinity of the AGN itself (Table \ref{obj_prop}). The uncertainties in the Stokes \textit{Q} and \textit{U} determining the uncertainties in the polarization angles used in the calculations of the RM fits were estimated in accordance with the recommendations of \citet{2012AJ....144..105H}, who showed that uncertainties in intensity images (either Stokes \textit{I}, \textit{Q} or \textit{U}) are greater in regions of source emission than the off-source rms fluctuations. 

When searching for transverse gradients in the RM, it is important to estimate the local jet direction as accurately as possible, in order to determine whether an observed gradient is close to perpendicular to the local jet direction, and the angle along which RM slices used to estimate the significance of a gradient should be taken. We did this by obtaining model fits to our highest-frequency (1665~MHz) intensity maps using the Cross-Entropy (CE) technique developed by \citet{2011ApJ...736...68C}, which uses the image as the input data and searches for the best parameters for the components, selecting the best candidates generated at each iteration and constructing new parameters from them.  We used the CE method to fit circular Gaussian components to our 1665-MHz maps and estimate the distances and position angles (PAs) of the components with respect to the core, together with their deconvolved sizes and fluxes. The results of our model-fitting are shown in Table~\ref{model_fitting}. The formal uncertainties in some of the model parameters calculated in the model fitting, particularly the component size and the PA relative to the core, are sometimes implausibly small; it has been suggested in the literature that realistic errors are of the order of a few tenths of mas for the position of the components, a few percent for their fluxes and few degrees for their PAs (\citealt{2002ApJ...568...99H}, \citeauthor{2009bAJ....138.1874L} 2009b). However, this is not important for our purposes, as we are essentially interested in using the position angles of the components relative to the core to estimate the local jet directions in the regions of observed transverse RM gradients, which we only need to know to within a few degrees. This means that details of the model fitting are not important for our analysis.

\begin{table}
 \scriptsize
 \begin{center}
  \caption{EVPA corrections for each VLBA observation date and frequency. \label{calib}
}

 \begin{tabular}{@{}cccc}
  \hline
  Frequency & 2010 February 2 & 2010 March 7 & 2010 August 23 \\
  (MHz) & $\Delta\chi^{\circ}$ & $\Delta\chi^{\circ}$ & $\Delta\chi^{\circ}$ \\
  \hline
  1358.46 & 130 & 91 & 12 \\
  
  1430.46 & 113 & 112 & 30 \\
  
  1493.46 & 147 & 84 & -7 \\
  
  1665.46 & 91 & 49 & -25 \\ 
  
  \hline
 \end{tabular}
 \end{center}

\end{table}

\begin{table}
 \scriptsize
  \caption{Model-fitting results. \label{model_fitting}
}
 \begin{threeparttable}
 \begin{center}
 \begin{tabular}{@{}cccccc}
  \hline
 Source & Component & \textit{r}\tnote{a} & \textit{F} & Major axis & PA\tnote{b} \\
  name & number & (mas) & (Jy) & (mas) & (deg) \\
  \hline
  OJ 287 & 1 & 0.0 & 0.938 & 4.21 & 0.0 \\
	     & 2 & 2.72 & 0.098 & 3.94 & -112.4 \\
	     & 3 & 6.36 & 0.043 & 5.07 & -111.2 \\
	     & 4 & 10.6 & 0.012 & 6.67 & -123.1 \\
  
  PKS 1510-089 & 1 & 0.0 & 1.092 & 5.05 & 0.0 \\
 			  & 2 & 2.53 & 0.222 & 4.46 & -26.7 \\
 			  & 3 & 5.90 & 0.092 & 6.14 & -28.0 \\
 			  & 4 & 17.1 & 0.028 & 11.3 & -24.5 \\
 			  & 5 & 35.4 & 0.015 & 14.4 & -31.8 \\
  
  3C 345 & 1 & 0.0 & 2.324 & 4.45 & 0.0 \\
  		 & 2 & 3.18 & 1.194 & 4.32 & -84.5 \\
  		 & 3 & 4.12 & 0.342 & 4.83 & -54.8 \\
  		 & 4 & 11.7 & 0.333 & 7.99 & -63.4 \\
  		 & 5 & 17.5 & 0.132 & 7.99 & -45.6 \\
  		 & 6 & 26.4 & 0.068 & 7.99 & -55.0 \\
  
  BL Lac & 1 & 0.0 & 1.723 & 4.36 & 0.0 \\
  		 & 2 & 2.42 & 0.560 & 4.12 & -174.3 \\
  		 & 3 & 3.75 & 0.324 & 7.57 & 175.4 \\
  		 & 4 & 12.4 & 0.086 & 8.00 & 156.6 \\
  		 & 5 & 20.0 & 0.053 & 8.00 & 153.4 \\
  	
  3C 454.3 & 1 & 0.0 & 2.705 & 5.06 & 0.0 \\
  		   & 2 & 3.26 & 2.511 & 5.26 & -82.4 \\
  		   & 3 & 7.38 & 1.884 & 4.37 & -62.6 \\
  		   & 4 & 11.5 & 1.111 & 7.16 & -54.6 \\
  		   & 5 & 19.0 & 0.143 & 7.41 & -54.1 \\
  		   & 6 & 32.2 & 0.168 & 7.84 & -56.2 \\
  		   & 7 & 44.2 & 0.084 & 7.78 & -55.6 \\
  		   
  \hline
  \end{tabular}
  \begin{tablenotes}
  \item[a] Mean distance from the core component em mas.
  \item[b] PA relative to the core component in degrees.
 \end{tablenotes}
 \end{center}
 \end{threeparttable}
\end{table} 	 

\begin{table}
 \scriptsize
  \caption{Properties of the 1358 MHz maps. \label{map_prop}
}
 \begin{center}
 \begin{tabular}{@{}lcccccc}
  \hline
  Source & Figure & Peak & Lowest contour & BMaj & BMin & BPA \\
  name &  & (Jy) & (per cent) & (mas) & (mas) & (deg) \\
  \hline
  OJ 287 & \ref{Fig1_OJ287_3C279_PKS1510}a & 0.91 & 0.125 & 11.71 & 6.02 & -2.78 \\
  
  3C 279 & \ref{Fig1_OJ287_3C279_PKS1510}d & 3.47 & 0.125 & 14.28 & 5.82 & -1.41 \\
  
  PKS 1510-089 & \ref{Fig1_OJ287_3C279_PKS1510}f & 1.50 & 0.125 & 14.90 & 5.58 & -1.24 \\
  
  3C 345 & \ref{Fig2_3C345_BLLac_3C454.3}a & 3.64 & 0.125 & 10.10 & 5.89 & 3.08 \\ 
  
  BL Lac & \ref{Fig2_3C345_BLLac_3C454.3}d & 2.05 & 0.125 & 9.83 & 5.79 & 8.88 \\ 
  
  3C 454.3 & \ref{Fig2_3C345_BLLac_3C454.3}f & 4.06 & 0.250 & 12.03 & 5.63 & 2.21 \\

  \hline
 \end{tabular}
 \end{center}
\end{table}

\begin{figure*}
\begin{center}

	\begin{minipage}{\textwidth}
	\begin{minipage}{.3\textwidth}
		\centering
		\includegraphics[width=0.9\textwidth]{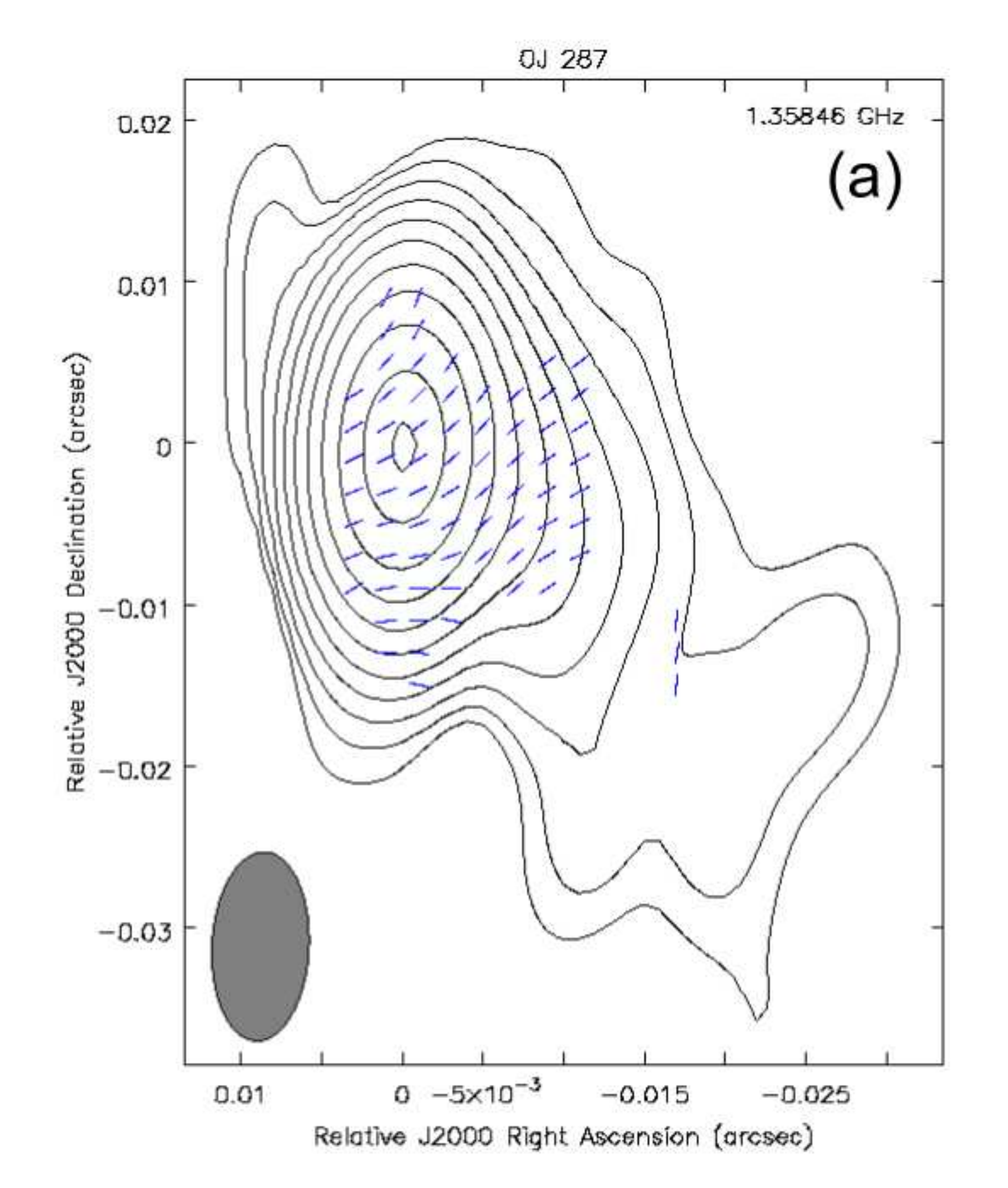}
	\end{minipage}
	\quad
	\begin{minipage}{.3\textwidth}
		\centering
		\includegraphics[width=\textwidth]{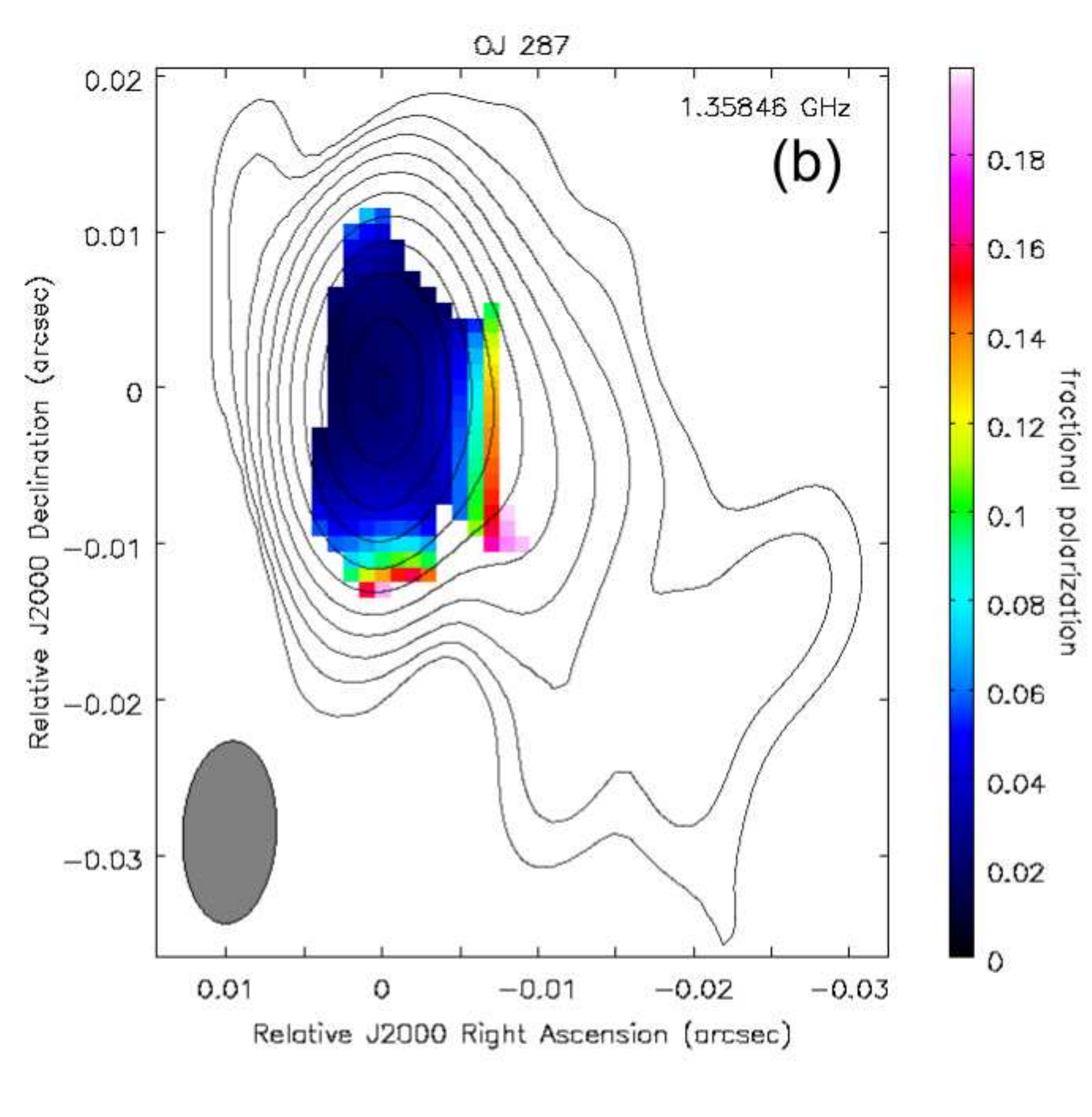}
	\end{minipage}
	\quad
	\begin{minipage}{.3\textwidth}
		\centering
		\includegraphics[width=\textwidth]{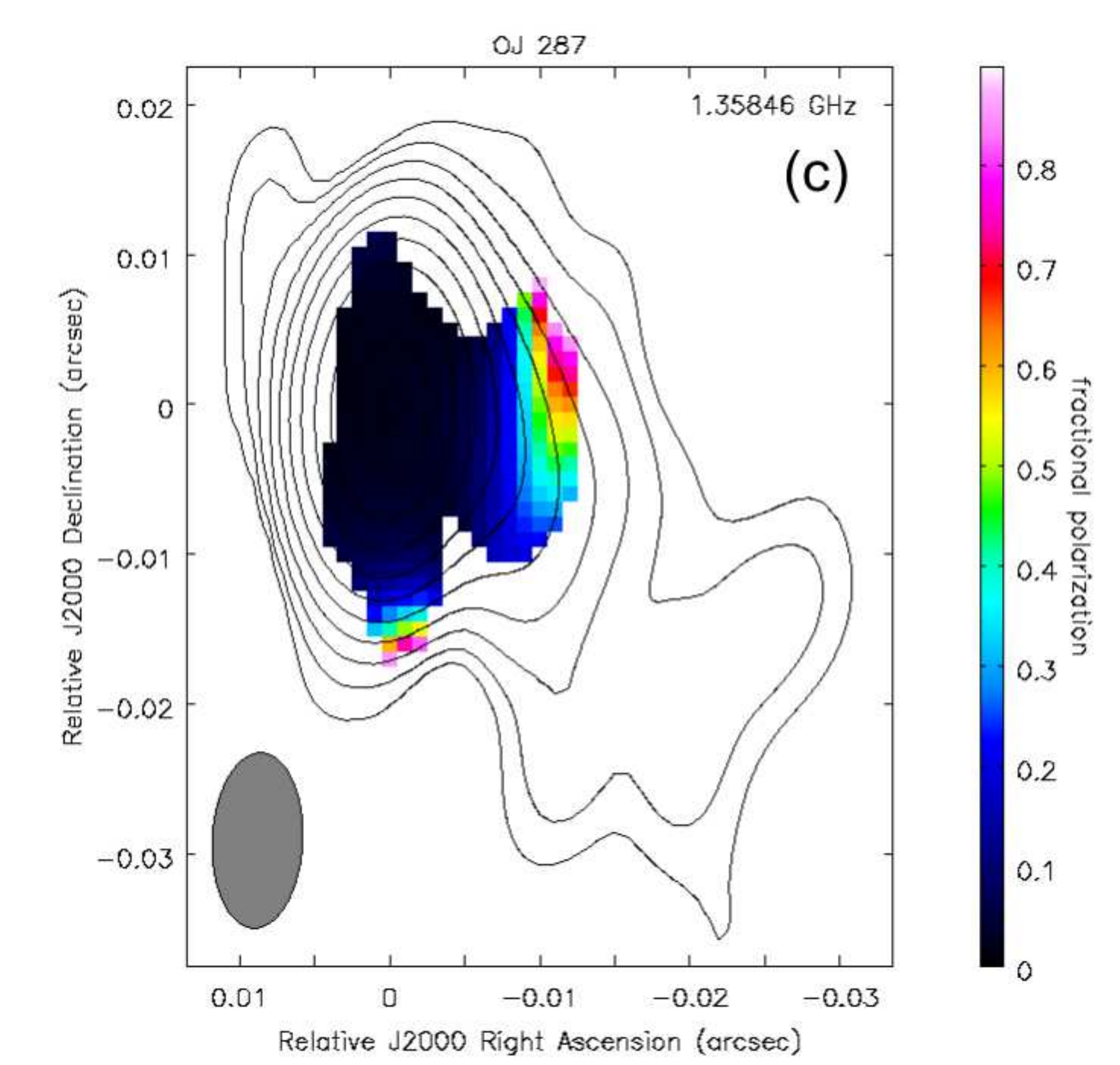}
	\end{minipage}
	\end{minipage}

	\begin{minipage}{\textwidth}
	\centering
	\begin{minipage}{.3\textwidth}
		\includegraphics[width=0.9\textwidth]{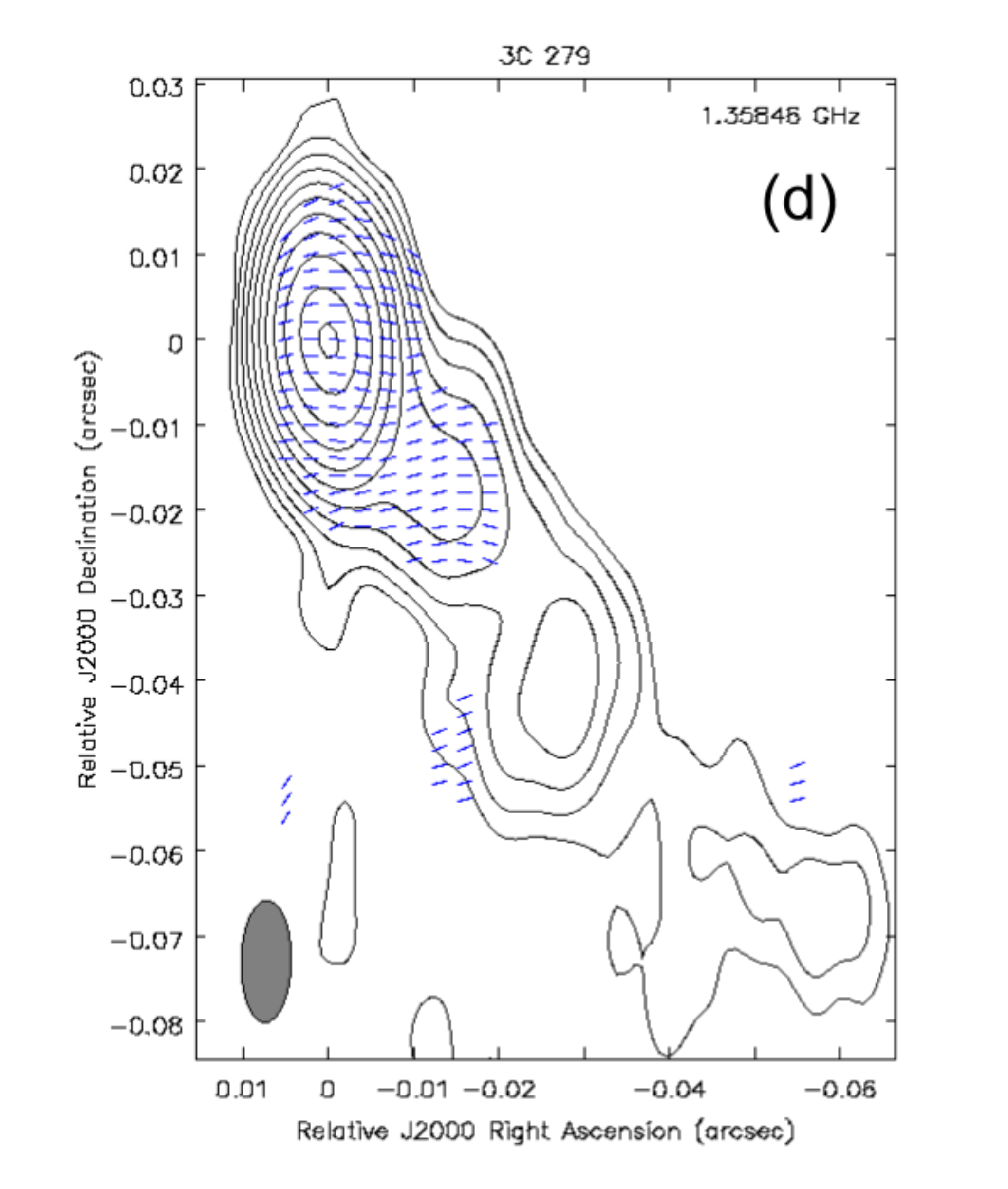}
	\end{minipage}
	\quad
	\begin{minipage}{.3\textwidth}
		\includegraphics[width=\textwidth]{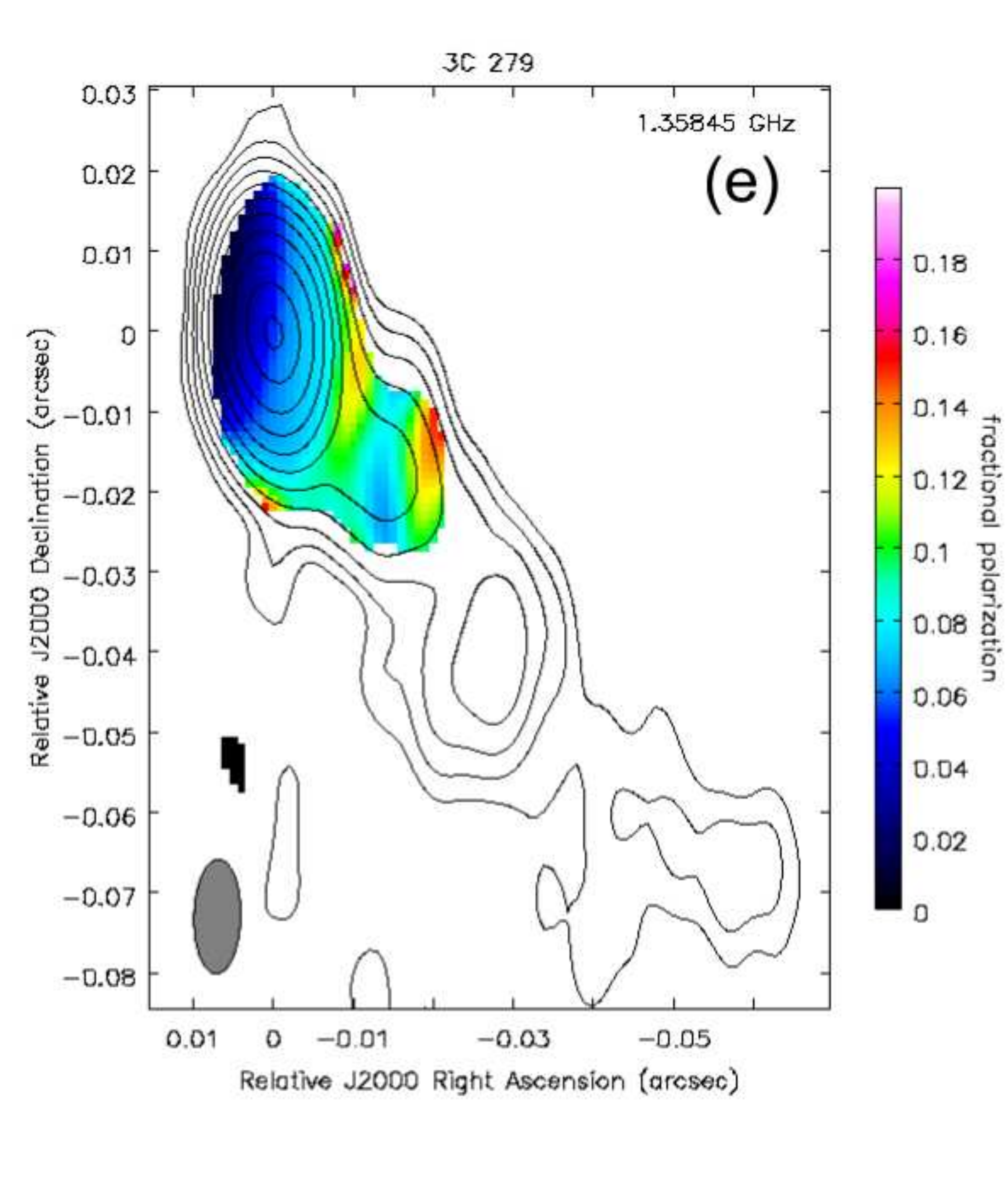}
	\end{minipage}
	\end{minipage}

	\begin{minipage}{\textwidth}
	\centering
	\begin{minipage}{.3\textwidth}
		\includegraphics[width=\textwidth]{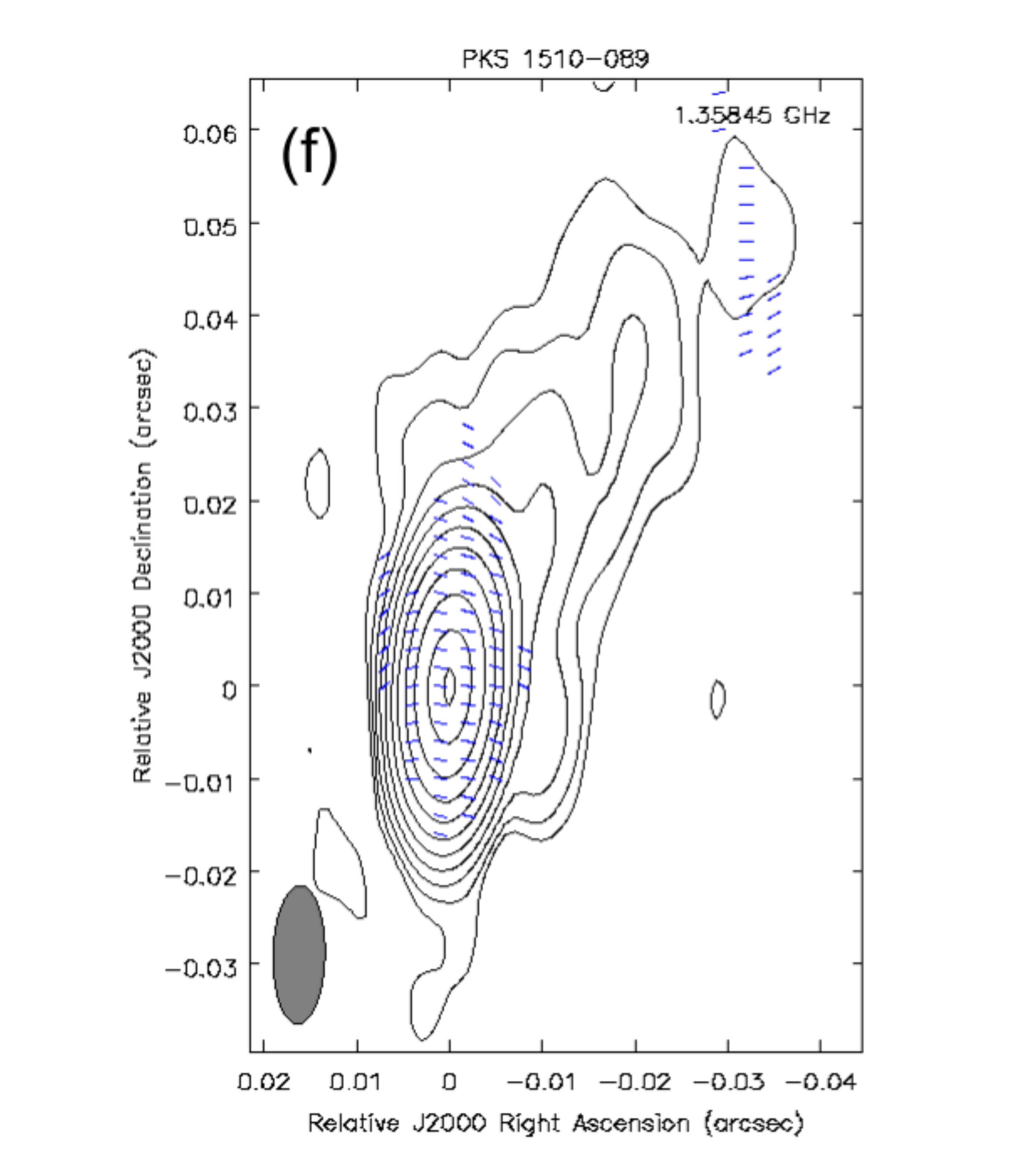}
	\end{minipage}
	\quad
	\begin{minipage}{.3\textwidth}
		\includegraphics[width=\textwidth]{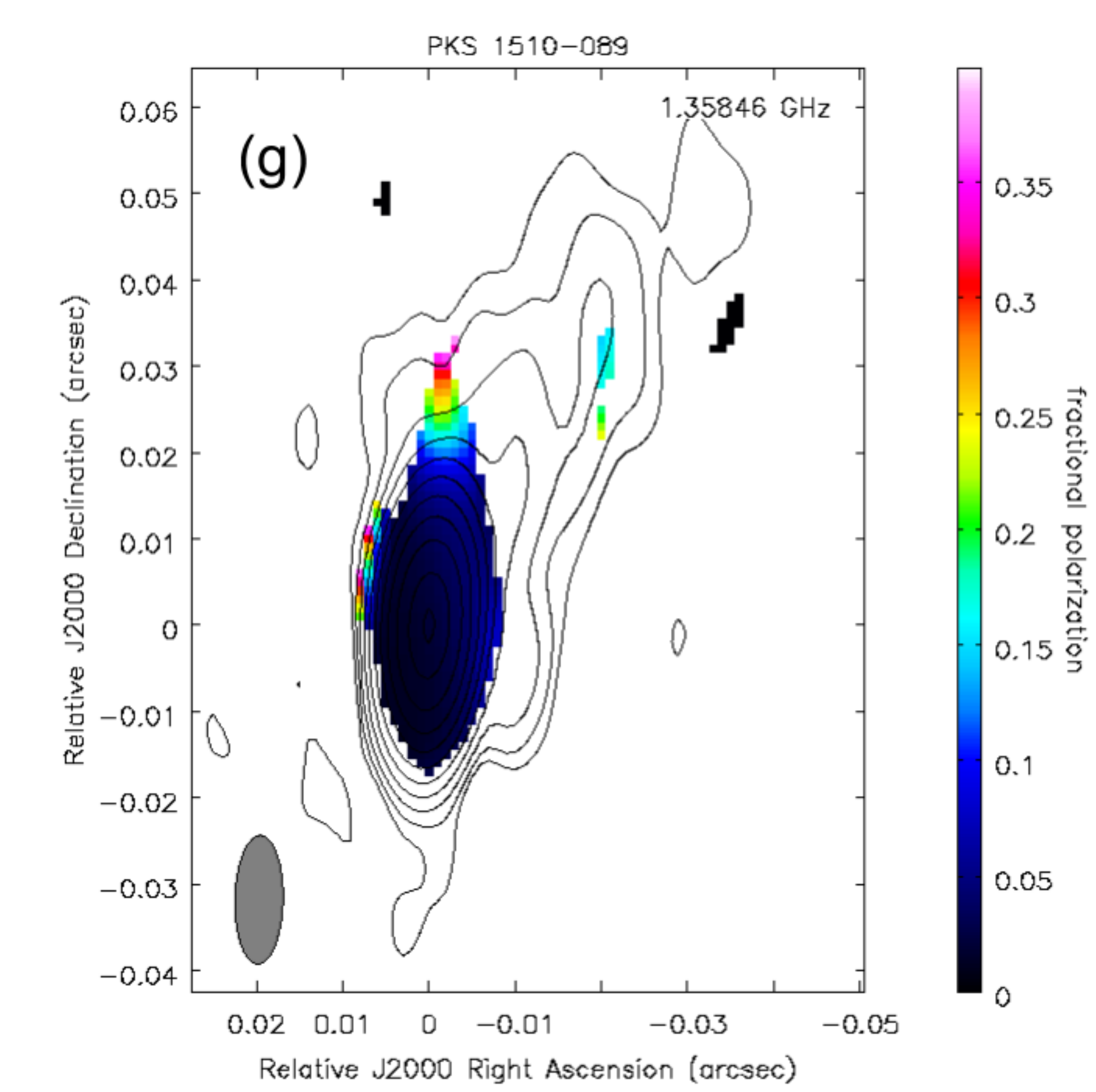}
	\end{minipage}
	\end{minipage}

\caption{Left: The ticks depict the EVPAs corrected for the integrated RMs of OJ~287 (top), 3C~279 (middle) and PKS~1510-089 (bottom) superimposed on the 1358 MHz total intensity contours. Middle and Right: 1358 MHz fractional polarization maps. The range of the degree of polarization is indicated by the colour bars.  \label{Fig1_OJ287_3C279_PKS1510}
 }
\end{center}
\end{figure*}

\begin{figure*}
\begin{center}

	\begin{minipage}{\textwidth}
	\begin{minipage}{.3\textwidth}
		\centering
		\includegraphics[width=0.9\textwidth]{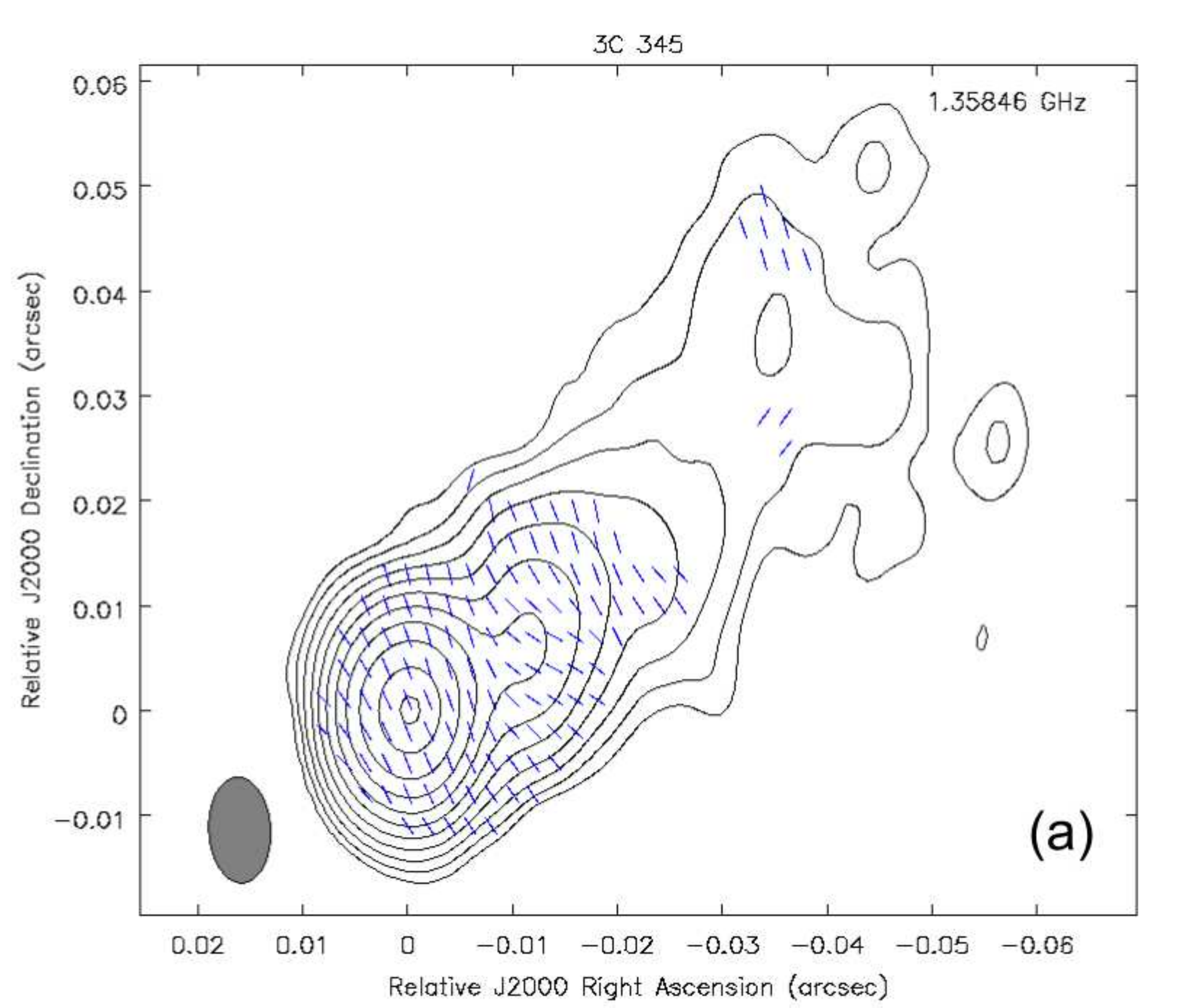}
	\end{minipage}
	\quad
	\begin{minipage}{.3\textwidth}
		\centering
		\includegraphics[width=\textwidth]{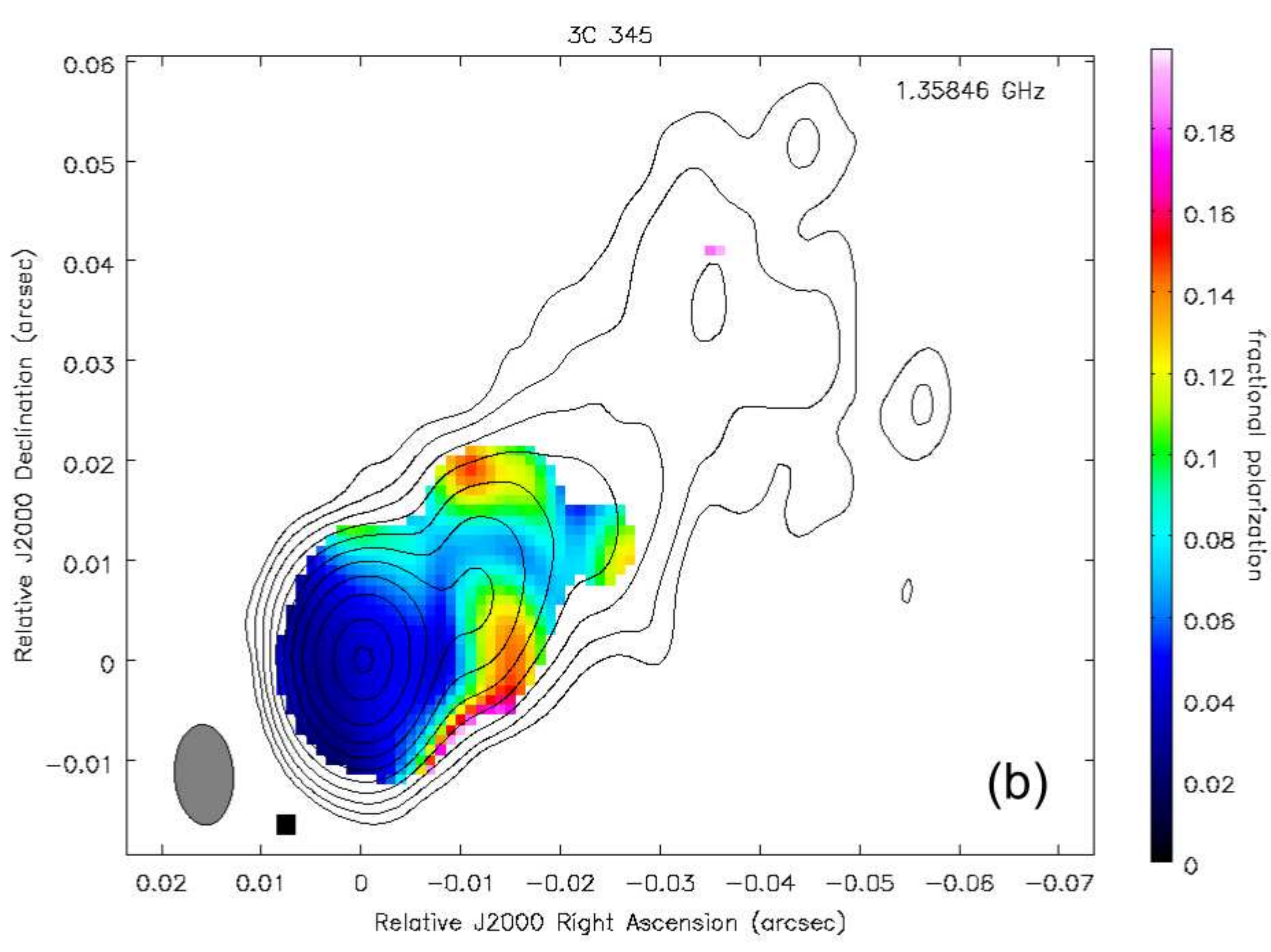}
	\end{minipage}
	\quad
	\begin{minipage}{.3\textwidth}
		\centering
		\includegraphics[width=\textwidth]{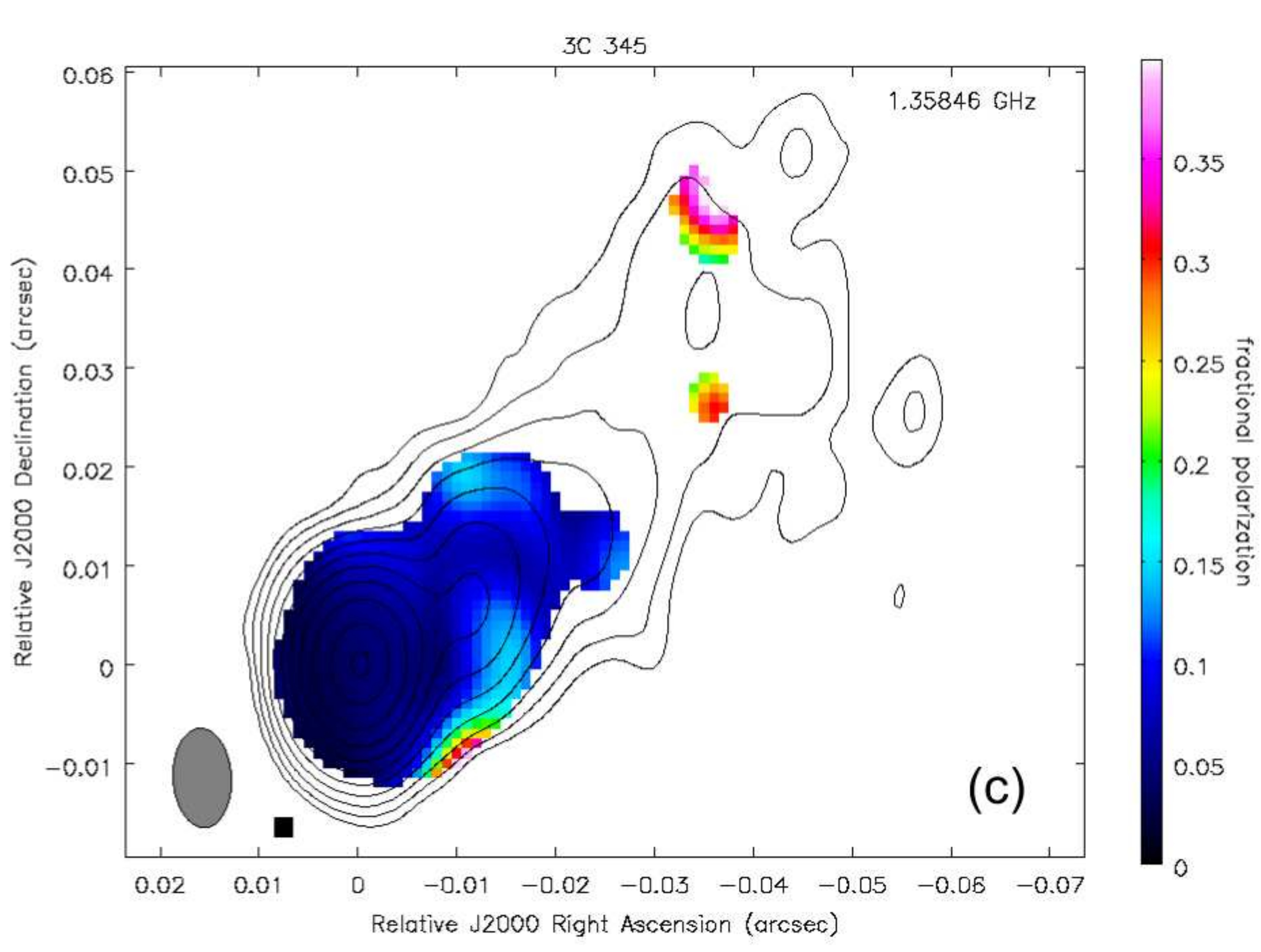}
	\end{minipage}
	\end{minipage}

	\begin{minipage}{\textwidth}
	\centering
	\begin{minipage}{.3\textwidth}
		\includegraphics[width=0.9\textwidth]{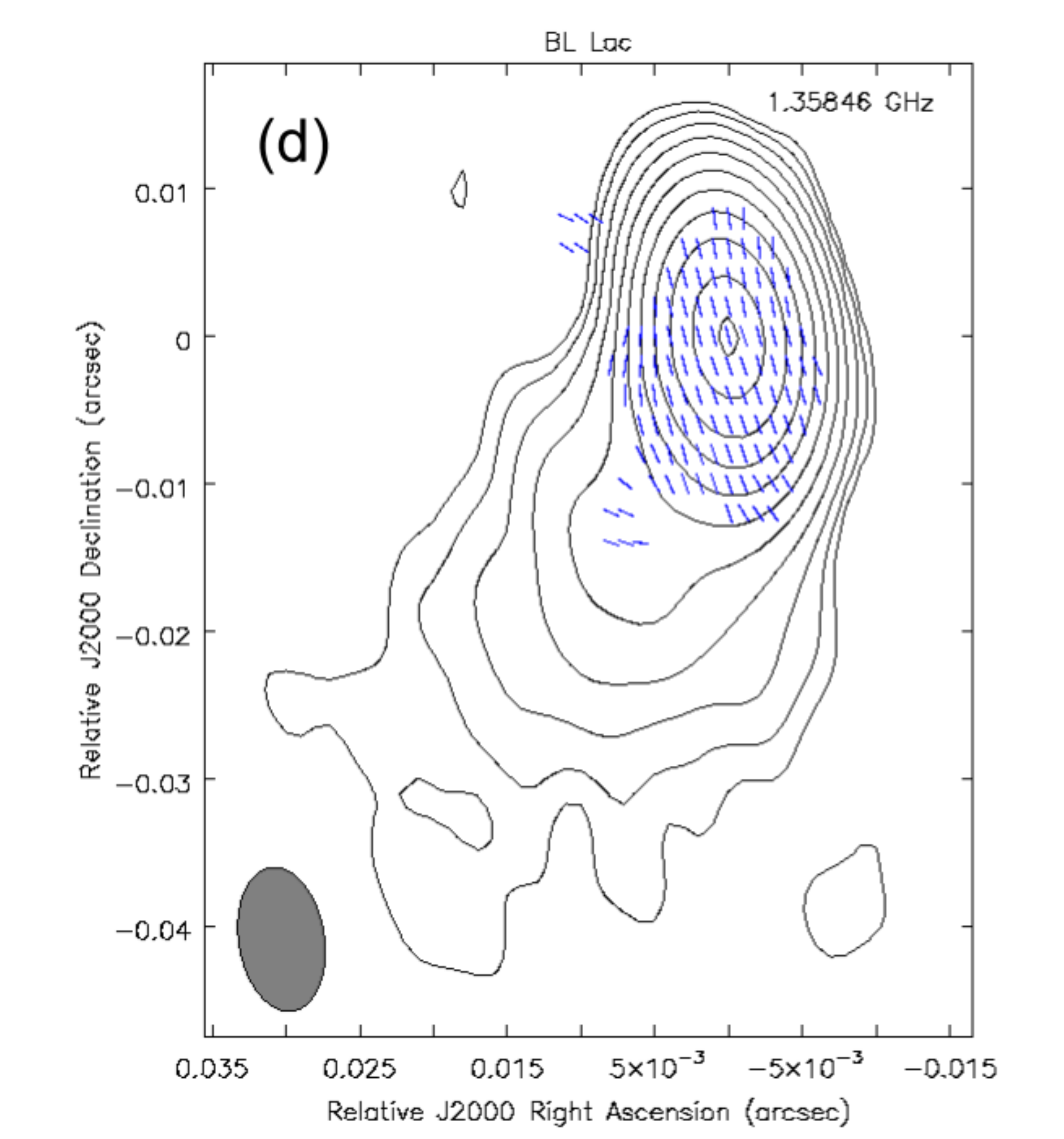}
	\end{minipage}
	\quad
	\begin{minipage}{.3\textwidth}
		\includegraphics[width=\textwidth]{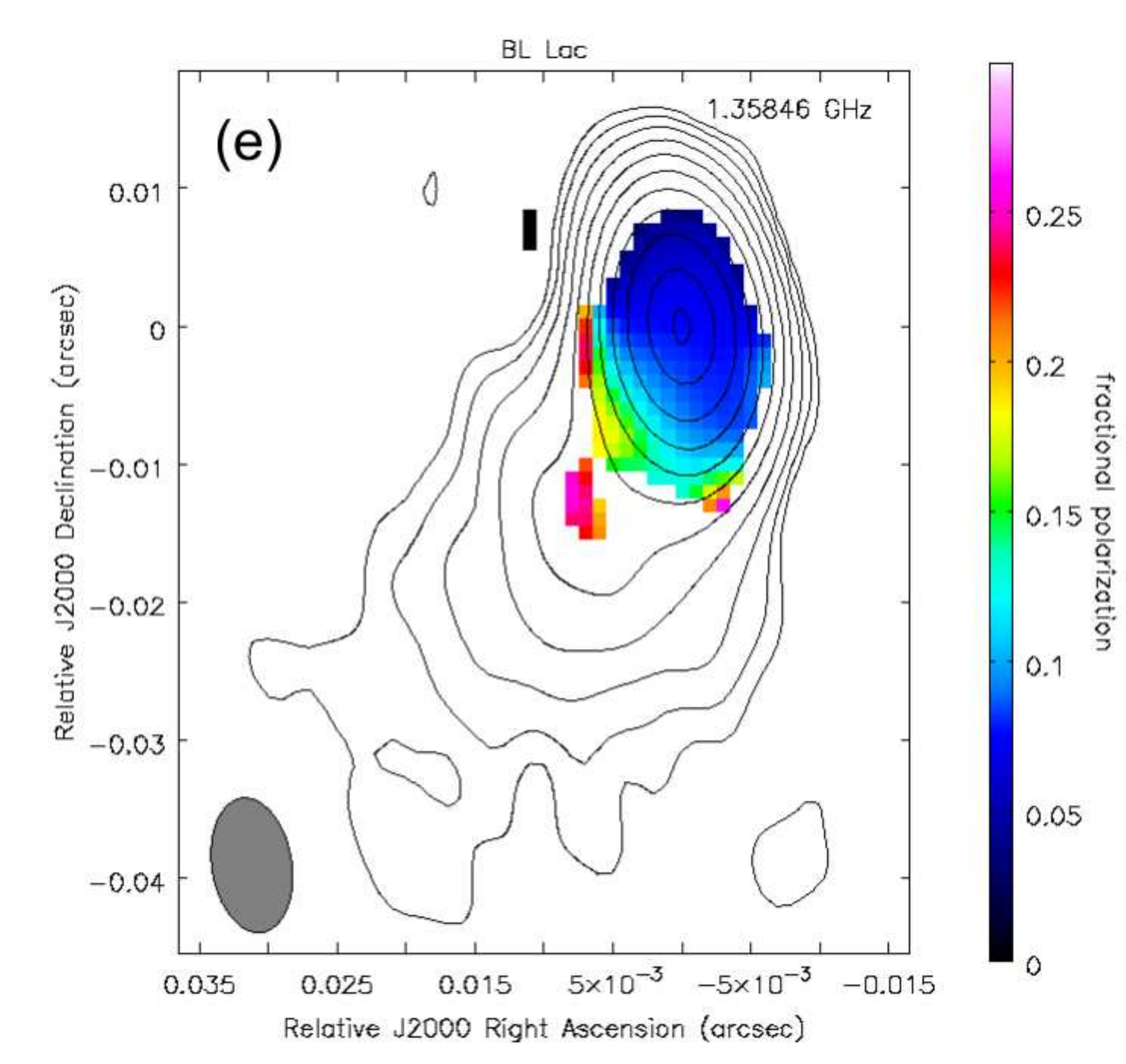}
	\end{minipage}
	\end{minipage}

	\begin{minipage}{\textwidth}
	\begin{minipage}{.3\textwidth}
		\centering
		\includegraphics[width=0.9\textwidth]{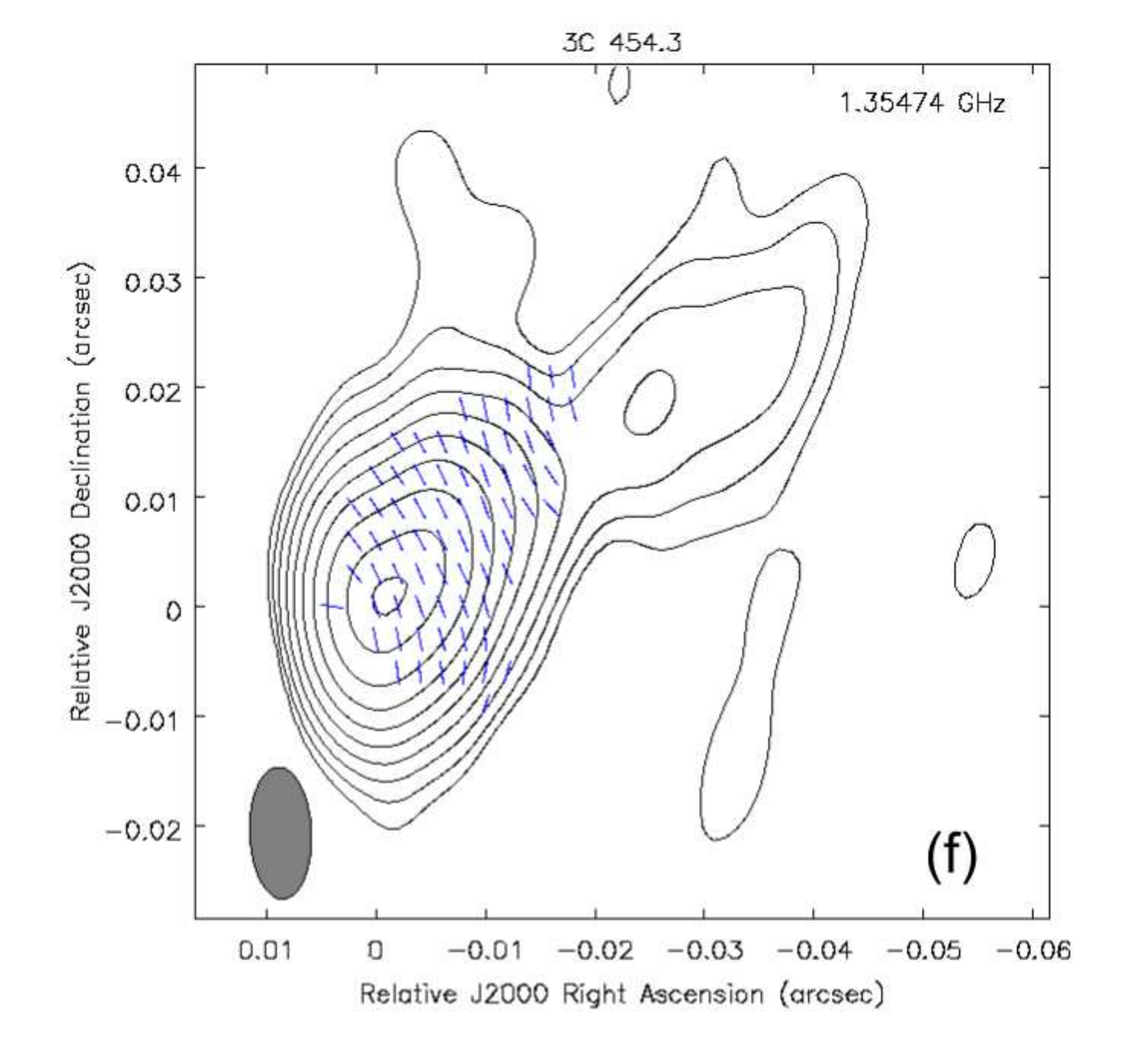}
	\end{minipage}
	\quad
	\begin{minipage}{.3\textwidth}
		\centering
		\includegraphics[width=\textwidth]{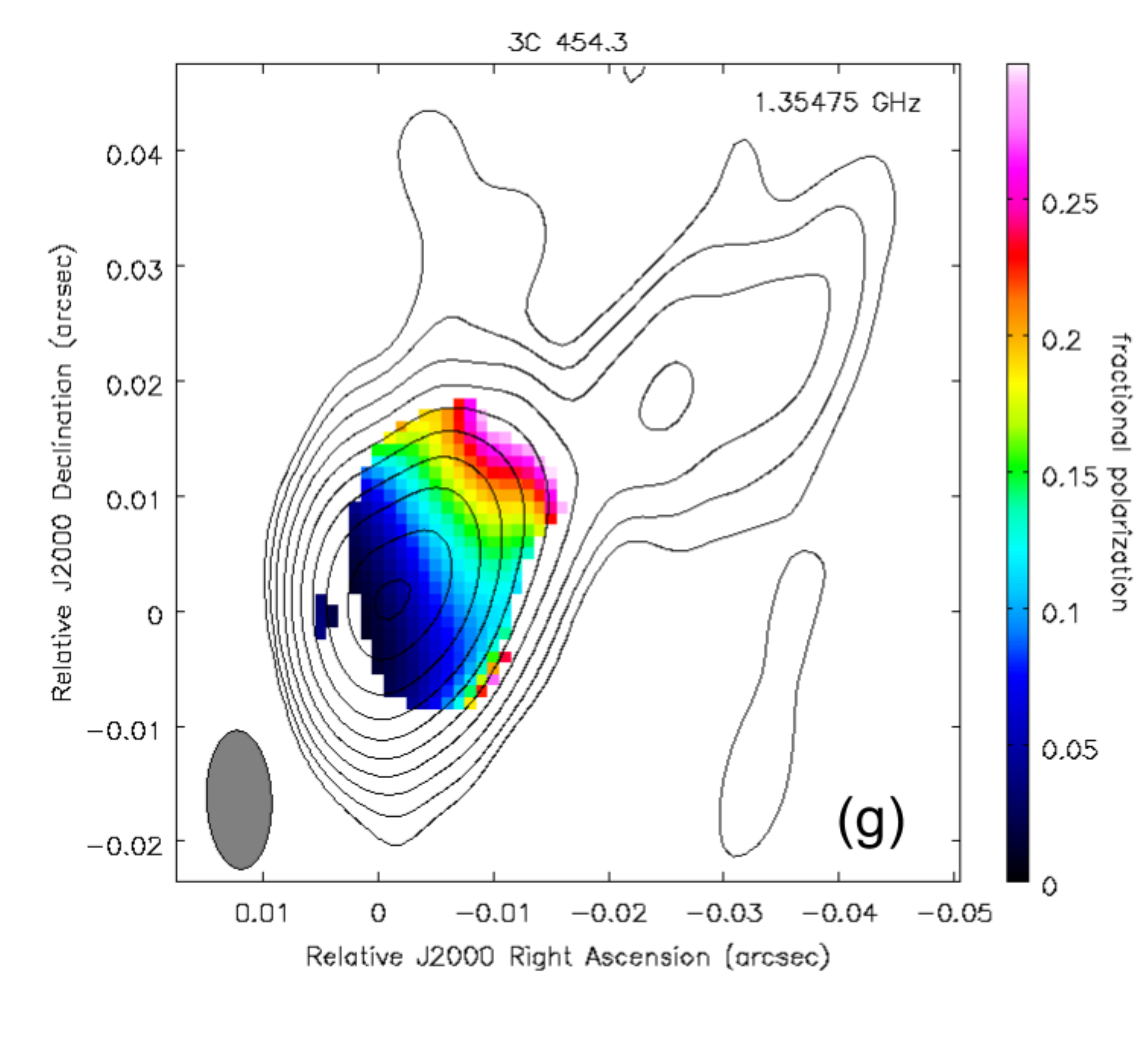}
	\end{minipage}
	\quad
	\begin{minipage}{.3\textwidth}
		\centering
		\includegraphics[width=\textwidth]{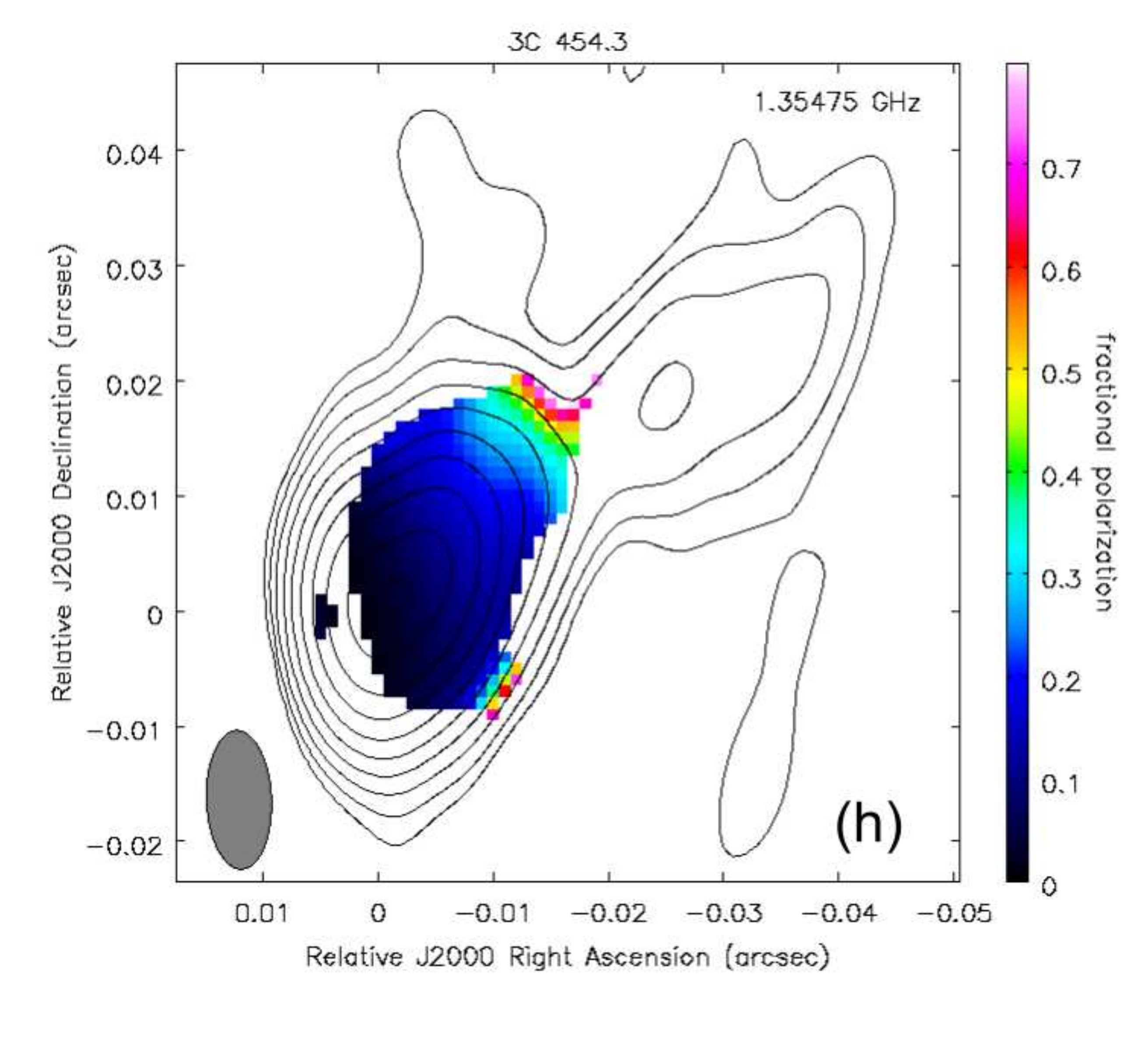}
	\end{minipage}
	\end{minipage}

\caption{Left: The ticks depict the EVPAs corrected for the integrated RMs of 3C~345 (top), BL~Lac (middle) and 3C~454.3 (bottom) superimposed on the 1358 MHz total intensity contours. Middle and Right: 1358 MHz fractional polarization maps. The range of the degree of polarization is indicated by the colour bars.  \label{Fig2_3C345_BLLac_3C454.3}
 }
\end{center}
\end{figure*}

\begin{figure*}
\begin{center}

	\begin{minipage}{\textwidth}
	\begin{minipage}{.45\textwidth}
		\centering
		\includegraphics[width=0.85\textwidth]{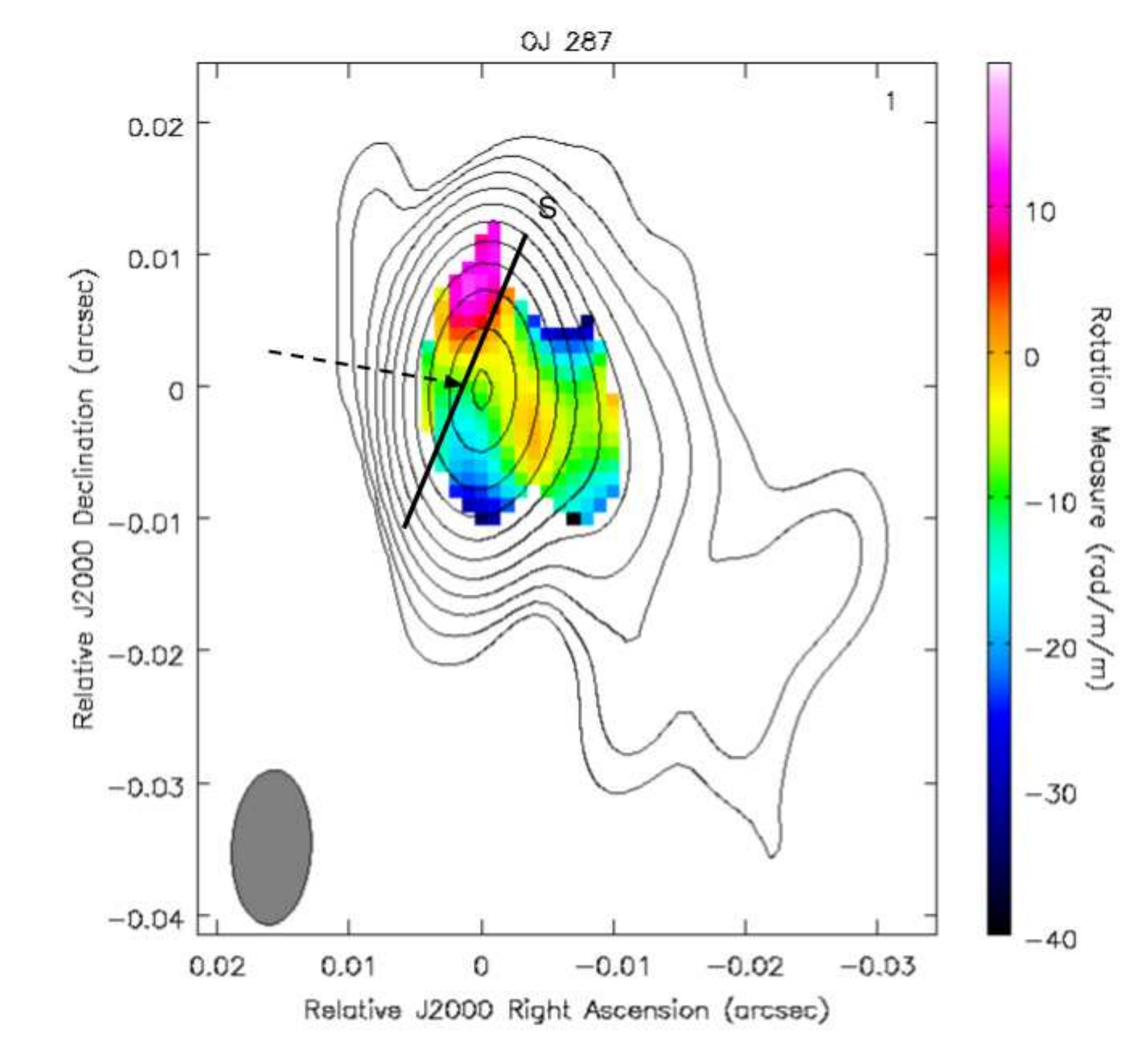}
	\end{minipage}
	\quad
	\begin{minipage}{.45\textwidth}
		\centering
		\begin{minipage}{\textwidth}
		\centering
		\includegraphics[width=0.8\textwidth]{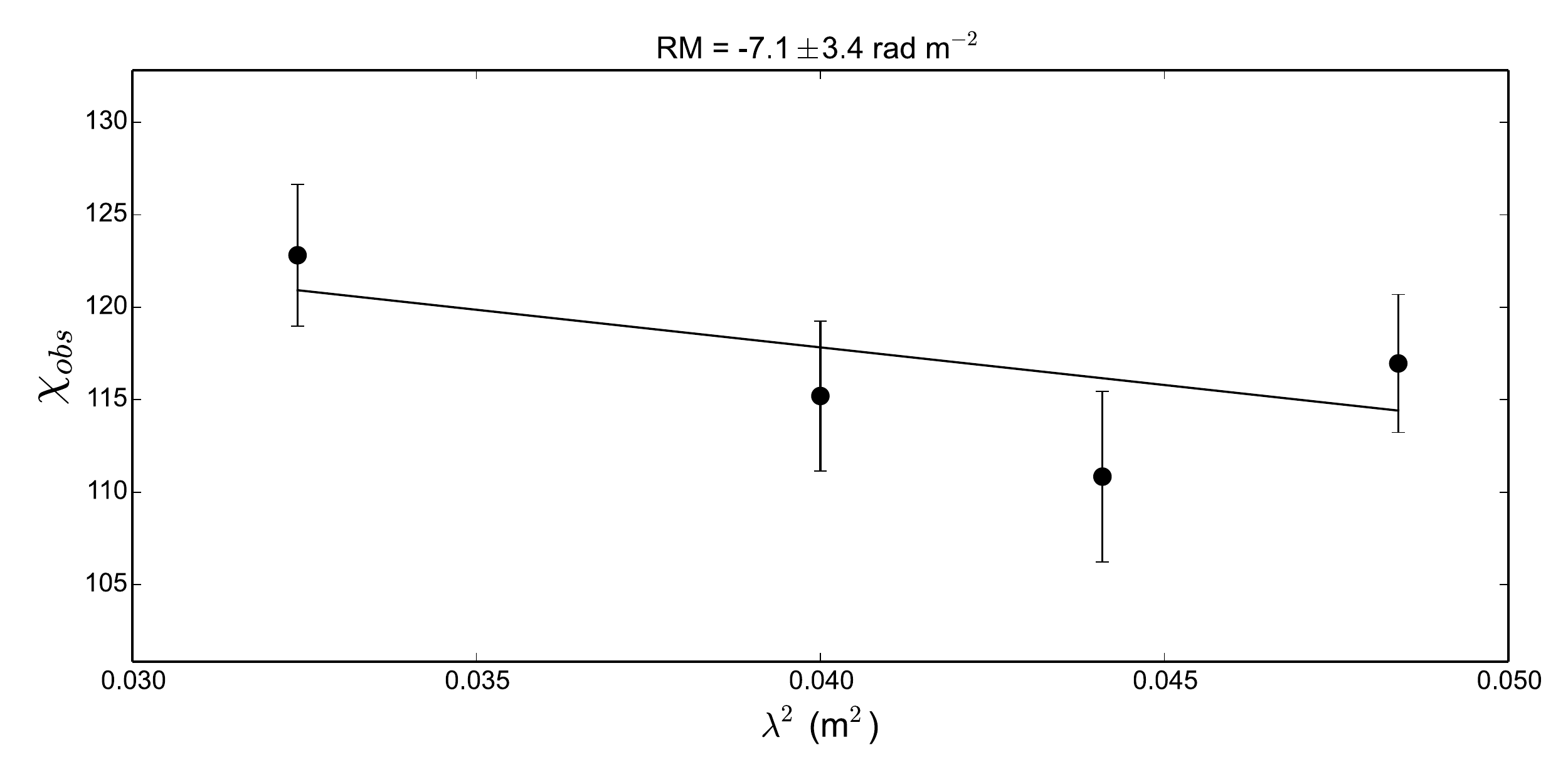}		
		\end{minipage}
	\end{minipage}
	\end{minipage}

	\begin{minipage}{\textwidth}
	\begin{minipage}{.45\textwidth}
		\centering
		\includegraphics[width=0.8\textwidth]{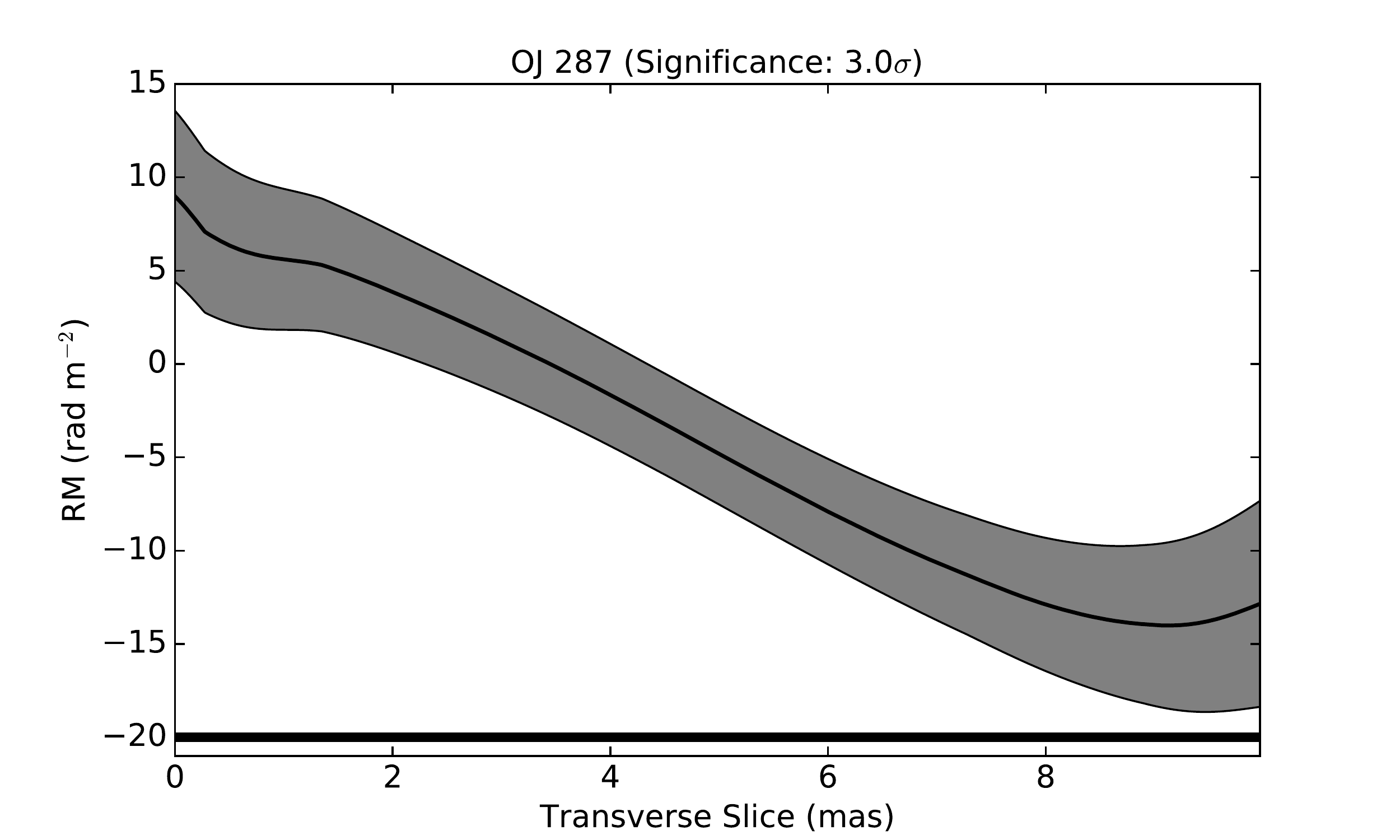}
	\end{minipage}
	\quad
	\begin{minipage}{.45\textwidth}
		\centering
		\includegraphics[width=0.8\textwidth]{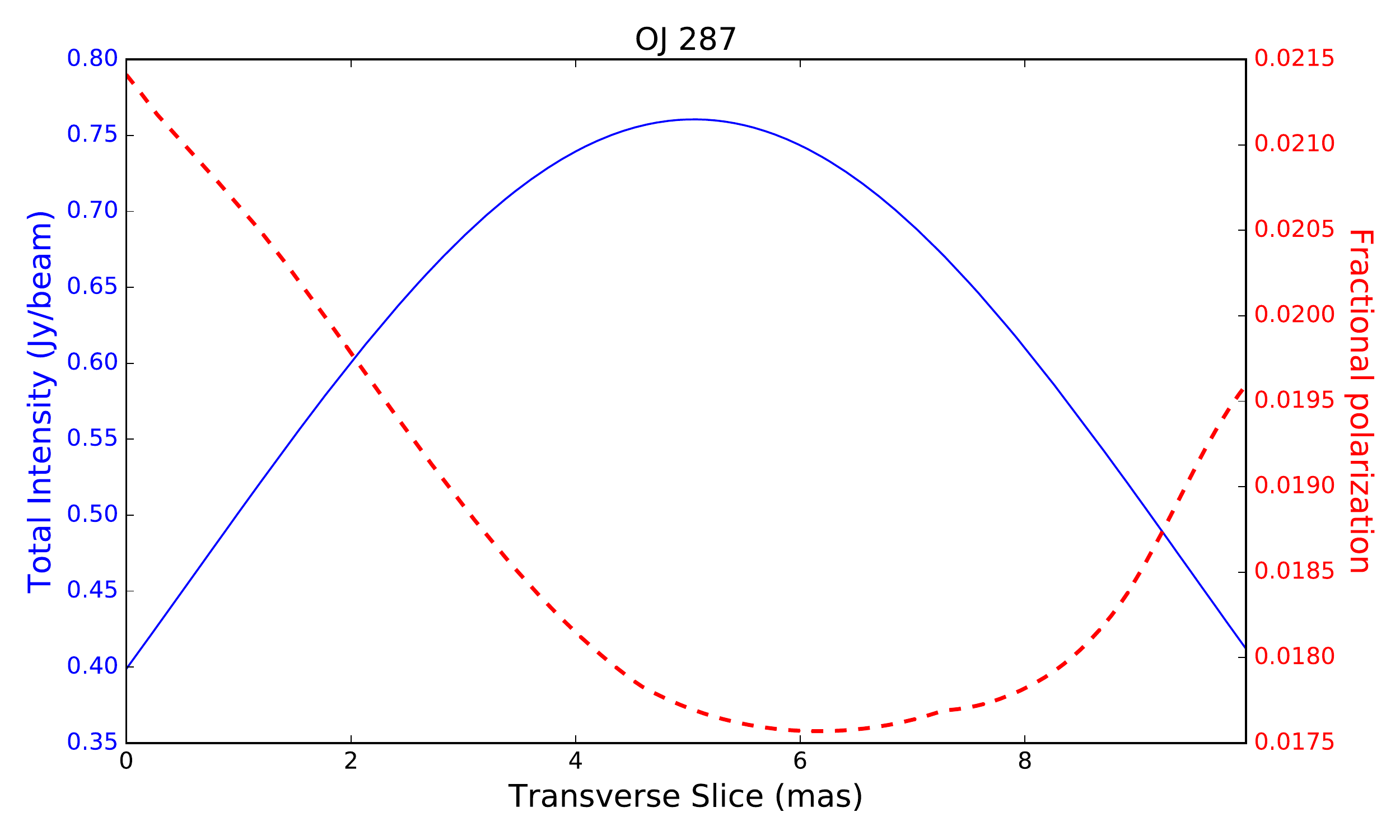}
	\end{minipage}
	\end{minipage}

	\begin{minipage}{\textwidth}
	\begin{minipage}{.45\textwidth}
		\centering
		\includegraphics[width=0.85\textwidth]{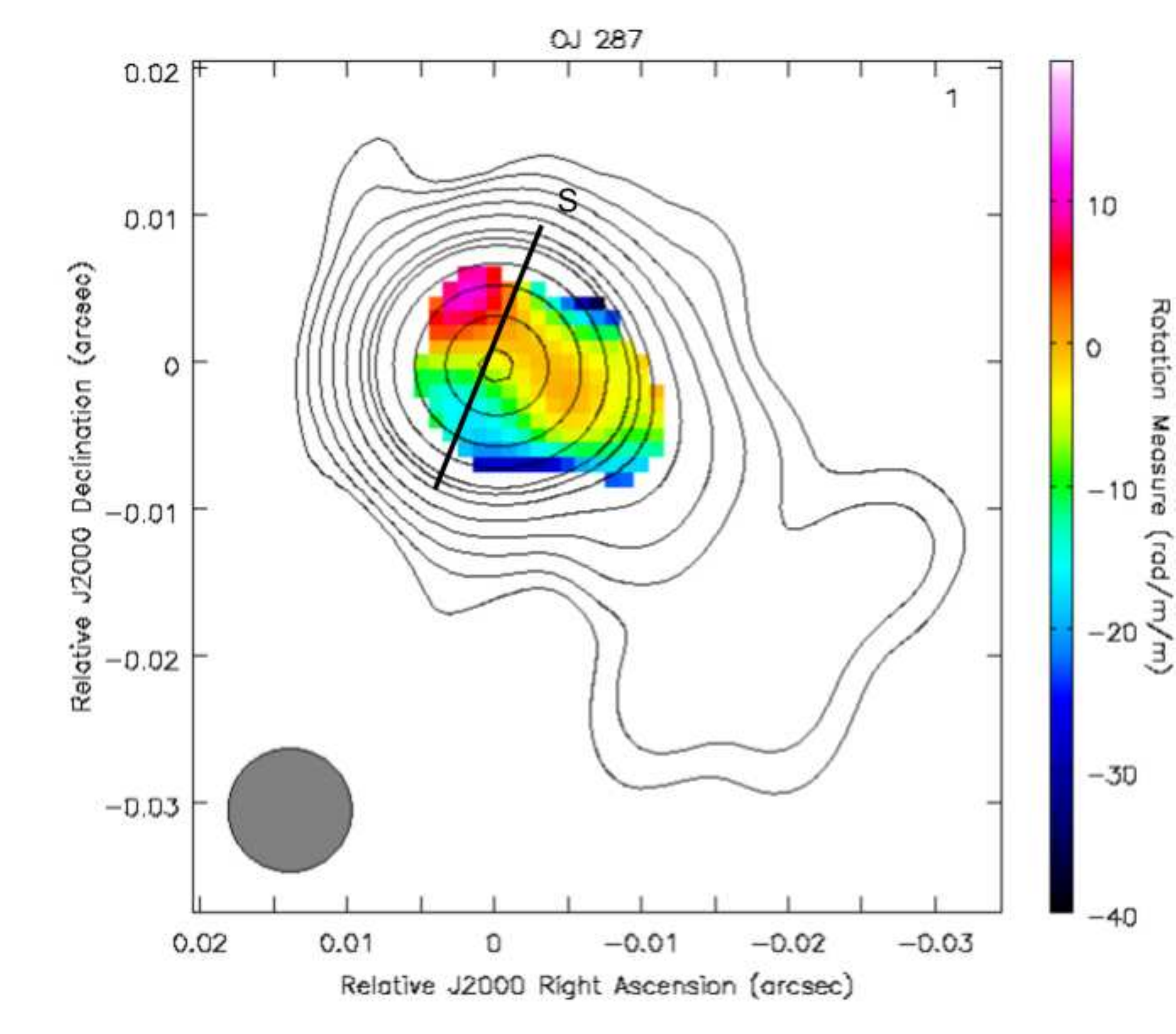}
	\end{minipage}
	\quad
	\begin{minipage}{.45\textwidth}
		\centering
		\includegraphics[width=0.8\textwidth]{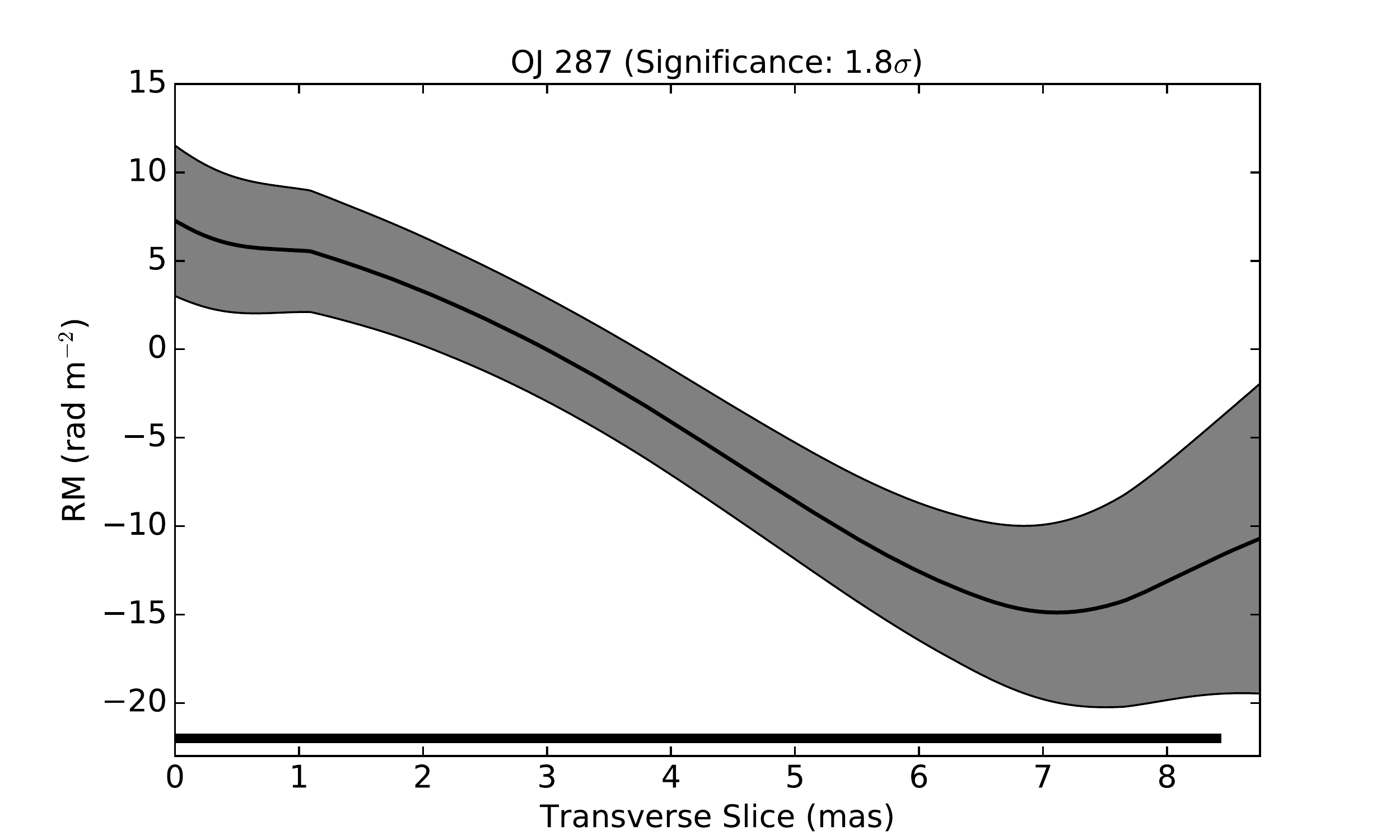}
	\end{minipage}
	\end{minipage}

\caption{RM distribution for OJ~287 superimposed on the 1358 MHz $I$ map for the intrinsic elliptical beam (top-left) and a circular beam of equivalent area (bottom-left) along with slices taken in a region where a transverse RM gradient is visible by eye, shown by the continuous black lines across the RM maps (middle-left and bottom-right). The ranges of the RM values are indicated by the colour bars. Output pixels were blanked for RM uncertainties exceeding 10 rad m$^{-2}$. The thick black horizontal lines  accompanying the transverse RM profiles indicate the projected size of the beam in the slice direction. An example of a $\chi_{obs}$ versus $\lambda^2$ fit in the region of the slice is shown in the top-right panel. The location of the pixel is indicated by the arrow in the RM map. Total intensity (solid line) and fractional polarization (dashed line) profiles at the same location of the transverse RM gradient are shown in the middle-right panel.  \label{OJ287_RM}
 }
\end{center}
\end{figure*}

\begin{figure*}
\begin{center}

	\begin{minipage}{\textwidth}
	\begin{minipage}{.45\textwidth}
		\centering
		\includegraphics[width=0.85\textwidth]{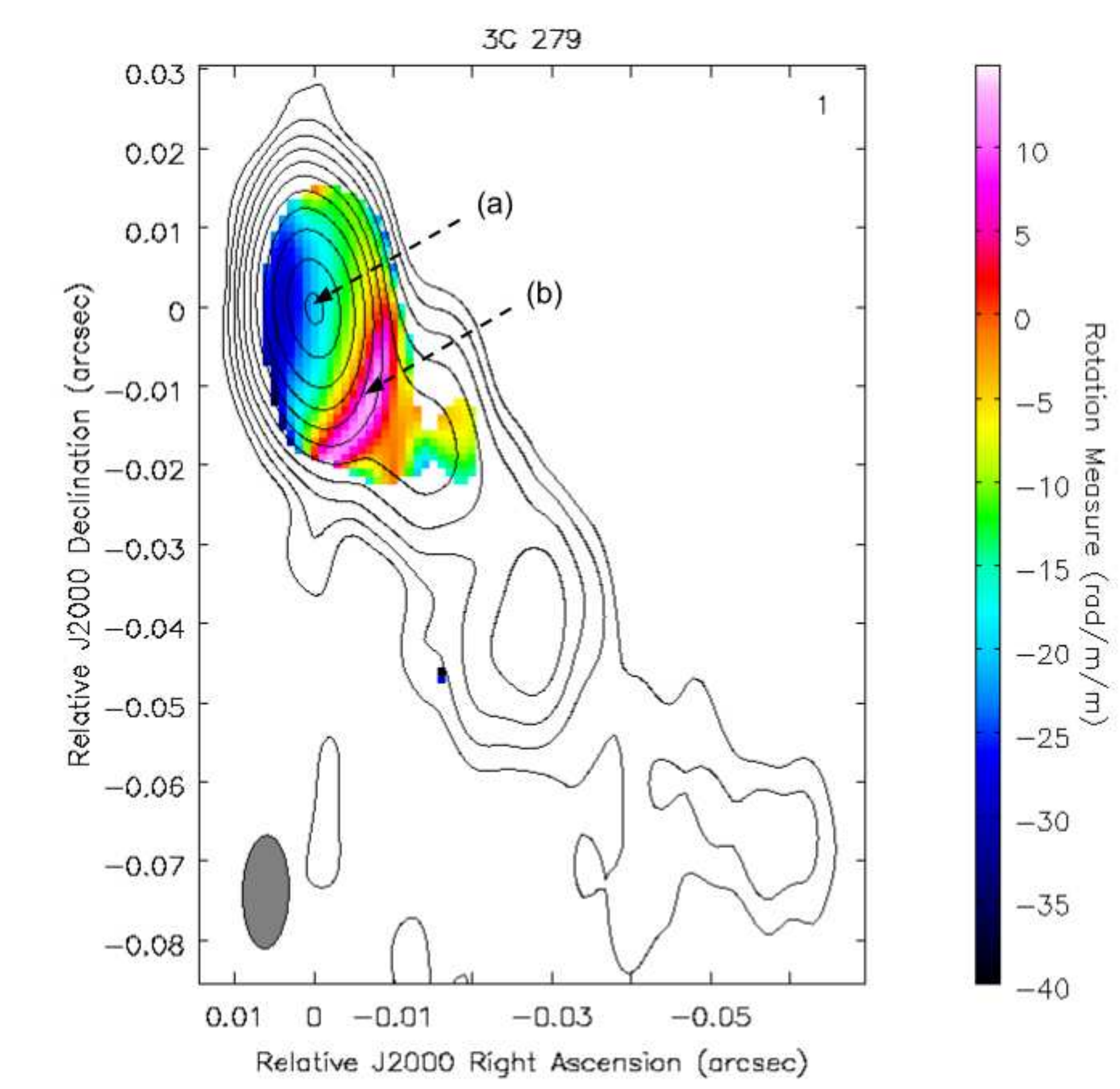}
	\end{minipage}
	\quad
	\begin{minipage}{.45\textwidth}
		\centering
		\includegraphics[width=0.8\textwidth]{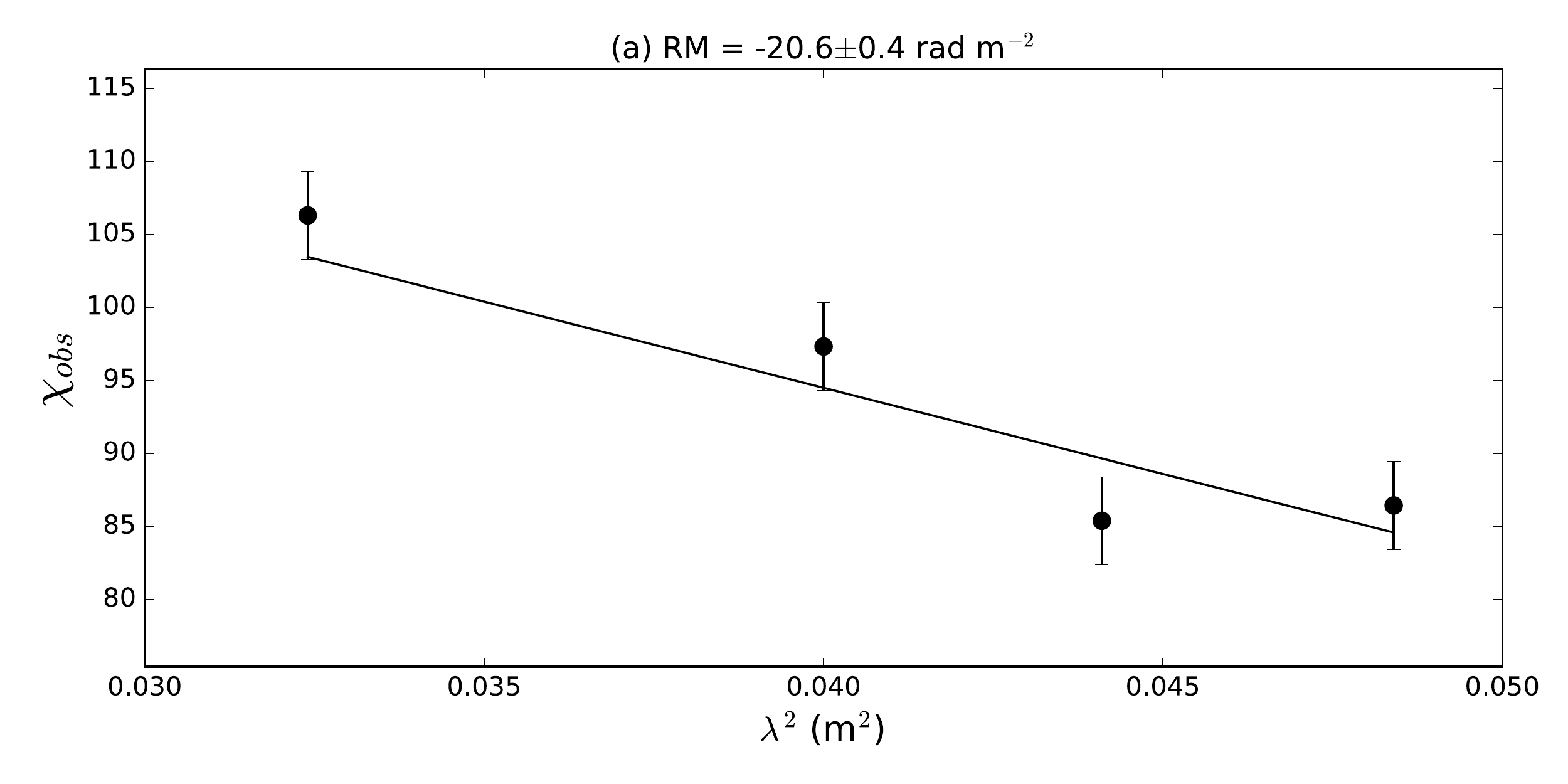}
		\includegraphics[width=0.8\textwidth]{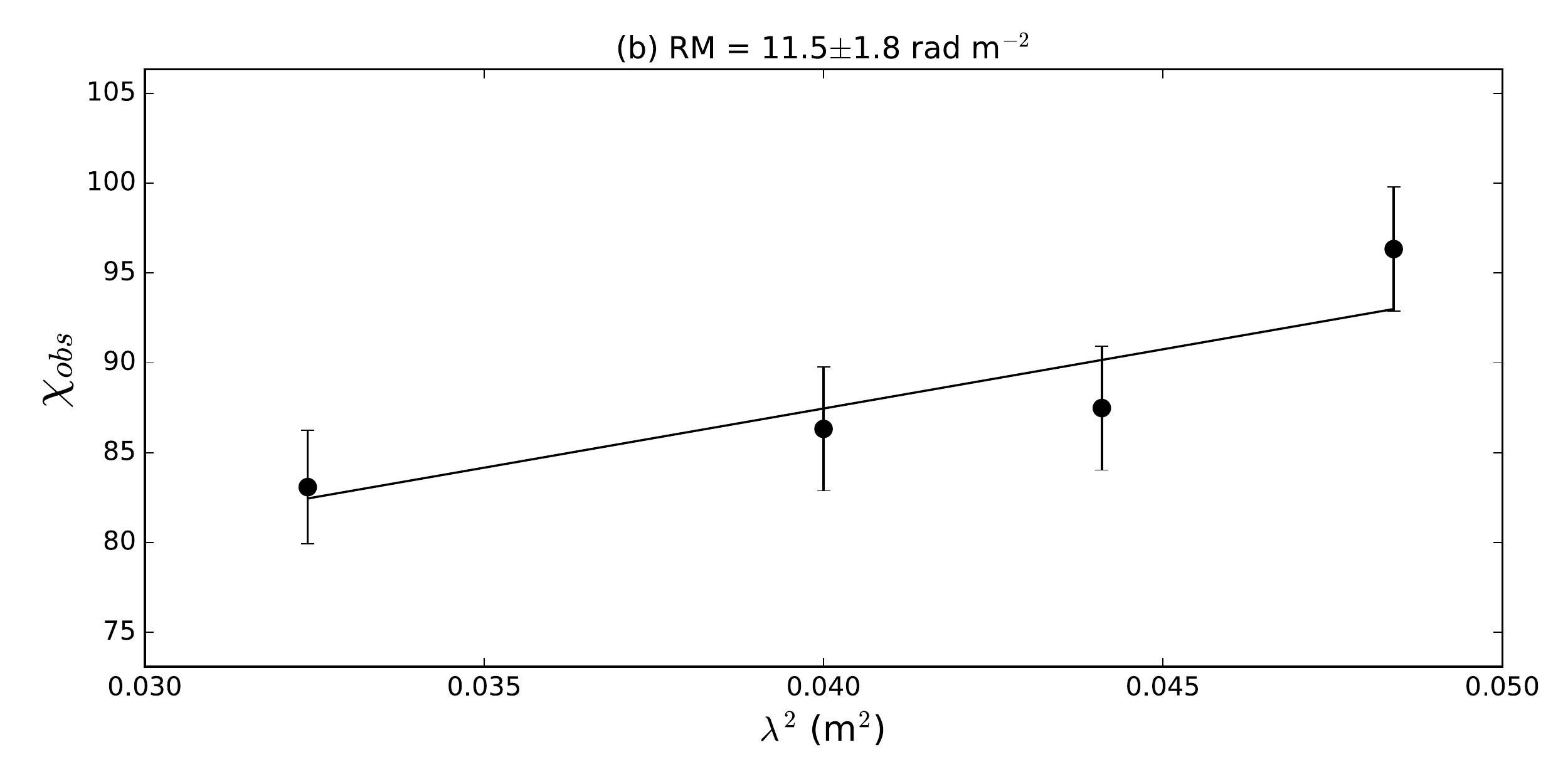}
	\end{minipage}
	\end{minipage}

	\begin{minipage}{\textwidth}
	\centering
	\begin{minipage}{.45\textwidth}
		\includegraphics[width=0.85\textwidth]{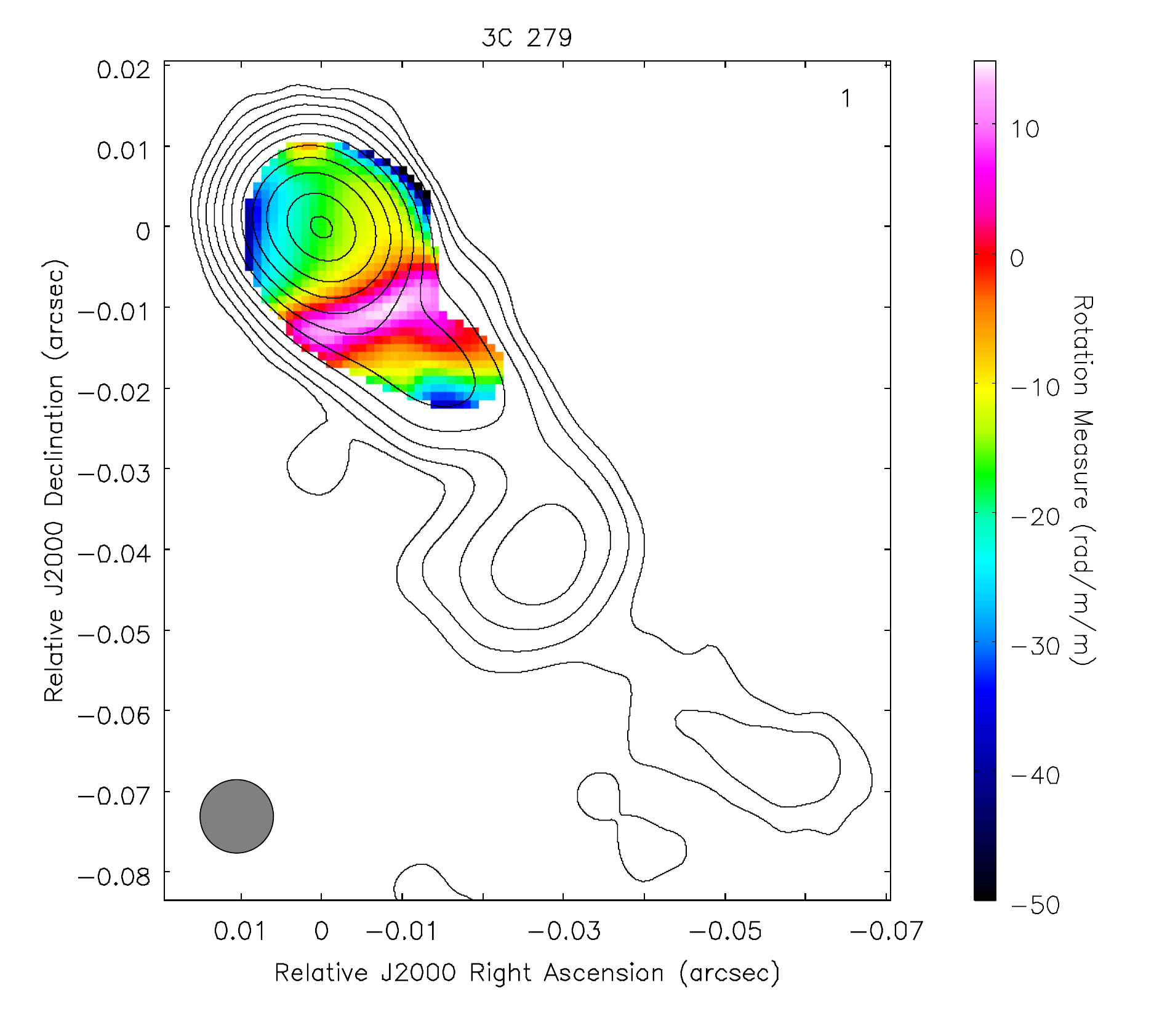}
	\end{minipage}
	\end{minipage}

\caption{RM distribution for 3C~279 superimposed on the 1358 MHz $I$ map for the intrinsic elliptical beam (top-left) and a circular beam of equivalent area (bottom). The ranges of the RM values are indicated by the colour bars. Output pixels were blanked for RM uncertainties exceeding 10 rad m$^{-2}$. Examples of the $\chi_{obs}$ versus $\lambda^2$ fits in the core and inner jet regions are shown in the two right panels. The locations of the pixels are indicated by the arrows in the RM map. \label{3C279_RM}
 }
\end{center}
\end{figure*}

\begin{figure*}
\begin{center}

	\begin{minipage}{\textwidth}
	\begin{minipage}{.45\textwidth}
		\centering
		\includegraphics[width=0.9\textwidth]{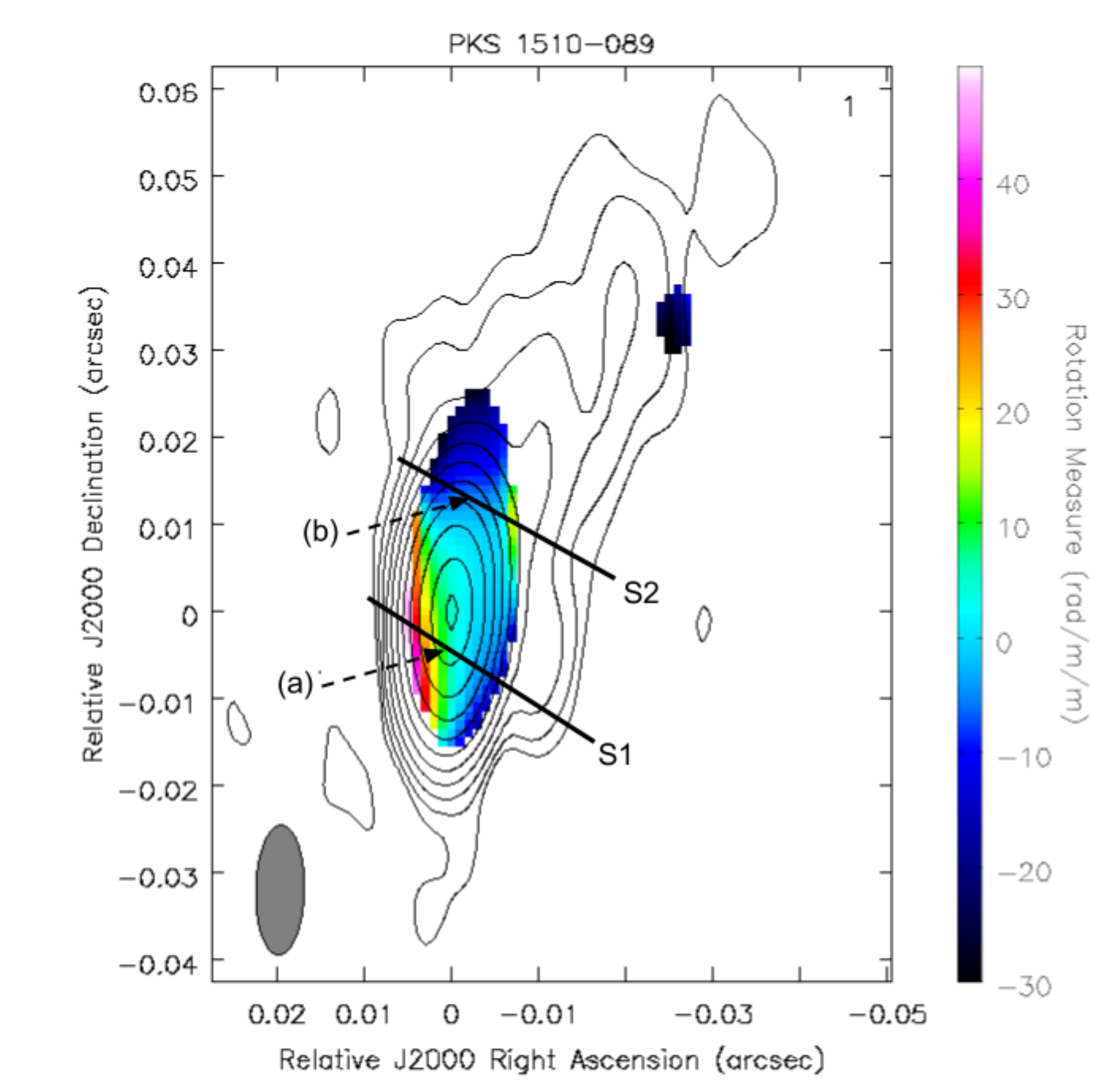}
	\end{minipage}
	\quad
	\begin{minipage}{.45\textwidth}
		\centering
		\includegraphics[width=0.8\textwidth]{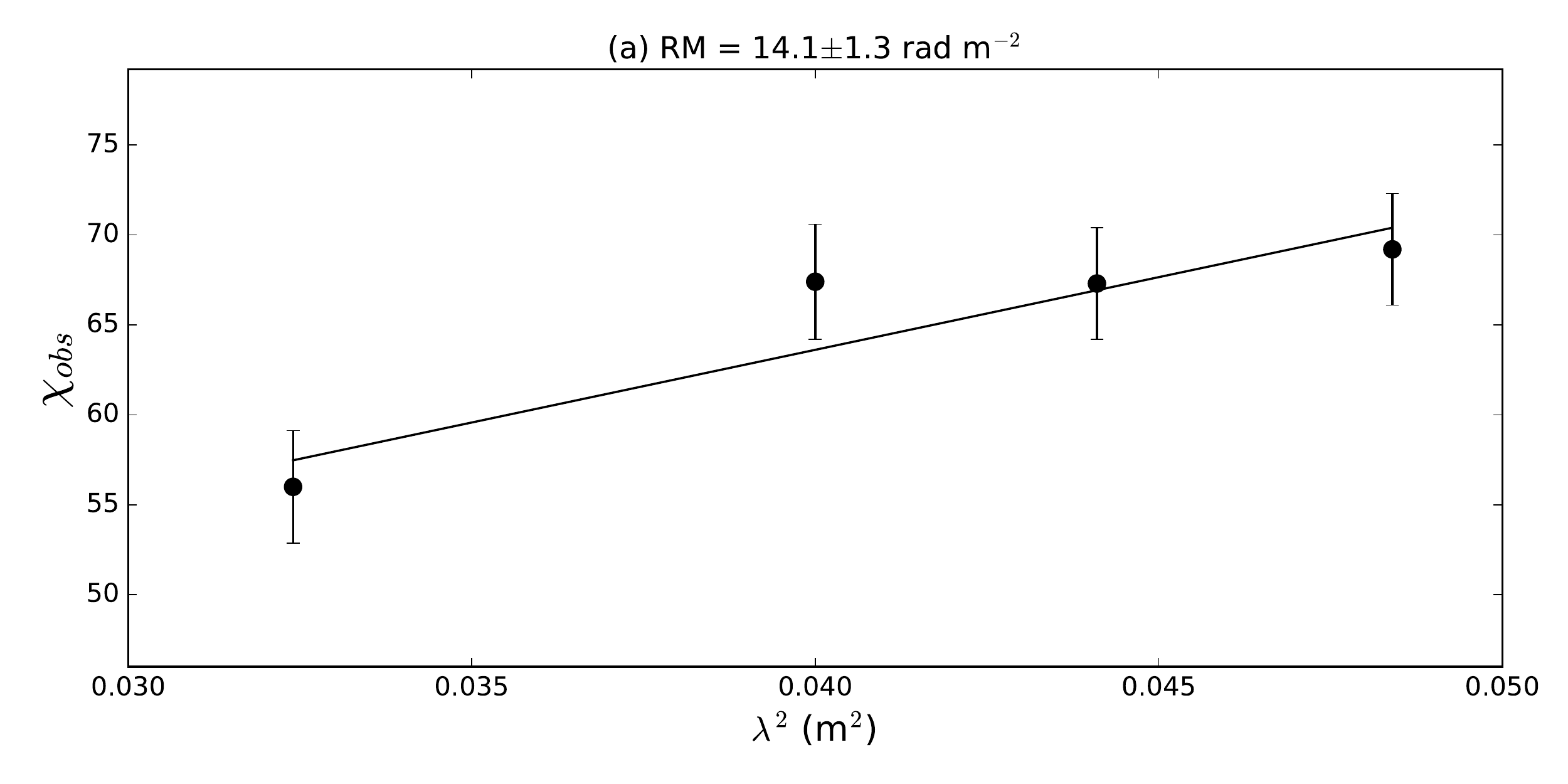}
		\includegraphics[width=0.8\textwidth]{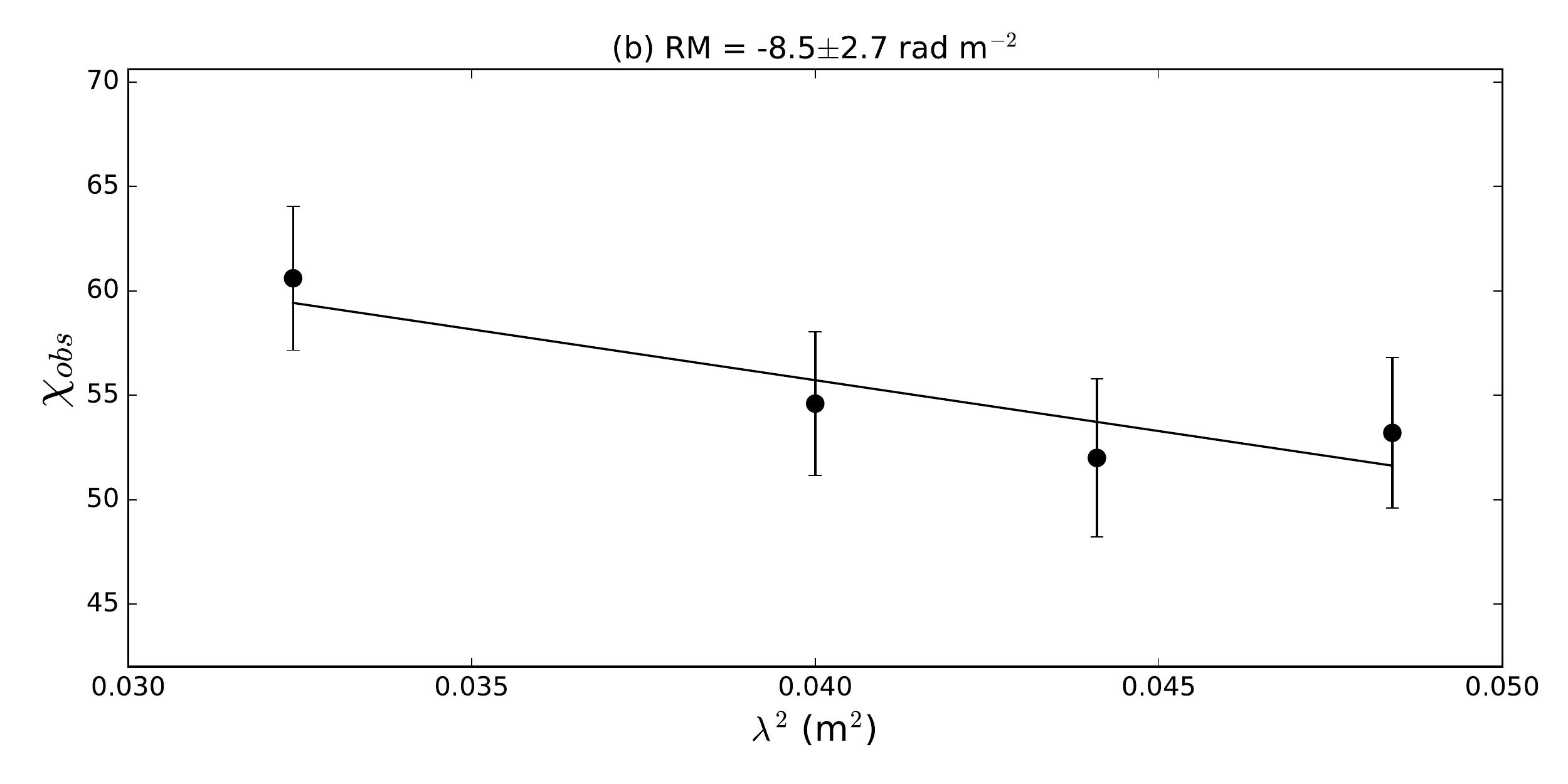}
	\end{minipage}
	\end{minipage}

	\begin{minipage}{\textwidth}
	\begin{minipage}{.45\textwidth}
		\centering
		\includegraphics[width=0.8\textwidth]{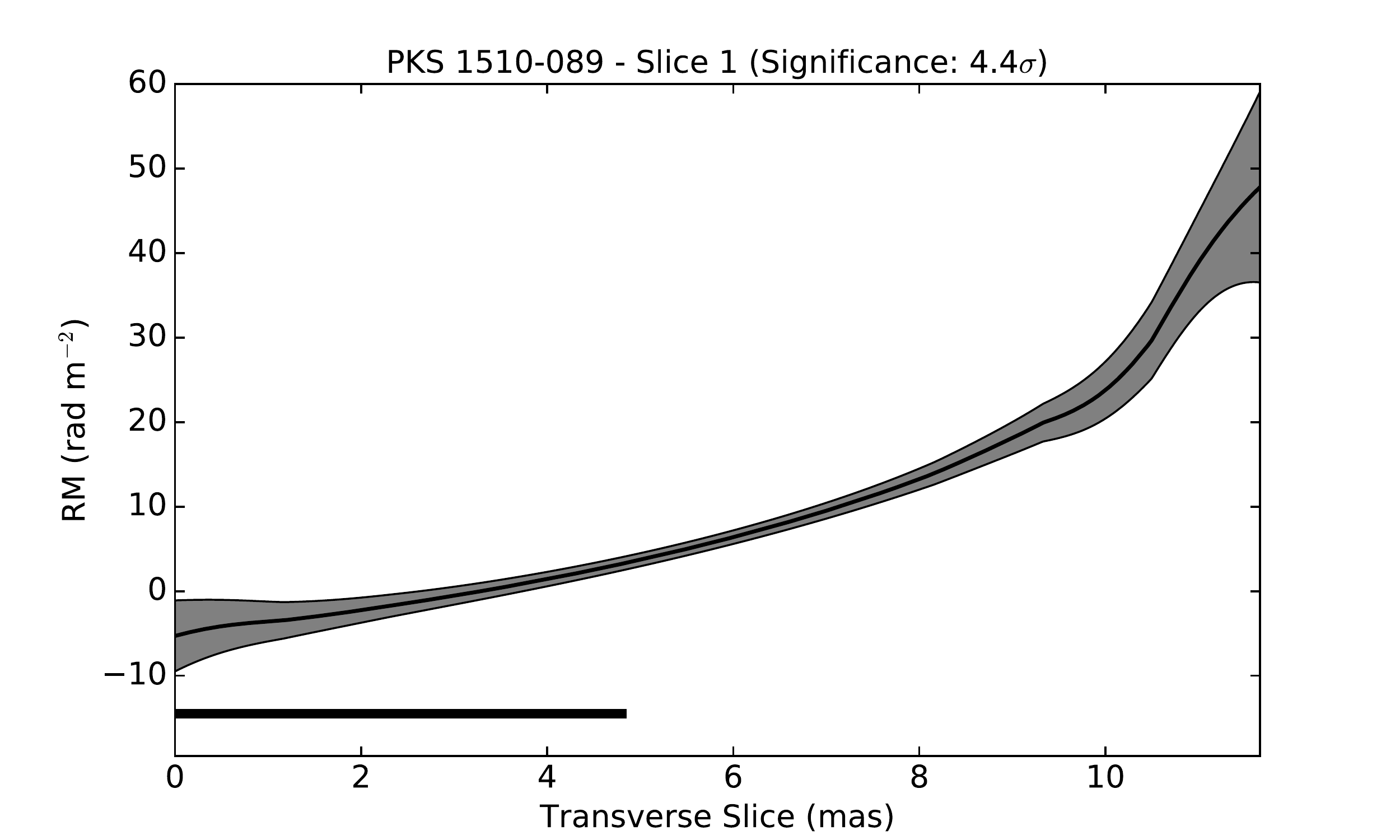}
	\end{minipage}
	\quad
	\begin{minipage}{.45\textwidth}
		\centering
		\includegraphics[width=0.8\textwidth]{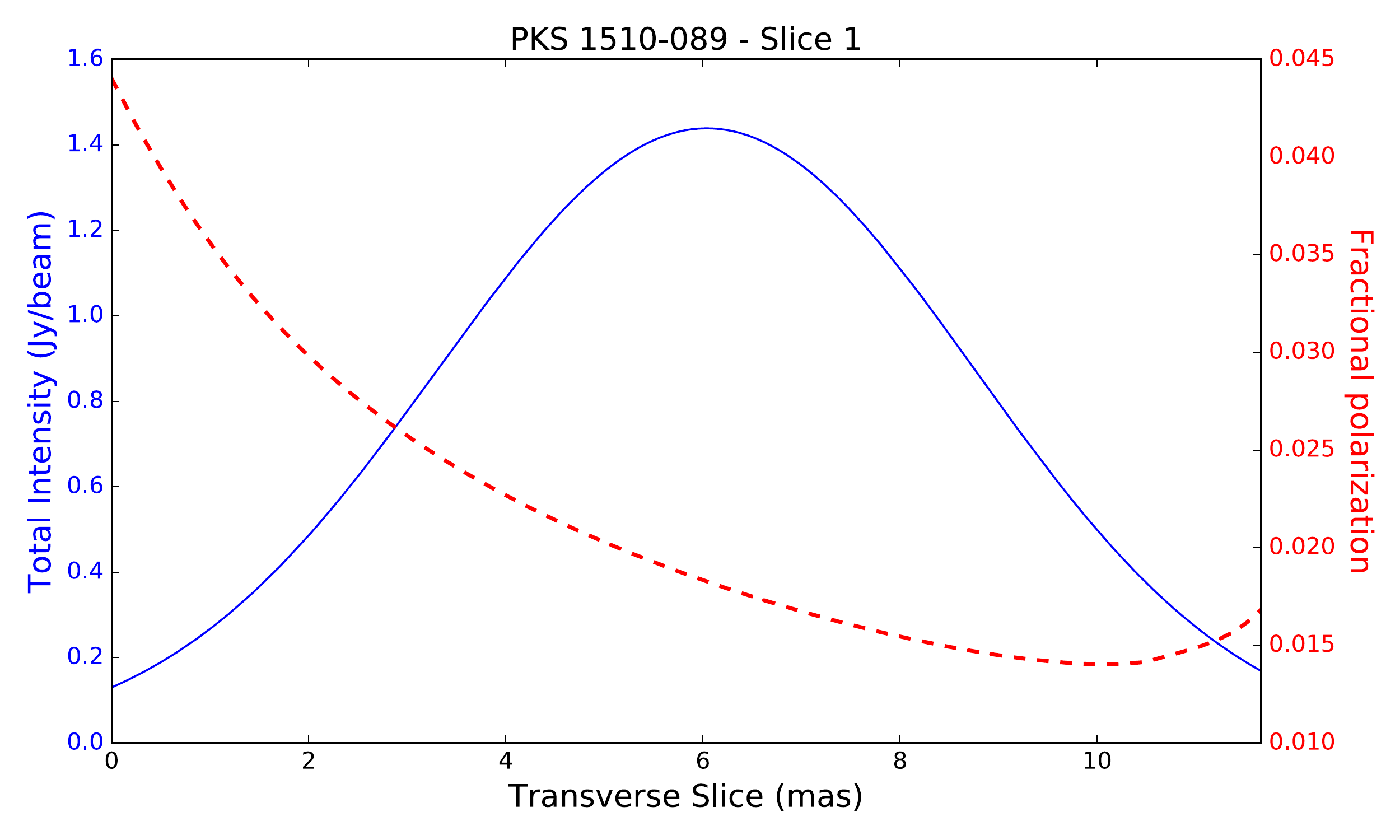}
	\end{minipage}
	\end{minipage}

	\begin{minipage}{\textwidth}
	\begin{minipage}{.45\textwidth}
		\centering
		\includegraphics[width=0.8\textwidth]{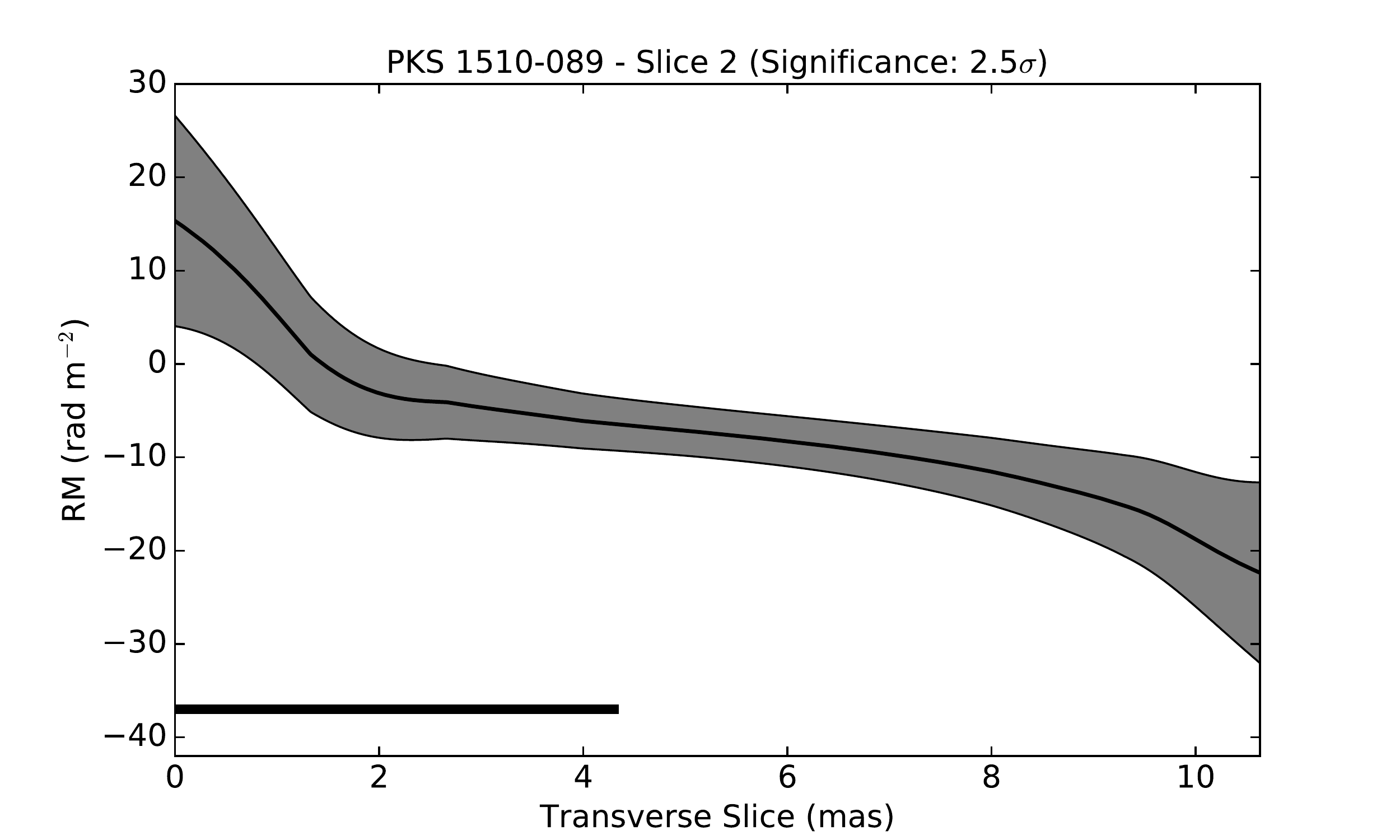}
	\end{minipage}
	\quad
	\begin{minipage}{.45\textwidth}
		\centering
		\includegraphics[width=0.8\textwidth]{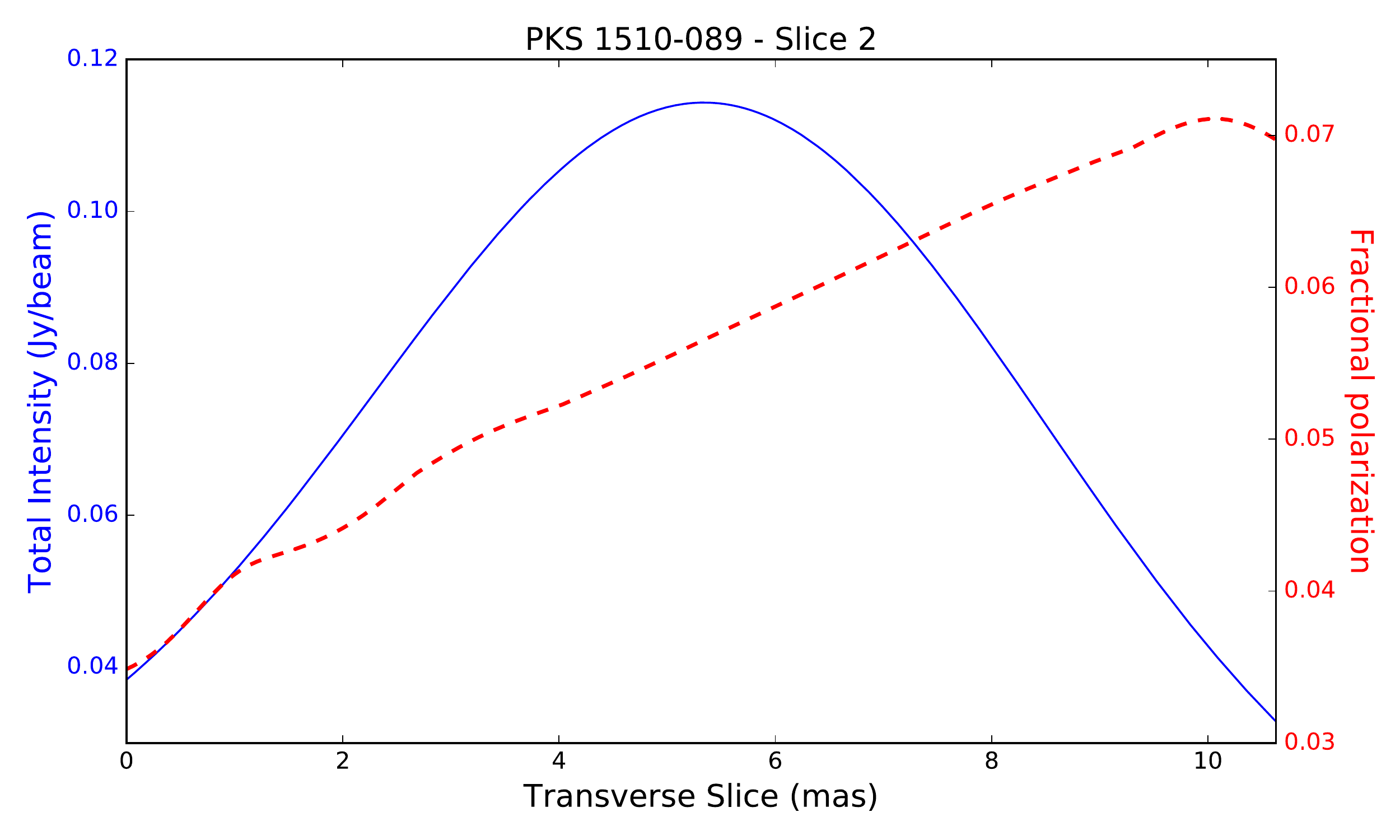}
	\end{minipage}
	\end{minipage}

	\begin{minipage}{\textwidth}
	\begin{minipage}{.45\textwidth}
		\centering
		\includegraphics[width=0.9\textwidth]{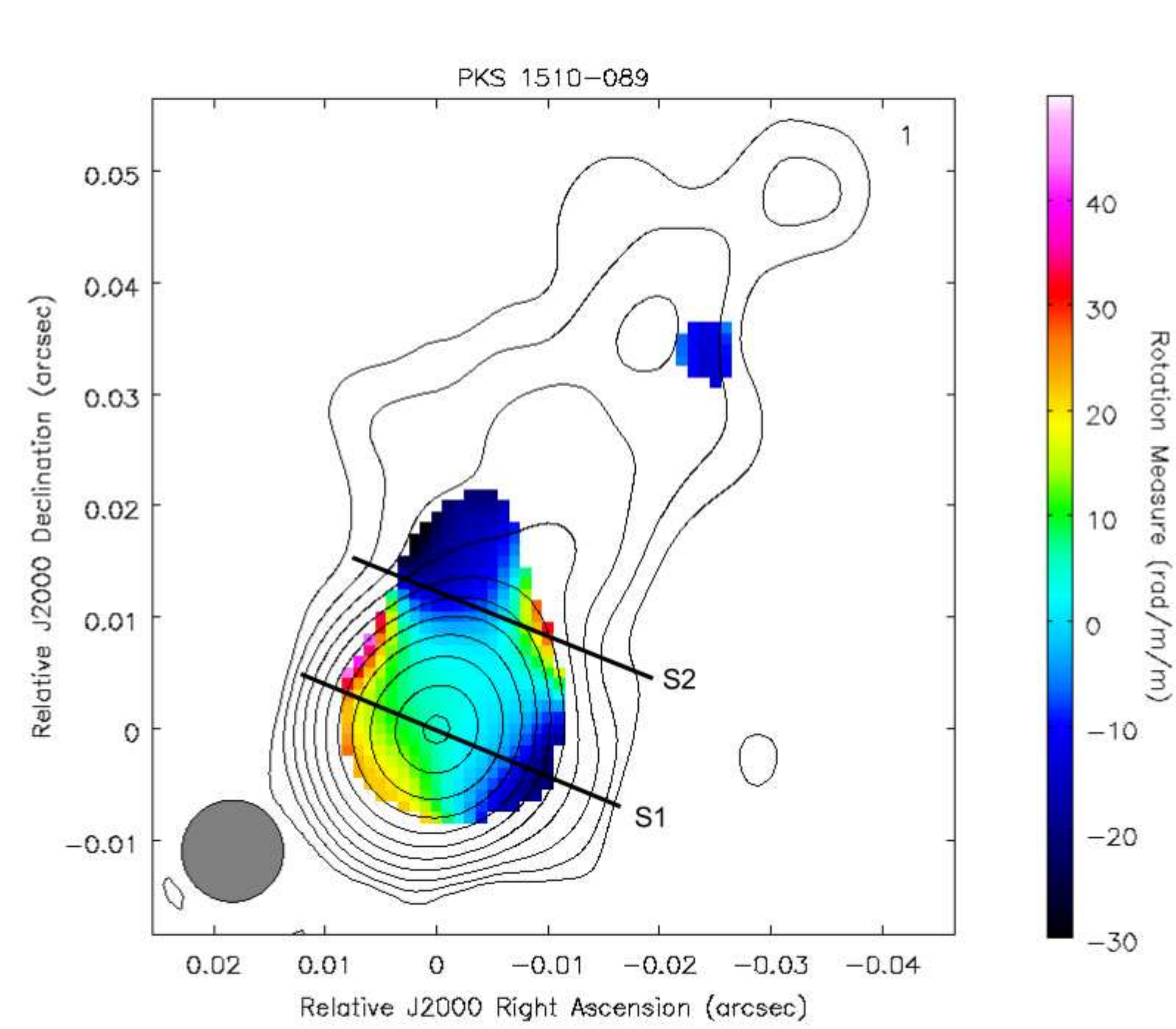}
	\end{minipage}
	\quad
	\begin{minipage}{.45\textwidth}
		\centering
		\includegraphics[width=0.7\textwidth]{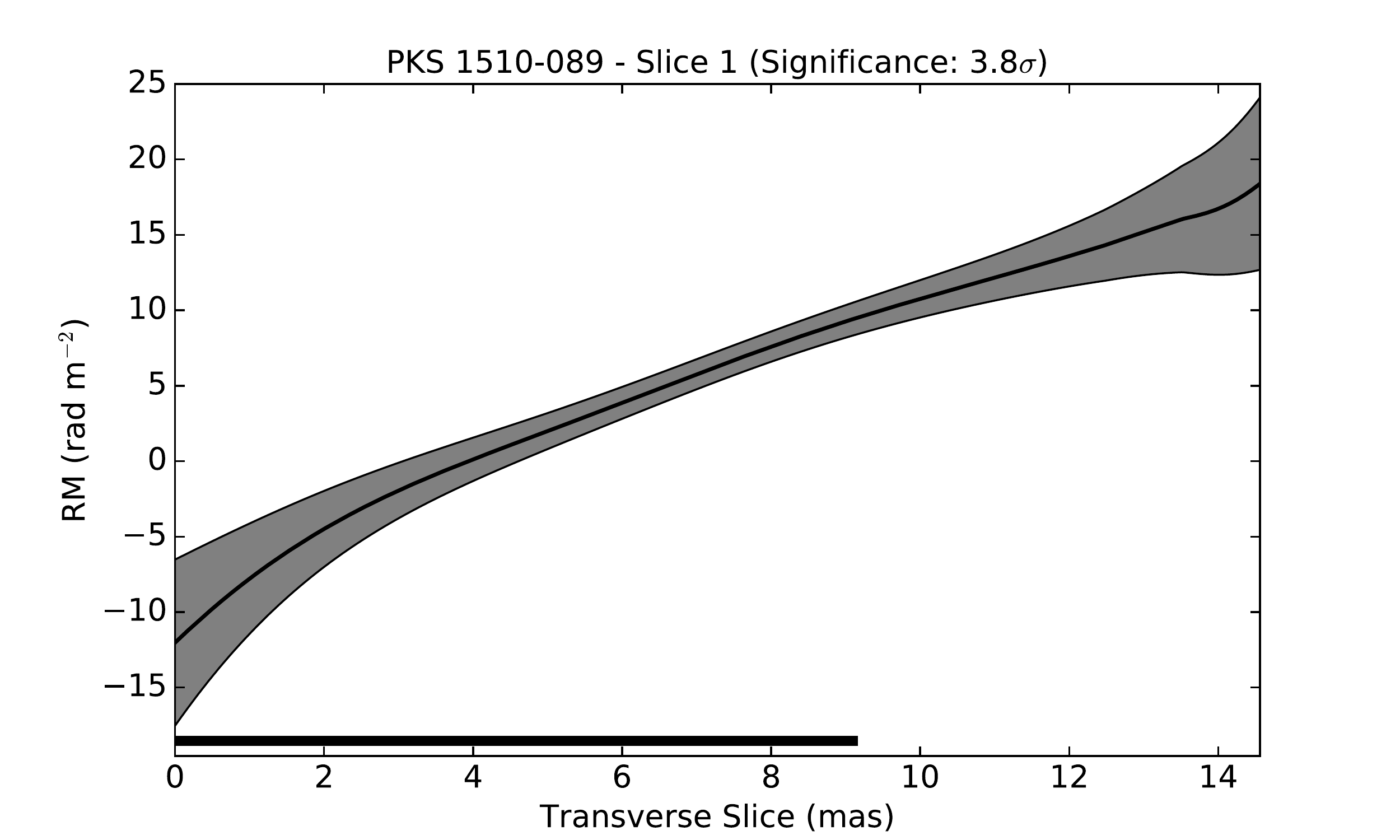}
		\includegraphics[width=0.7\textwidth]{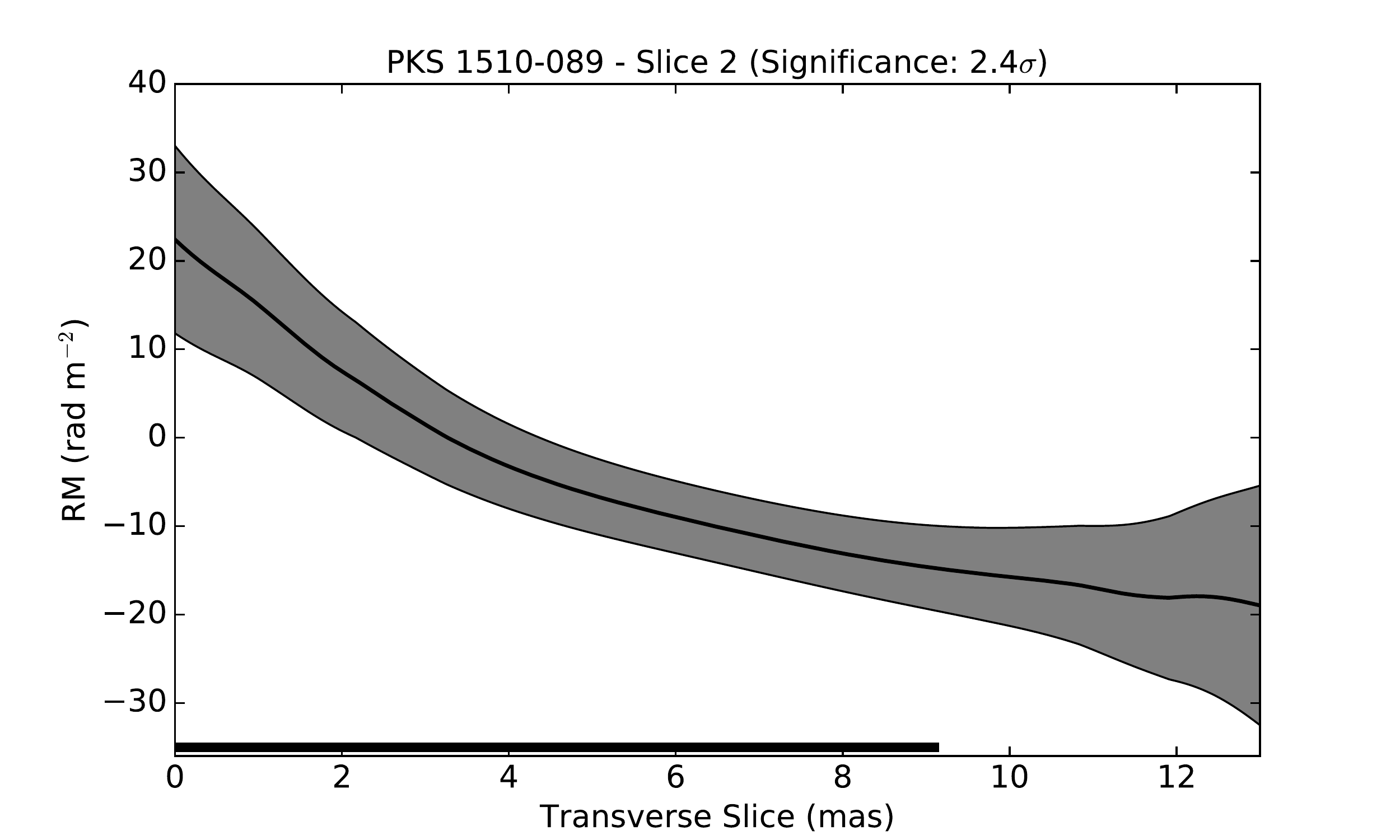}
	\end{minipage}
	\end{minipage}

\caption{RM distribution for PKS~1510-089 superimposed on the 1358 MHz $I$ map for the intrinsic elliptical beam (top-left) and a circular beam of equivalent area (bottom-left) along with slices taken in regions where transverse RM gradients are visible by eye, shown by the black lines across the RM maps (middle-left and bottom-right). The ranges of the RM values are indicated by the colour bars. Output pixels were blanked for RM uncertainties exceeding 15 rad m$^{-2}$. The thick black horizontal lines  accompanying the transverse RM profiles indicate the projected sizes of the beams in the slice direction. Examples of $\chi_{obs}$ versus $\lambda^2$ fits in the regions of the slices are shown in the top-right panels. The locations of the pixels are indicated by the arrows in the RM map. Total intensity (solid line) and fractional polarization (dashed line) profiles at the same location of the transverse RM gradients are shown in the middle-right panels. \label{PKS1510_RM}
 }
\end{center}
\end{figure*}

\begin{figure*}
\begin{center}

	\begin{minipage}{\textwidth}
	\begin{minipage}{.45\textwidth}
		\centering
		\includegraphics[width=0.9\textwidth]{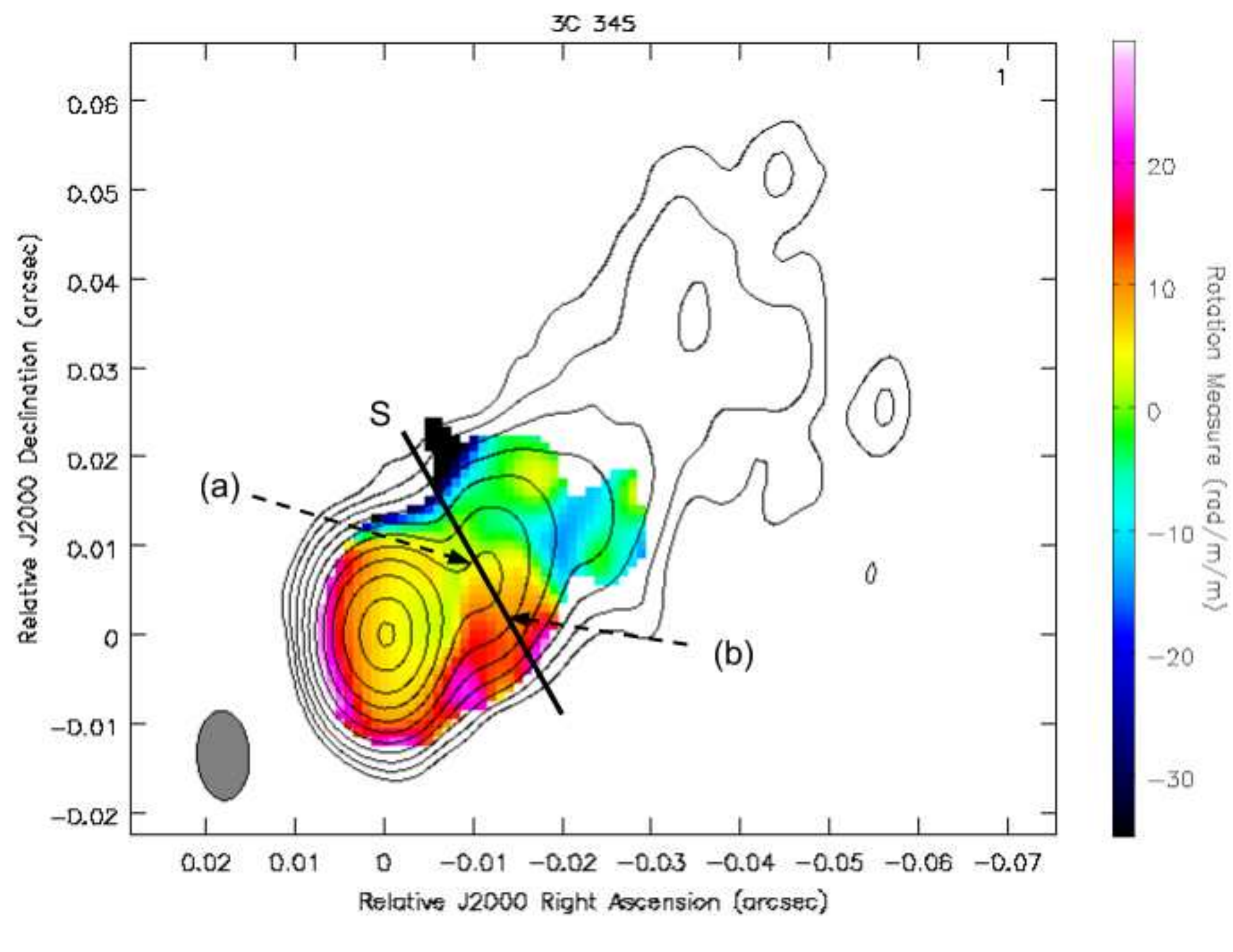}
	\end{minipage}
	\quad
	\begin{minipage}{.45\textwidth}
		\centering
		\includegraphics[width=0.8\textwidth]{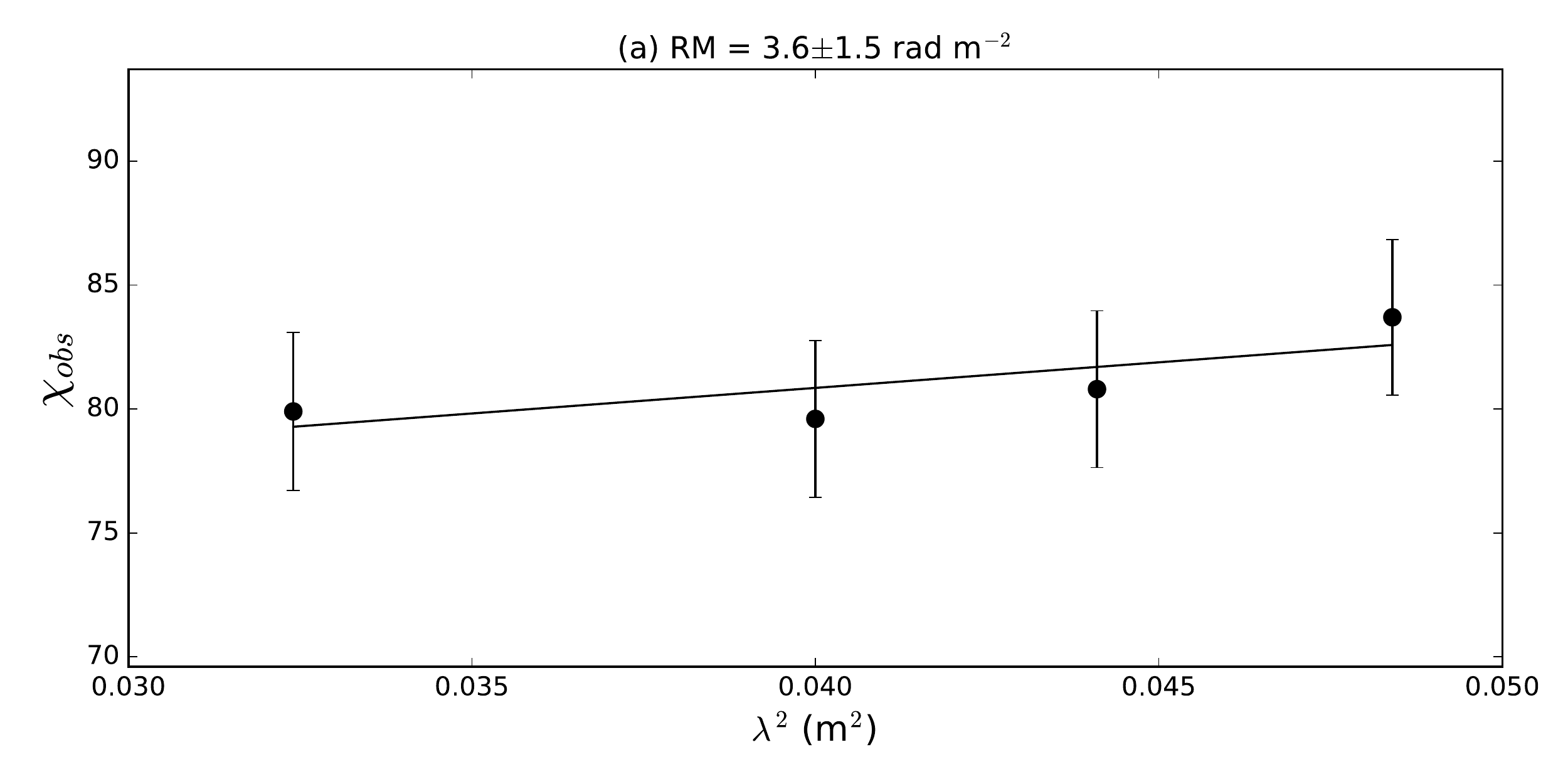}
		\includegraphics[width=0.8\textwidth]{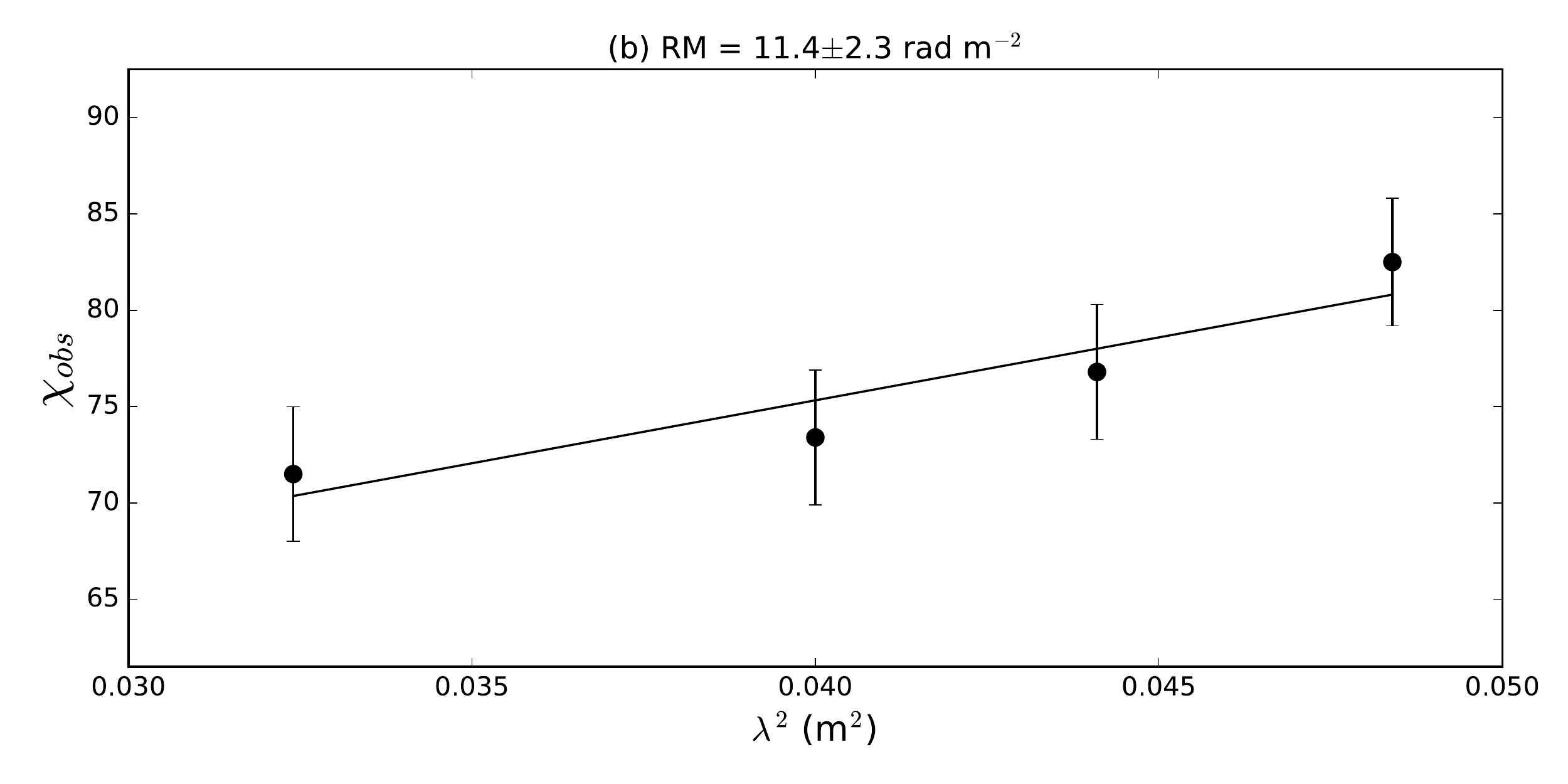}

	\end{minipage}
	\end{minipage}

	\begin{minipage}{\textwidth}
	\begin{minipage}{.45\textwidth}
		\centering
		\includegraphics[width=0.8\textwidth]{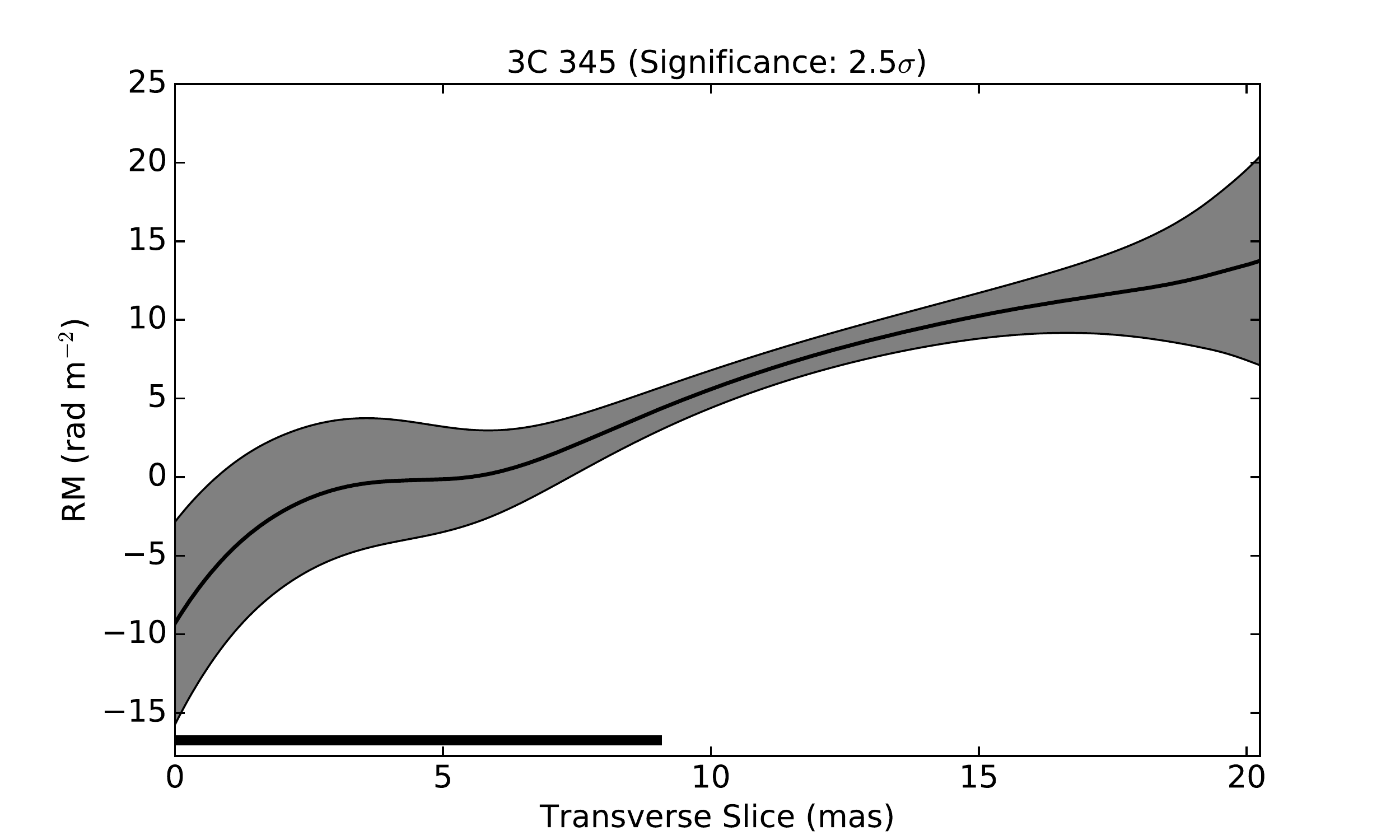}
	\end{minipage}
	\quad
	\begin{minipage}{.45\textwidth}
		\centering
		\includegraphics[width=0.8\textwidth]{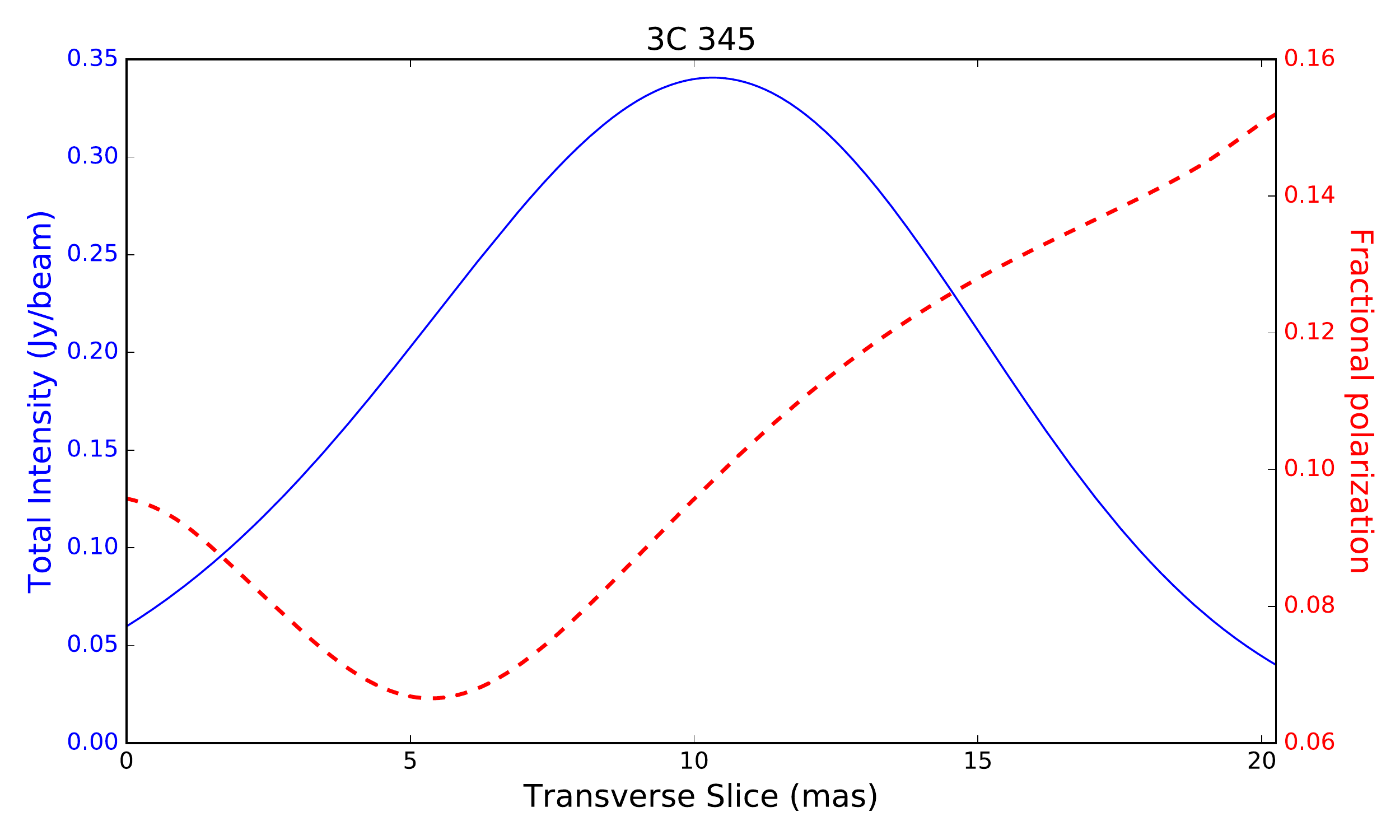}
	\end{minipage}
	\end{minipage}

	\begin{minipage}{\textwidth}
	\centering
	\begin{minipage}{.45\textwidth}
		\includegraphics[width=0.9\textwidth]{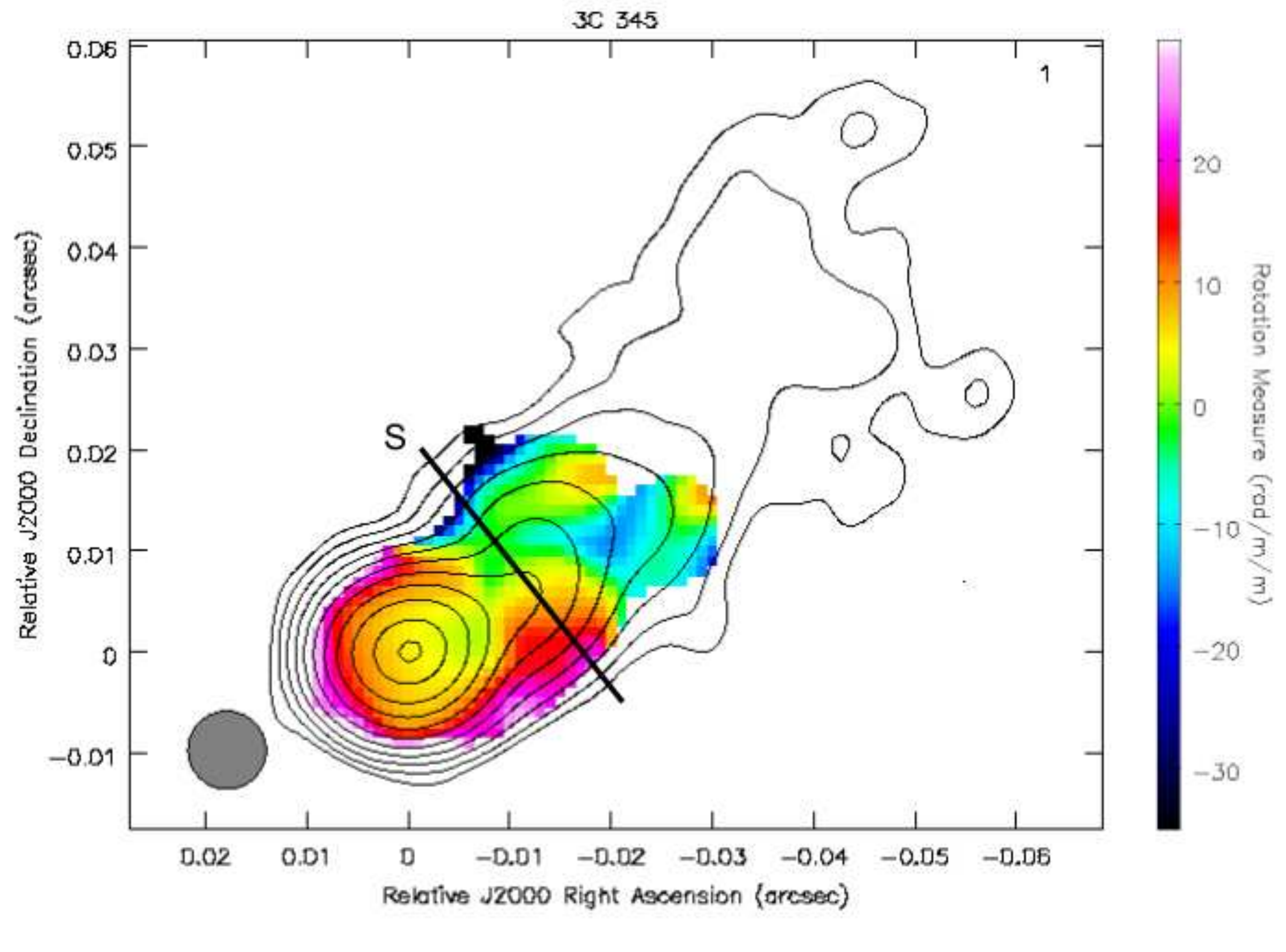}
	\end{minipage}
	\quad
	\begin{minipage}{.45\textwidth}
		\includegraphics[width=0.8\textwidth]{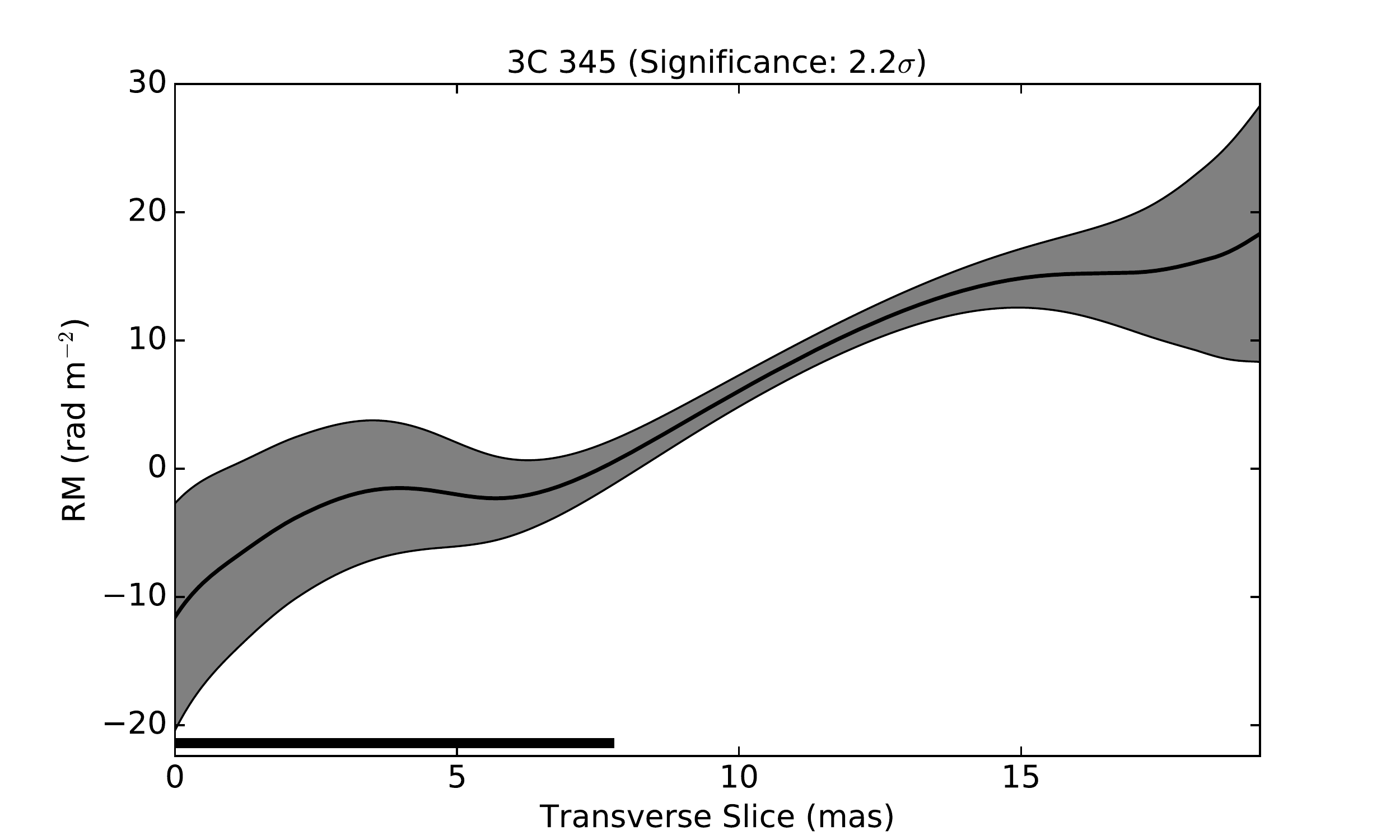}
	\end{minipage}
	\end{minipage}

\caption{RM distribution for 3C~345 superimposed on the 1358 MHz $I$ map for the intrinsic elliptical beam (top-left) and a circular beam of equivalent area (bottom-left) along with slices taken in a region where a transverse RM gradient is visible by eye, shown by the black line across the RM map (middle-left and bottom-right). The ranges of the RM values are indicated by the colour bars. Output pixels were blanked for RM uncertainties exceeding rad m$^{-2}$. The thick black horizontal lines  accompanying the transverse RM profiles indicate the projected sizes of the beams in the slice direction. Examples of the $\chi_{obs}$ versus $\lambda^2$ fits in the region of the slice are shown in the top-right panels. The locations of the pixels are indicated by the arrows in the RM map. Total intensity (solid line) and fractional polarization (dashed line) profiles at the same location of the transverse RM gradient are shown in the middle-right panel. \label{3C345_RM}
 }
\end{center}
\end{figure*}

\begin{figure*}
\begin{center}

	\begin{minipage}{\textwidth}
	\begin{minipage}{.45\textwidth}
		\centering
		\includegraphics[width=0.85\textwidth]{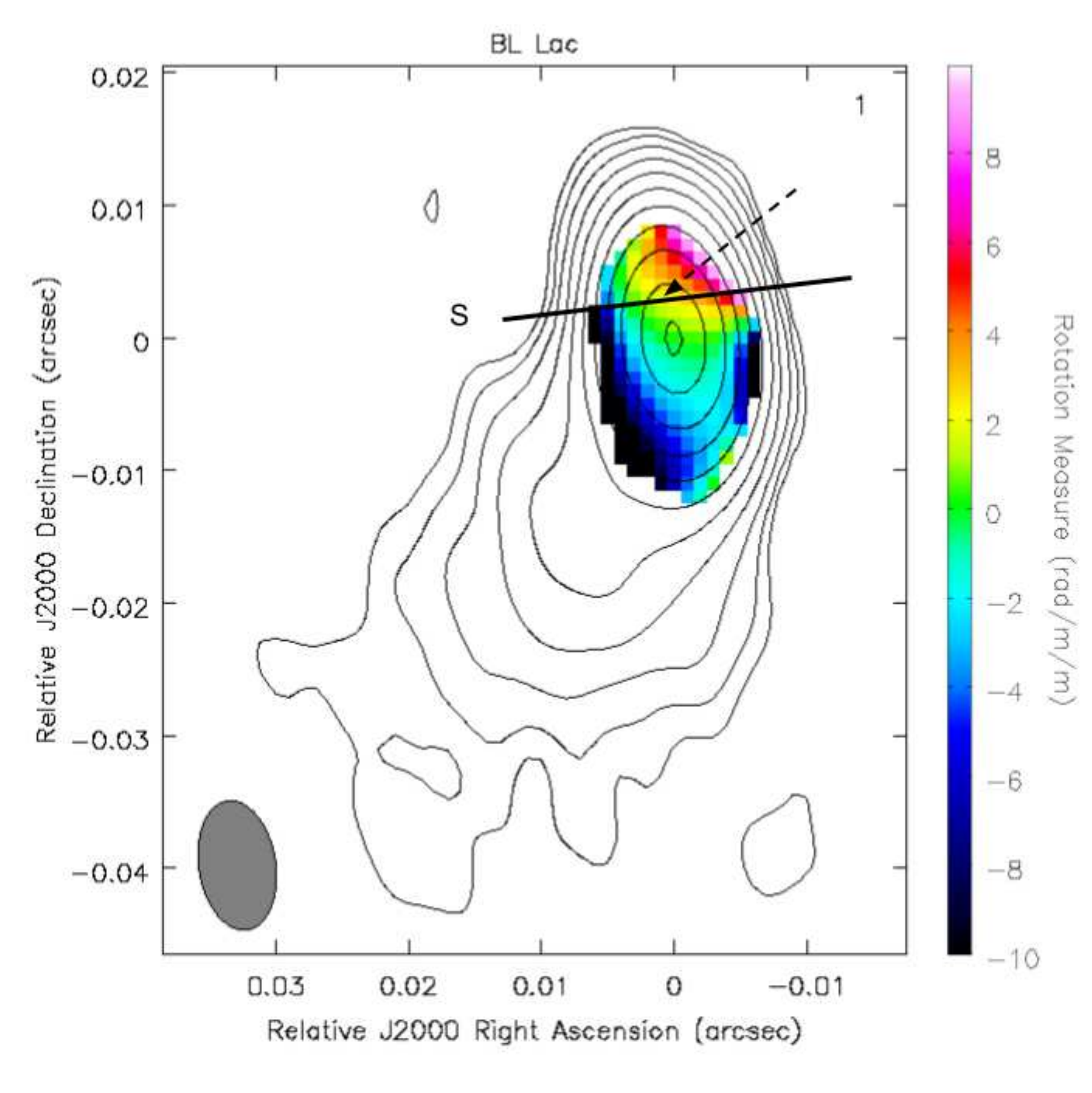}
	\end{minipage}
	\quad
	\begin{minipage}{.45\textwidth}
		\centering
		\includegraphics[width=0.8\textwidth]{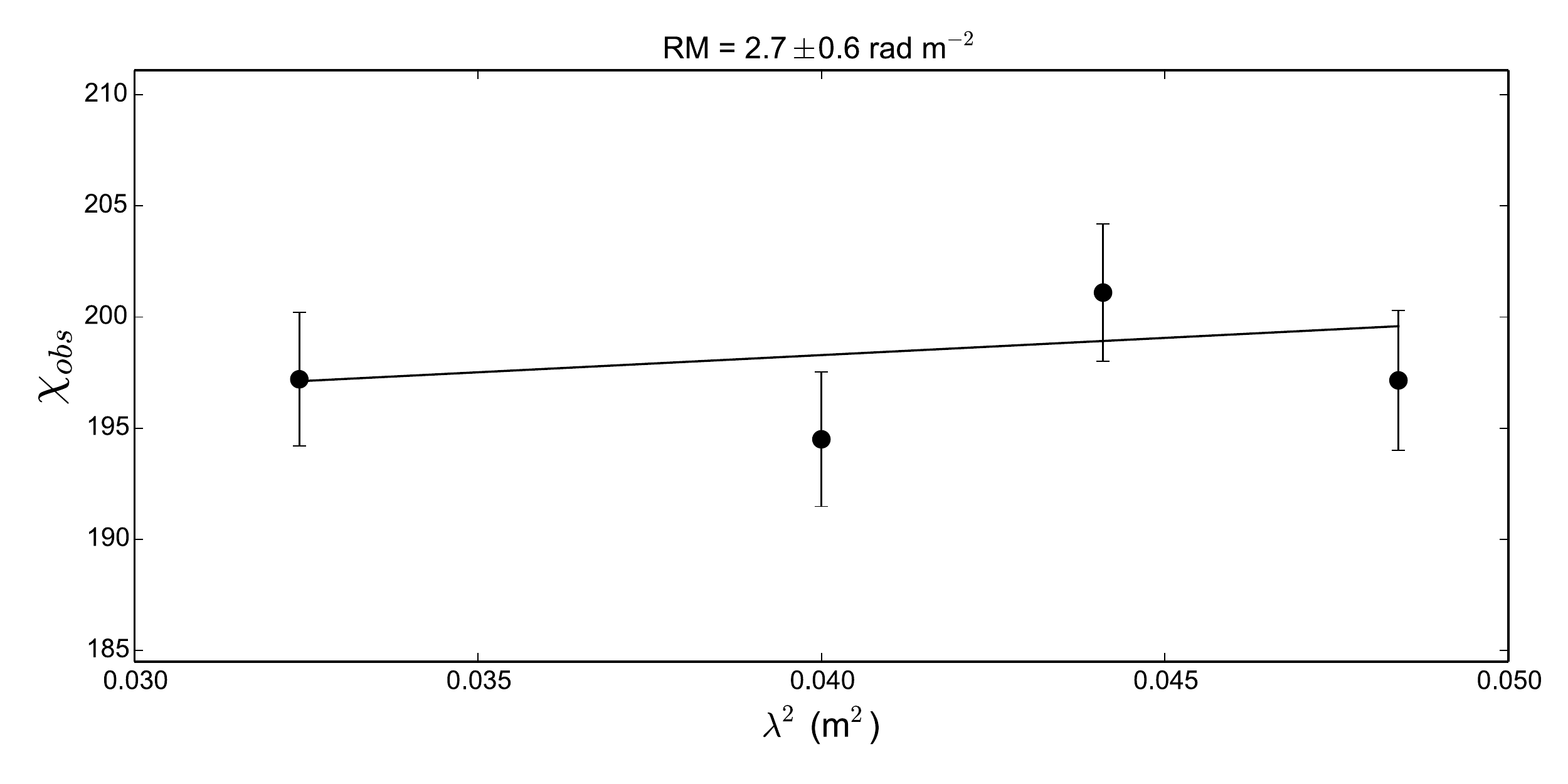}
	\end{minipage}
	\end{minipage}

	\begin{minipage}{\textwidth}
	\begin{minipage}{.45\textwidth}
		\centering
		\includegraphics[width=0.8\textwidth]{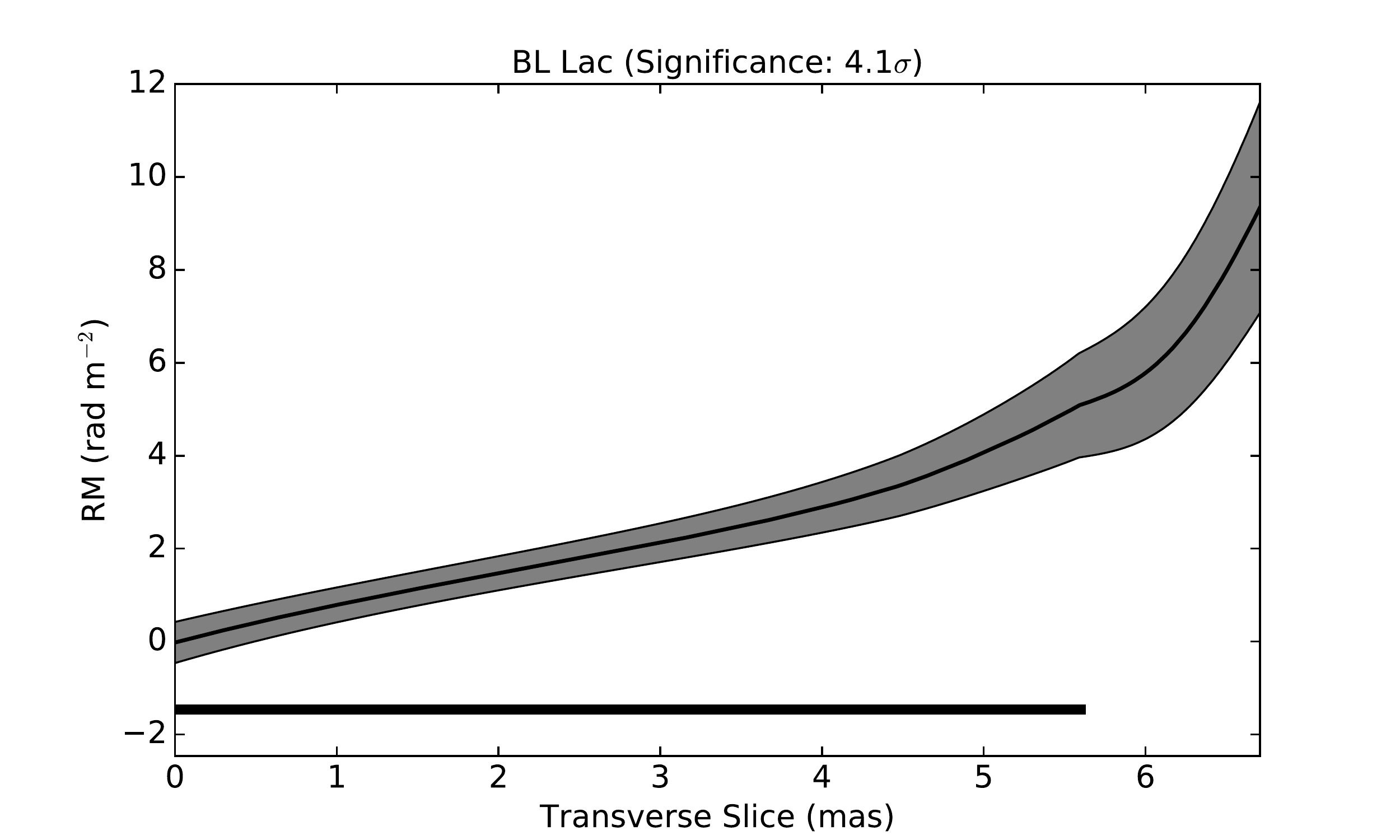}
	\end{minipage}
	\quad
	\begin{minipage}{.45\textwidth}
		\centering
		\includegraphics[width=0.8\textwidth]{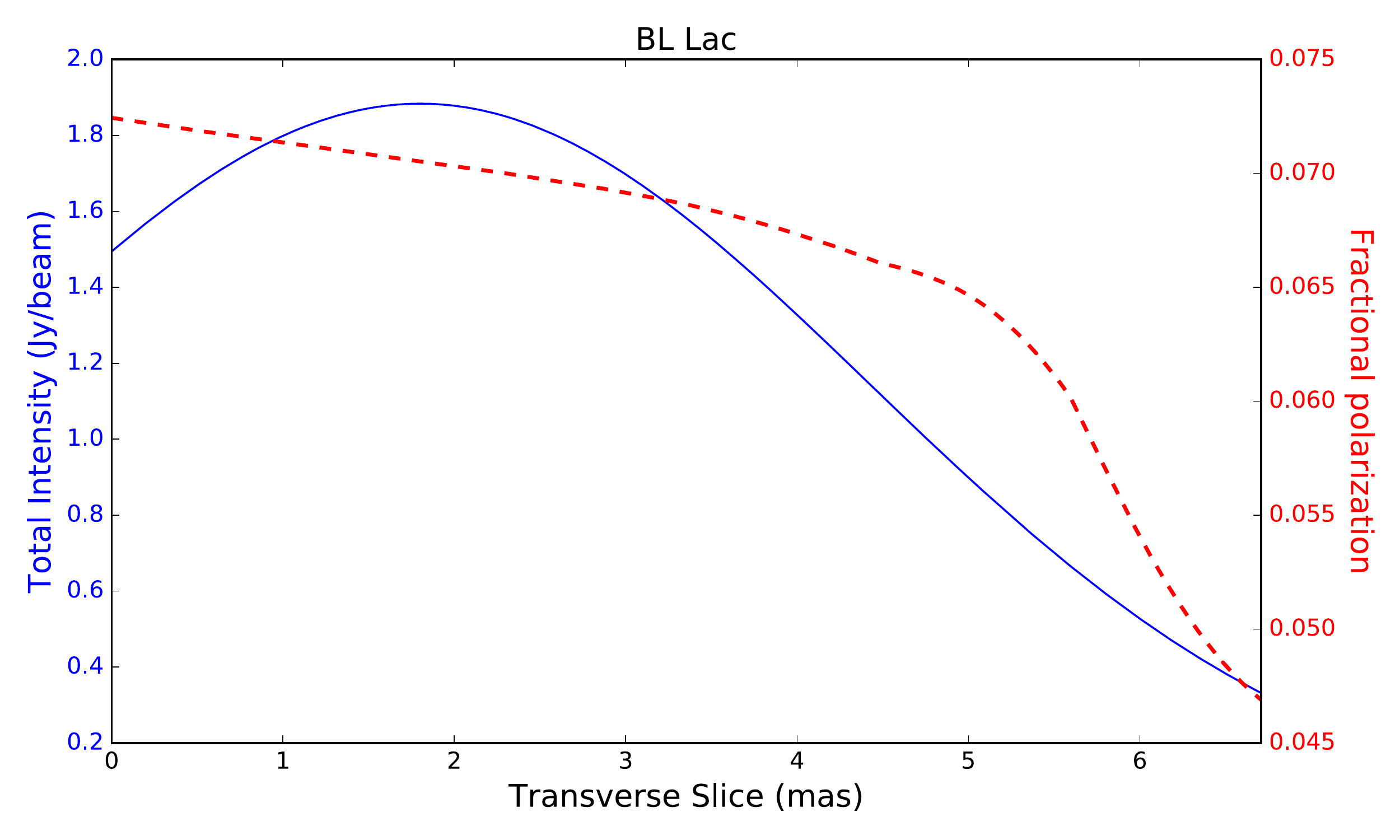}
	\end{minipage}
	\end{minipage}

	\begin{minipage}{\textwidth}
	\centering
	\begin{minipage}{.45\textwidth}
		\includegraphics[width=0.85\textwidth]{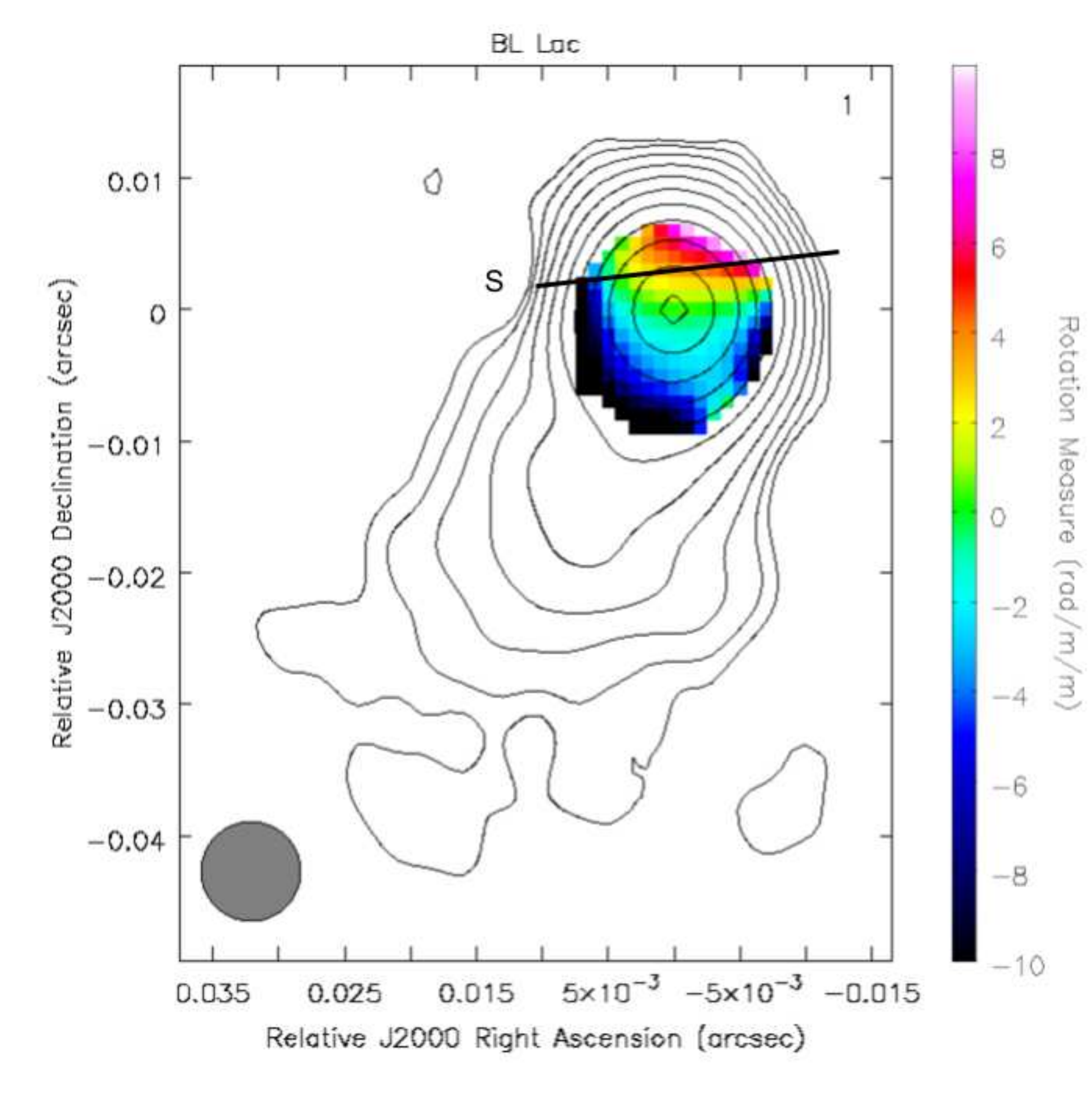}
	\end{minipage}
	\quad
	\begin{minipage}{.45\textwidth}
		\includegraphics[width=0.8\textwidth]{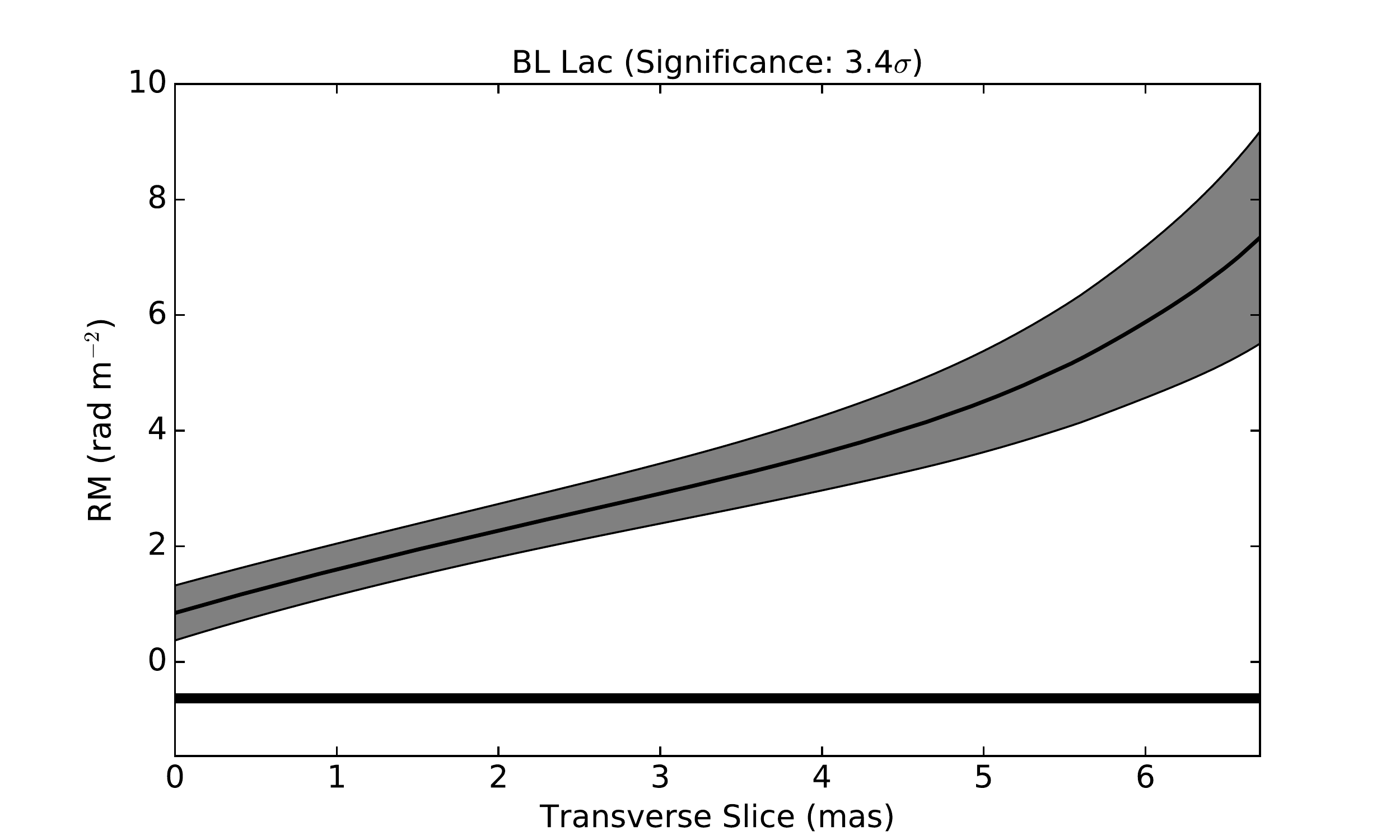}
	\end{minipage}
	\end{minipage}

\caption{RM distribution for BL~Lac superimposed on the 1358 MHz $I$ map for the intrinsic elliptical beam (top-left) and a circular beam of equivalent area (bottom-left) along with slices taken in a region where a transverse RM gradient is visible by eye, shown by the black line across the RM map (middle-left and bottom-right). The ranges of the RM values are indicated by the colour bars. Output pixels were blanked for RM uncertainties exceeding 5 rad m$^{-2}$. The thick black horizontal lines  accompanying the transverse RM profiles indicate the projected sizes of the beams in the slice direction. An example of a $\chi_{obs}$ versus $\lambda^2$ fits in the region of the slice is shown in the top-right panel. The location of the pixel is indicated by the arrow in the RM map. Total intensity (solid line) and fractional polarization (dashed line) profiles at the same location of the transverse RM gradient are shown in the middle-right panel. \label{BLLAC_RM}
 }
\end{center}
\end{figure*}

\begin{figure*}
\begin{center}

	\begin{minipage}{\textwidth}
	\begin{minipage}{.45\textwidth}
		\centering
		\includegraphics[width=0.9\textwidth]{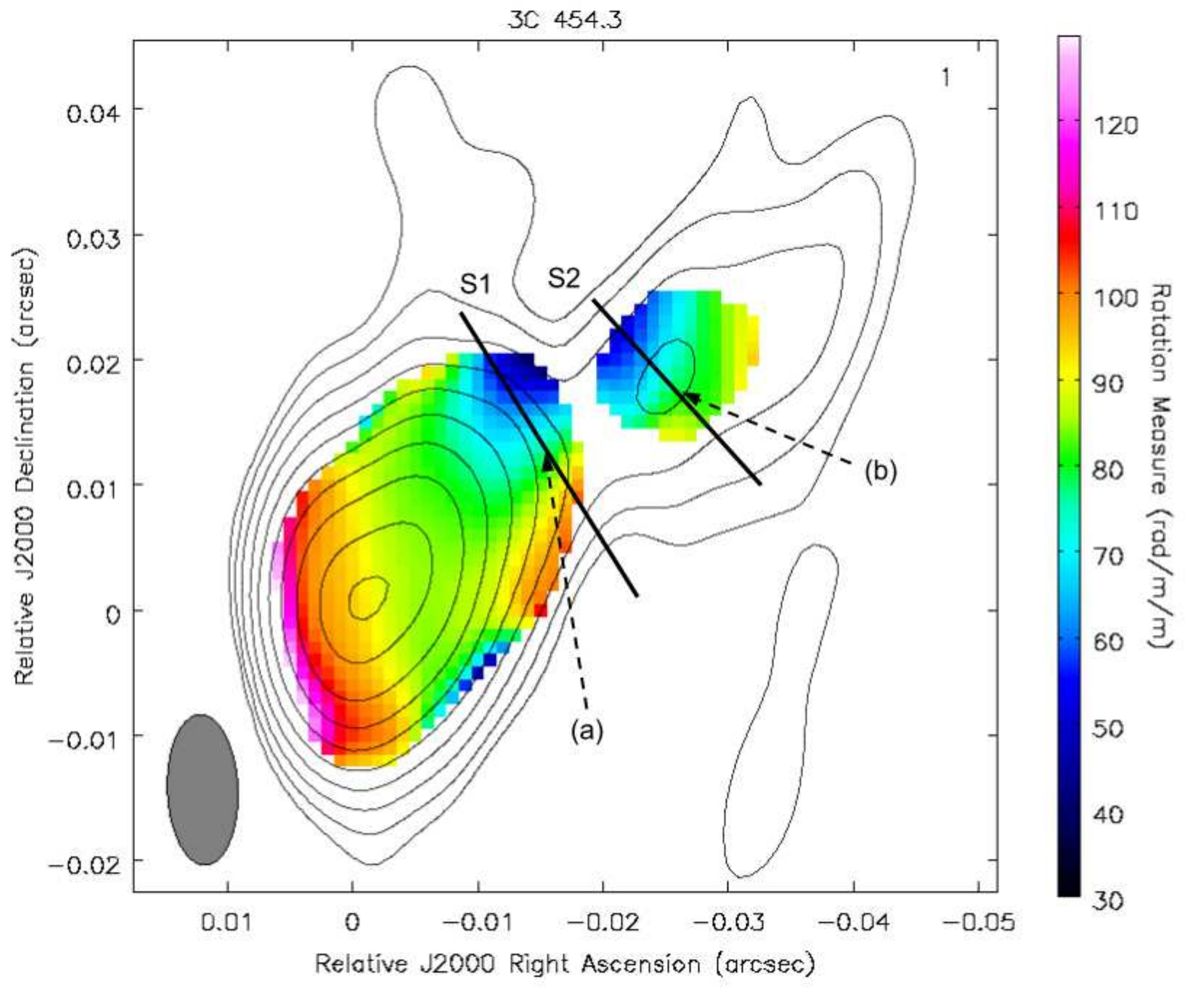}
	\end{minipage}
	\quad
	\begin{minipage}{.45\textwidth}
		\centering
		\includegraphics[width=0.8\textwidth]{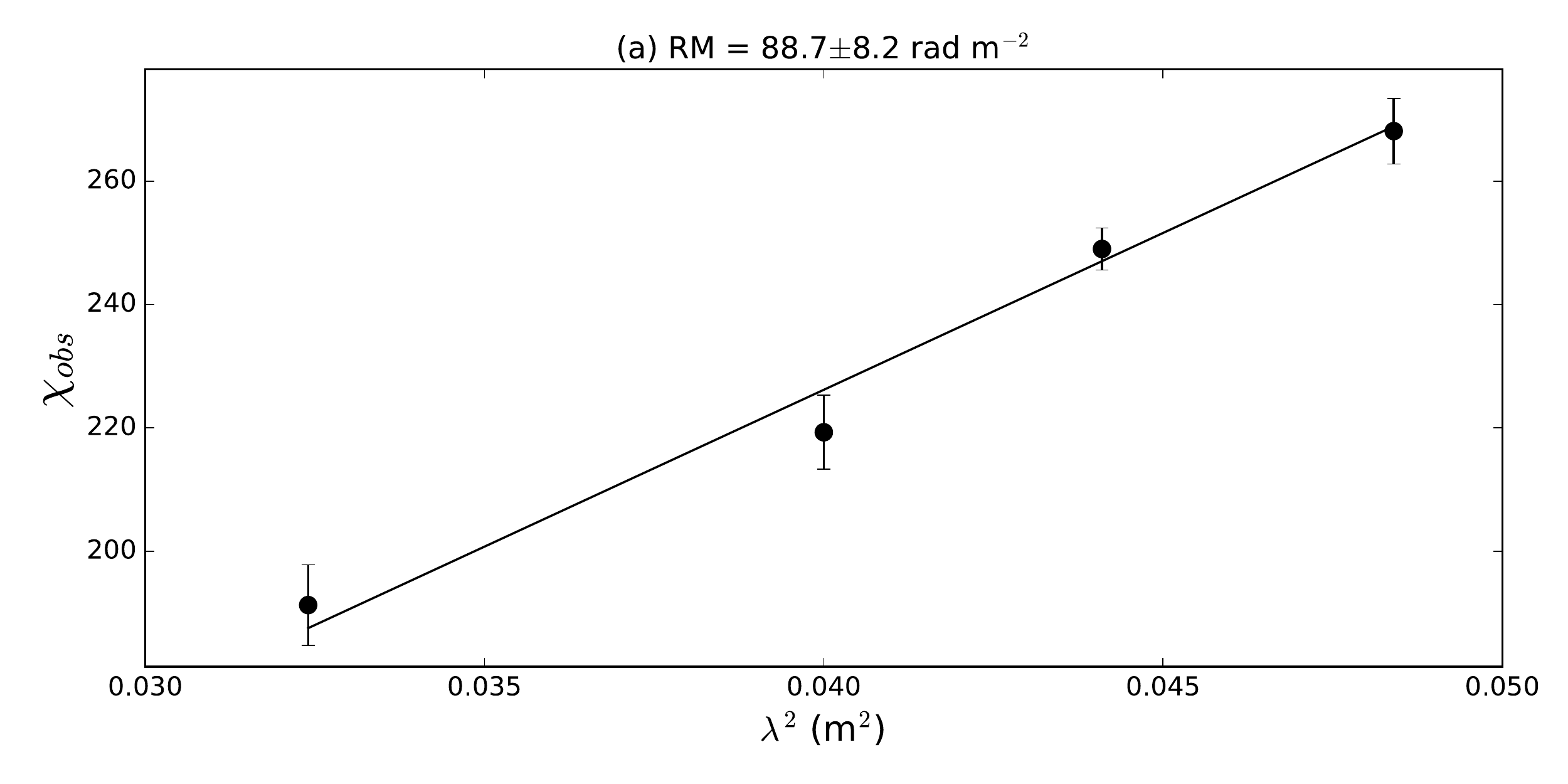}
		\includegraphics[width=0.8\textwidth]{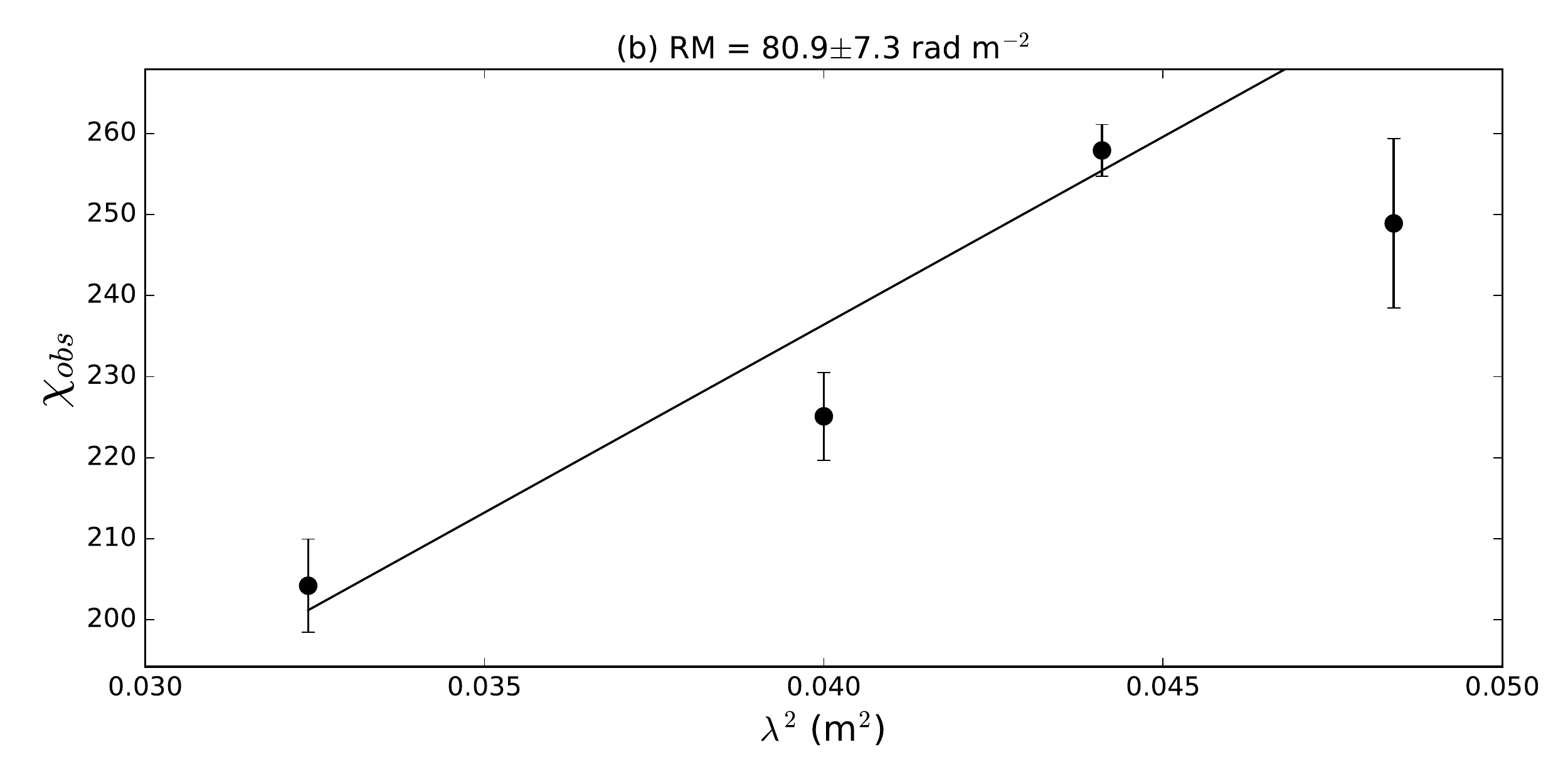}
	\end{minipage}
	\end{minipage}

	\begin{minipage}{\textwidth}
	\begin{minipage}{.45\textwidth}
		\centering
		\includegraphics[width=0.8\textwidth]{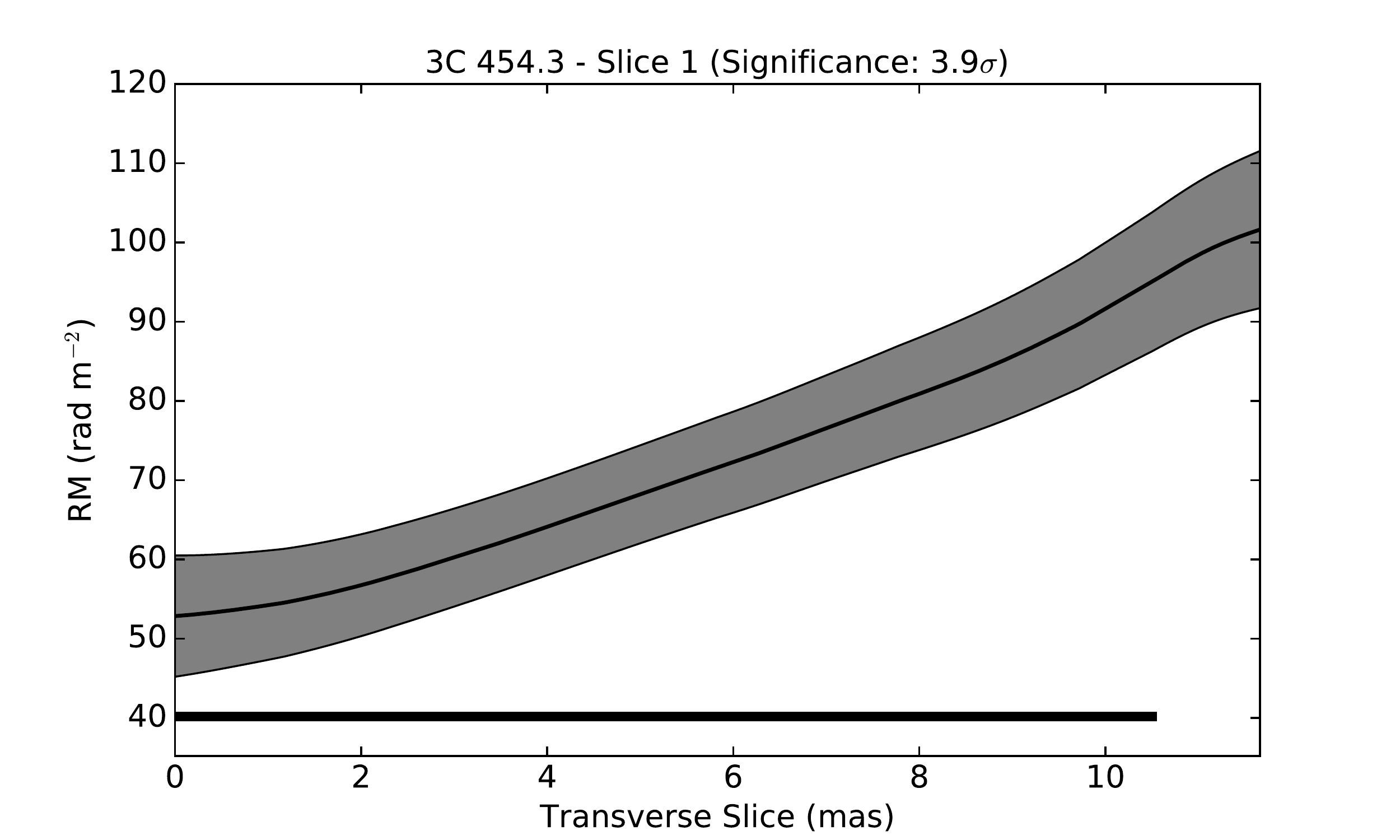}
	\end{minipage}
	\quad
	\begin{minipage}{.45\textwidth}
		\centering
		\includegraphics[width=0.8\textwidth]{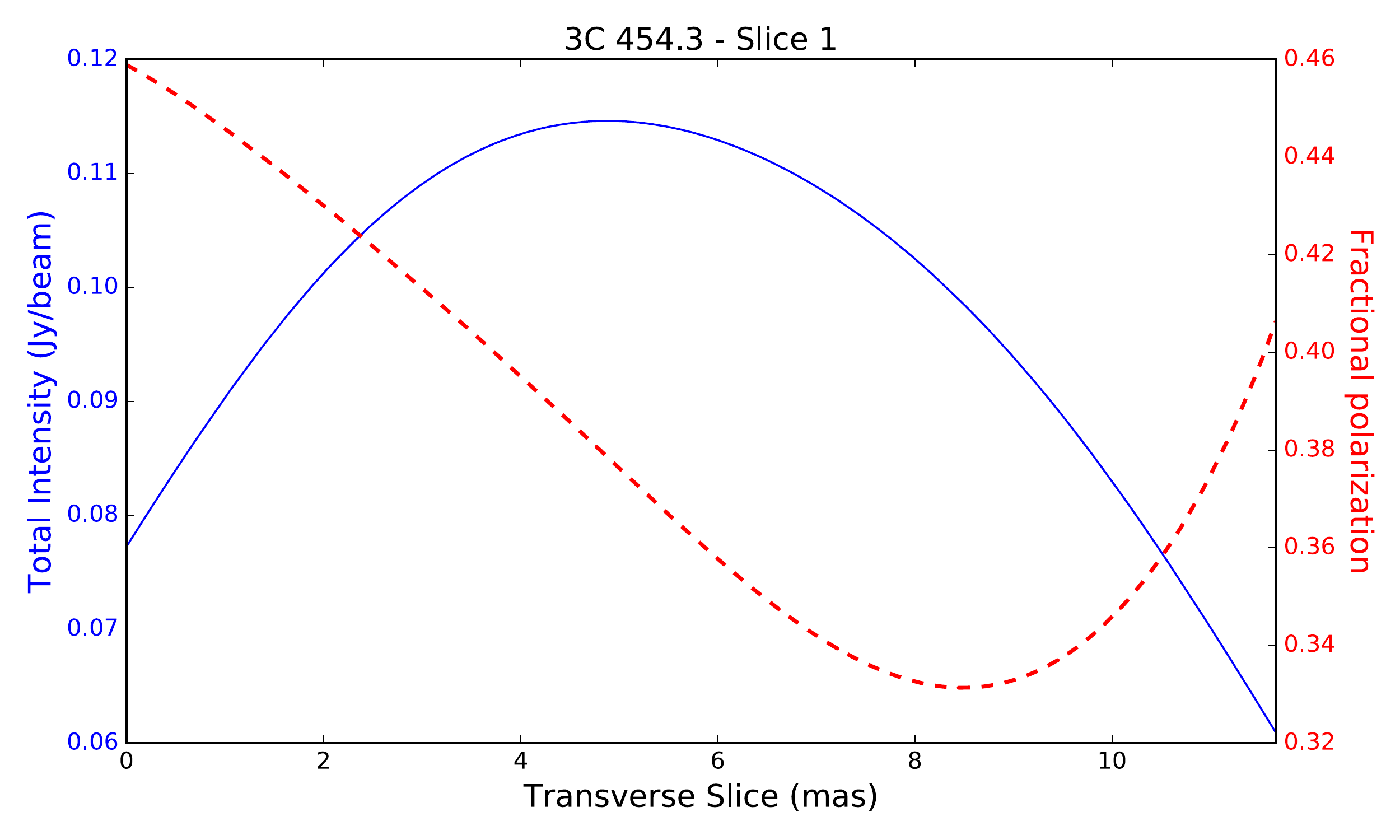}
	\end{minipage}
	\end{minipage}

	\begin{minipage}{\textwidth}
	\begin{minipage}{.45\textwidth}
		\centering
		\includegraphics[width=0.8\textwidth]{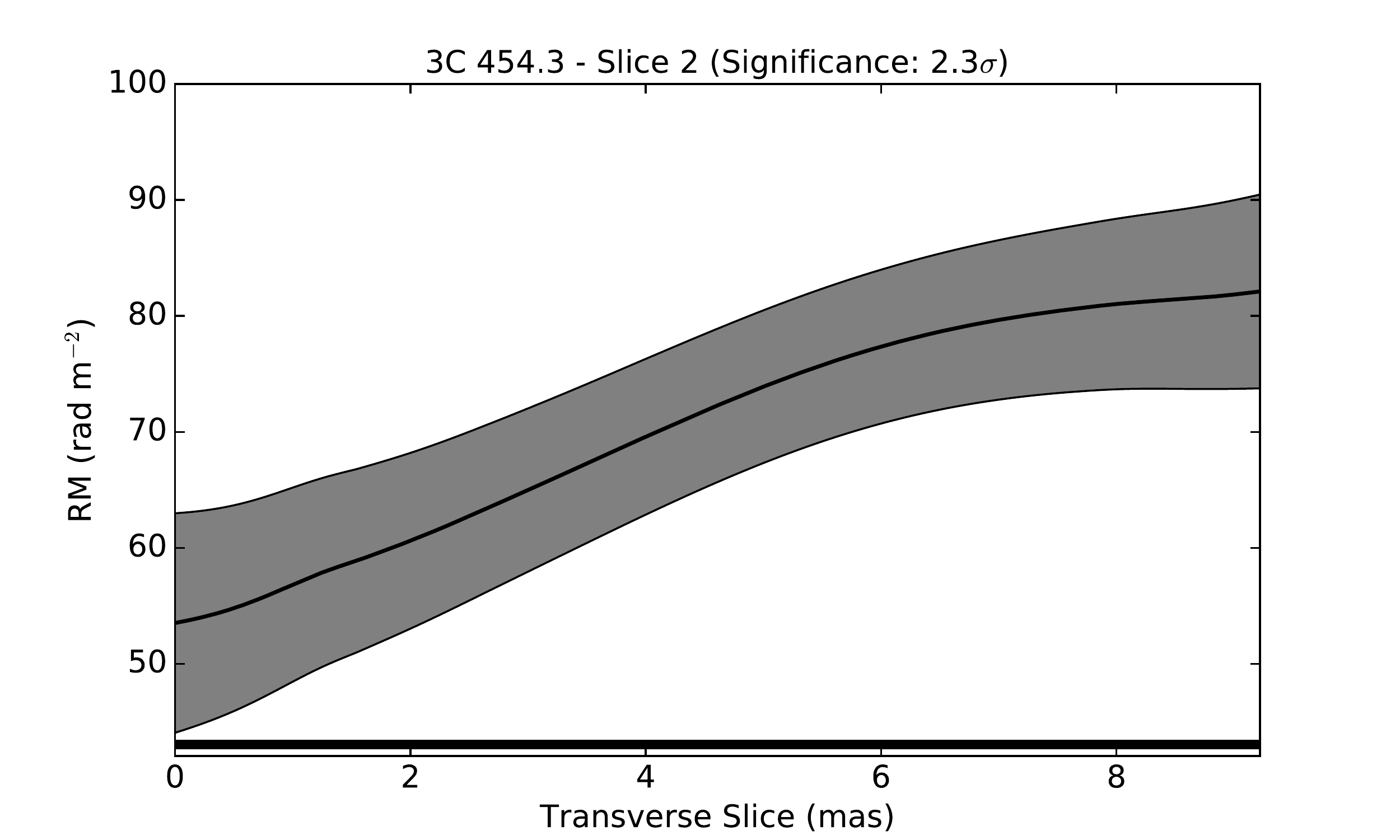}
	\end{minipage}
	\quad
	\begin{minipage}{.45\textwidth}
		\centering
		\includegraphics[width=0.8\textwidth]{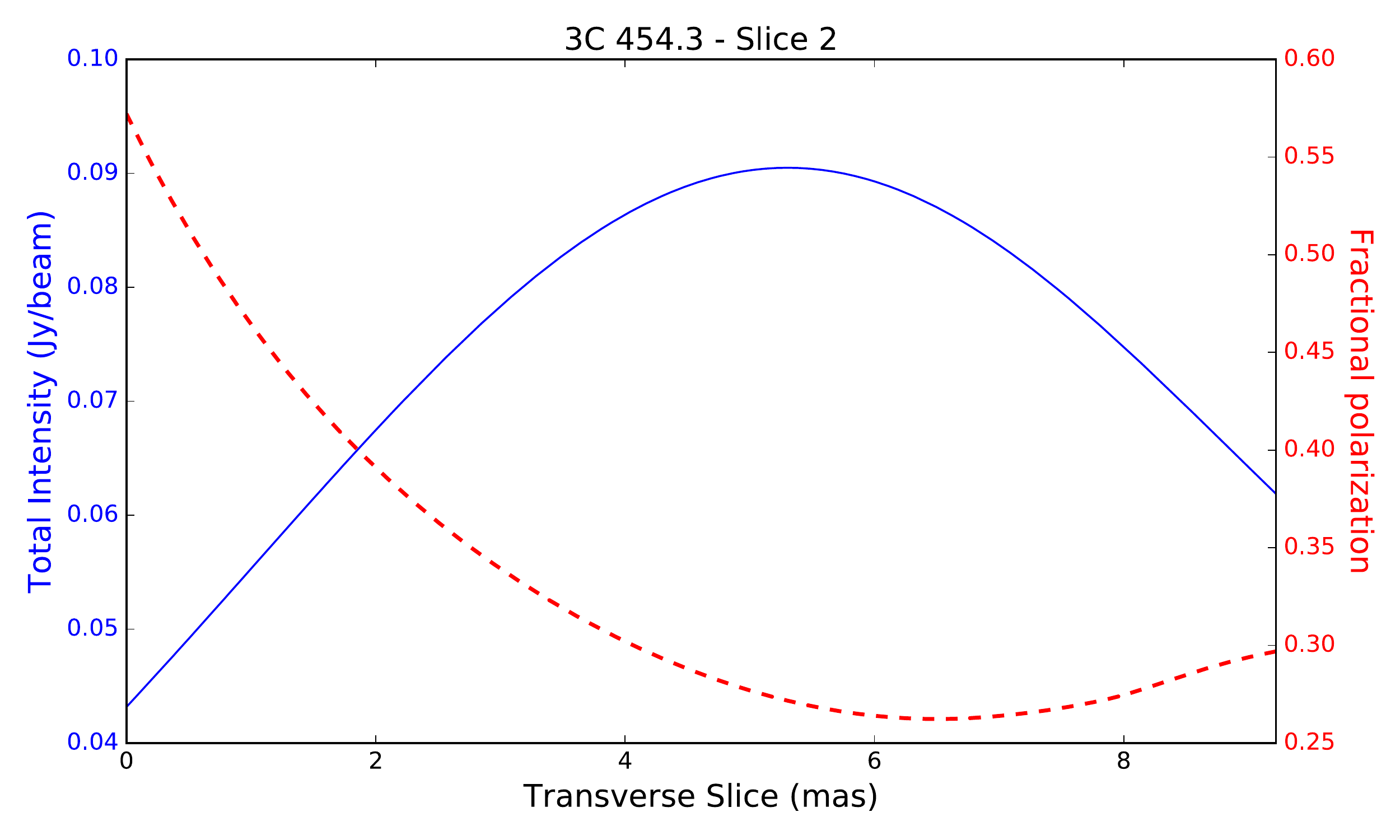}
	\end{minipage}
	\end{minipage}

	\begin{minipage}{\textwidth}
	\begin{minipage}{.45\textwidth}	
		\centering
		\includegraphics[width=0.9\textwidth]{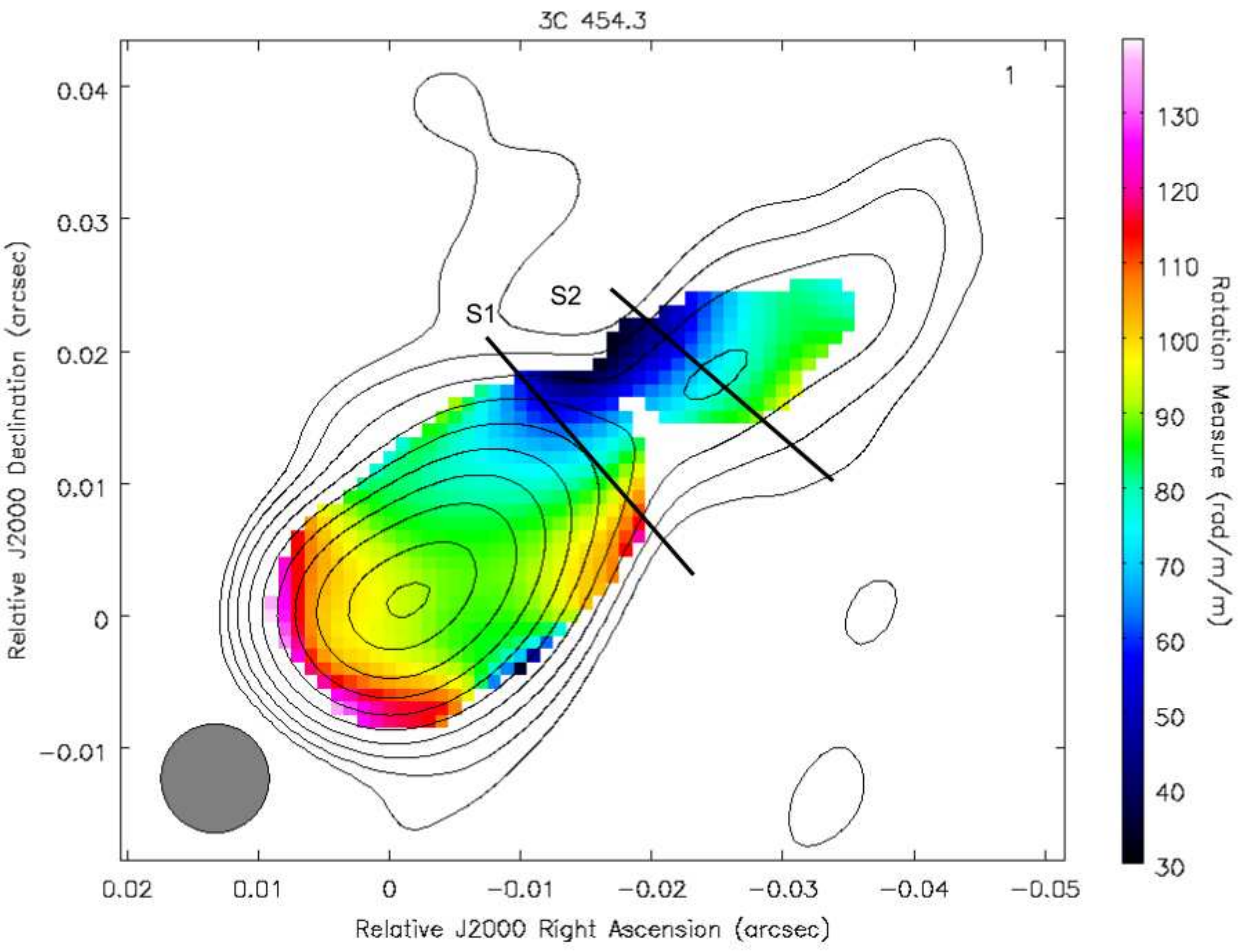}
	\end{minipage}
	\quad
	\begin{minipage}{.45\textwidth}
		\centering
		\includegraphics[width=0.7\textwidth]{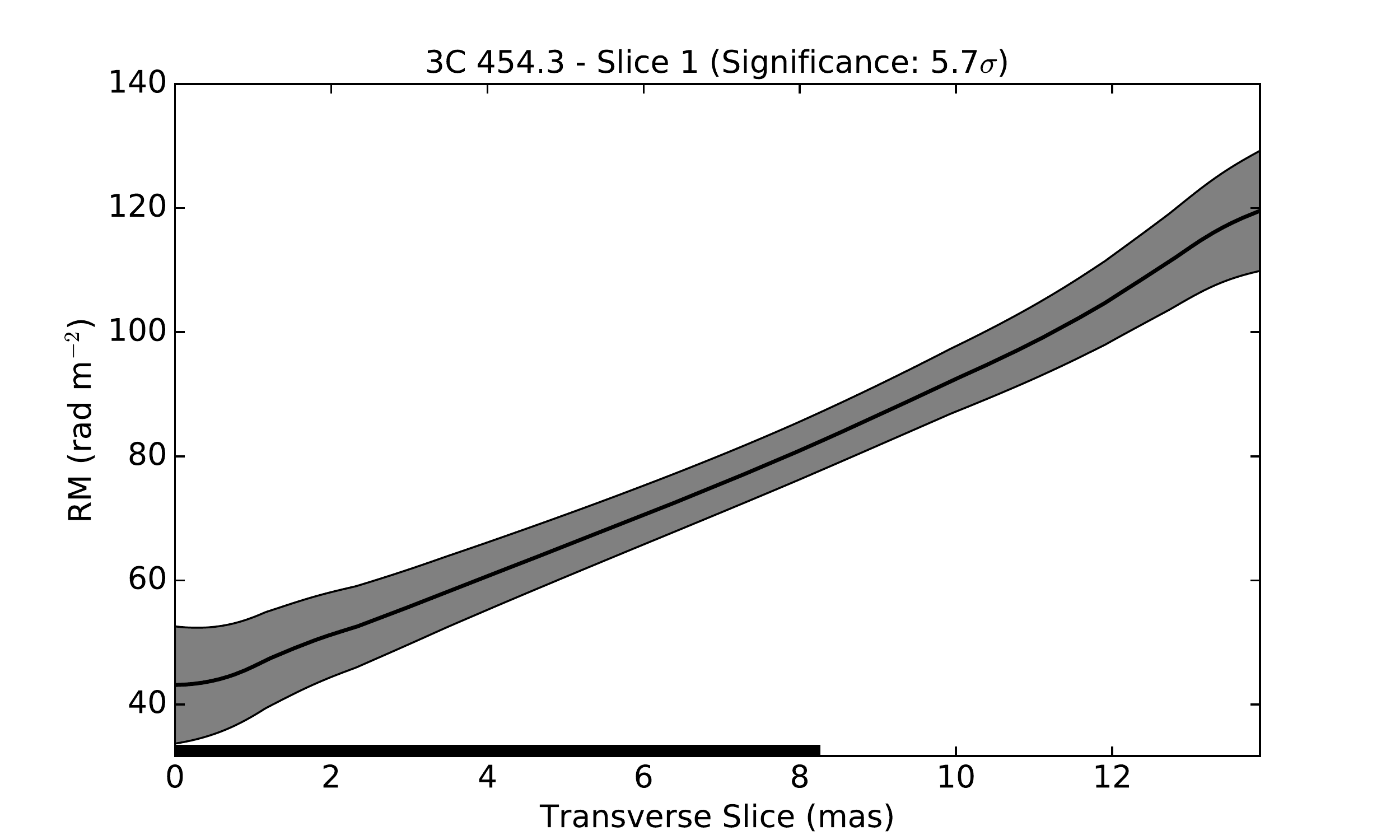}
		\includegraphics[width=0.7\textwidth]{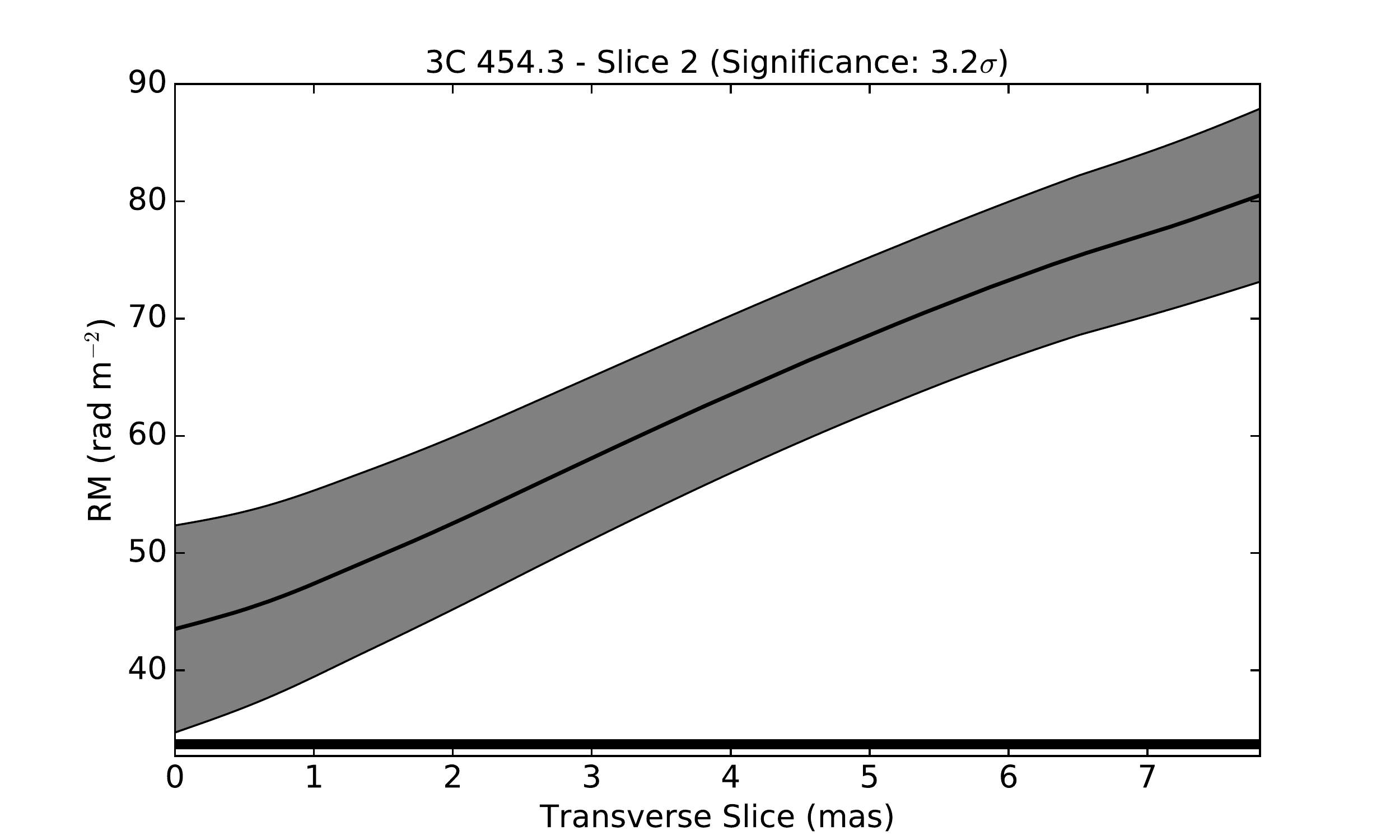}
	\end{minipage}
	\end{minipage}

\caption{RM distribution for 3C~454.3 superimposed on the 1358 MHz $I$ map for the intrinsic elliptical beam (top-left) and a circular beam of equivalent area (bottom-left) along with slices taken in a region where a transverse RM gradient is visible by eye, shown by the black lines across the RM map (middle-left and bottom-right). The ranges of the RM values are indicated by the colour bars. Output pixels were blanked for RM uncertainties exceeding 10 rad m$^{-2}$. The thick black horizontal lines  accompanying the transverse RM profiles indicate the projected sizes of the beams in the slice direction. Examples of the $\chi_{obs}$ versus $\lambda^2$ fits in the regions of the slices are shown in the top-right panels. The locations of the pixels are indicated by the arrows in the RM map. Total intensity (solid line) and fractional polarization (dashed line) profiles at the same locations of the transverse RM gradients are shown in the middle-right panels. \label{3C454.3_RM}
 }
\end{center}
\end{figure*}

\section{Results} \label{res}

In the left-hand panels of Figs. \ref{Fig1_OJ287_3C279_PKS1510} and \ref{Fig2_3C345_BLLac_3C454.3} we present the EVPAs corrected for the integrated RM superimposed on the 1358 MHz total intensity contours for all six sources. In the middle and right-hand panels of Figs. \ref{Fig1_OJ287_3C279_PKS1510} and \ref{Fig2_3C345_BLLac_3C454.3}, we present the 1358 MHz fractional polarization maps with two different ranges, if necessary, indicated by the colour bars, for better visualization of the variations in the degree of polarization. The properties of the 1358 MHz maps are summarized in Table \ref{map_prop}, and in all cases the contour levels increase in steps of a factor of two. The top-left panels of Figs. \ref{OJ287_RM}-\ref{3C454.3_RM} show the RM maps constructed using the intrinsic (elliptical) 1358~MHz beams superimposed on the corresponding 1358~MHz total intensity contours, along with slices in regions where transverse RM gradients were observed. In all cases, the slices were taken perpendicular to the local jet direction, as indicated by our model fits to the 1665 MHz intensity data. For gradients present in the core region in our observations, we took the local jet direction to be the direction from the core to the innermost jet component. The ranges of the RM maps are indicated by the colour bars accompanying them. In order to avoid the appearance of spurious features in the RM distributions in regions of off-source emission, we blanked the output pixels when the uncertainty in the RM exceeded the values listed in the corresponding figure caption.  Examples of the $\chi_{obs}$ versus $\lambda^2$ fits are shown in the top-right panels of the figures; the uncertainty in $\chi$ includes the uncertainty in the EVPA calibration ($\pm 3^{\circ}$) added in quadrature. Versions of the RM maps made with an equal-area circular beam are shown in the bottom-left panels, with corresponding RM slices shown in the bottom-right panels. The middle-right panels of Figs. \ref{OJ287_RM}-\ref{3C454.3_RM} show total intensity (solid line) and fractional polarization (dashed line) profiles taken at the same locations and along the same directions as of the transverse RM gradients. In all cases, the convolving beams used to produce the RM maps are shown in the lower left-hand corners of the figures. The black lines drawn across the RM distributions show the locations of the slices, and the letter `S' indicates the side corresponding to their starting points.

\begin{figure*}
\begin{center}

	\begin{minipage}{\textwidth}
	\begin{minipage}{.45\textwidth}
		\centering
		\includegraphics[width=0.8\textwidth]{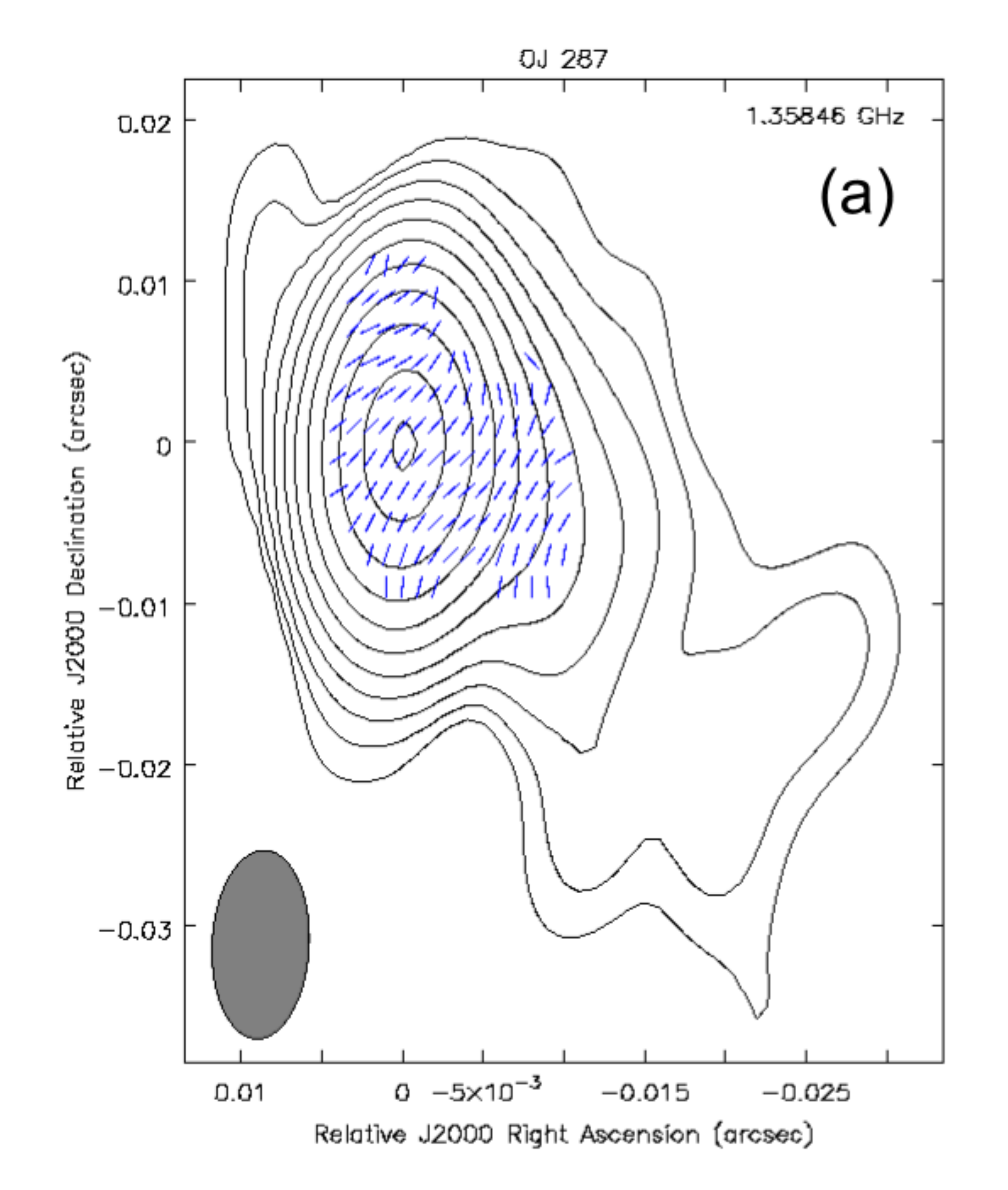}
	\end{minipage}
	\quad
	\begin{minipage}{.45\textwidth}
		\centering
		\includegraphics[width=0.8\textwidth]{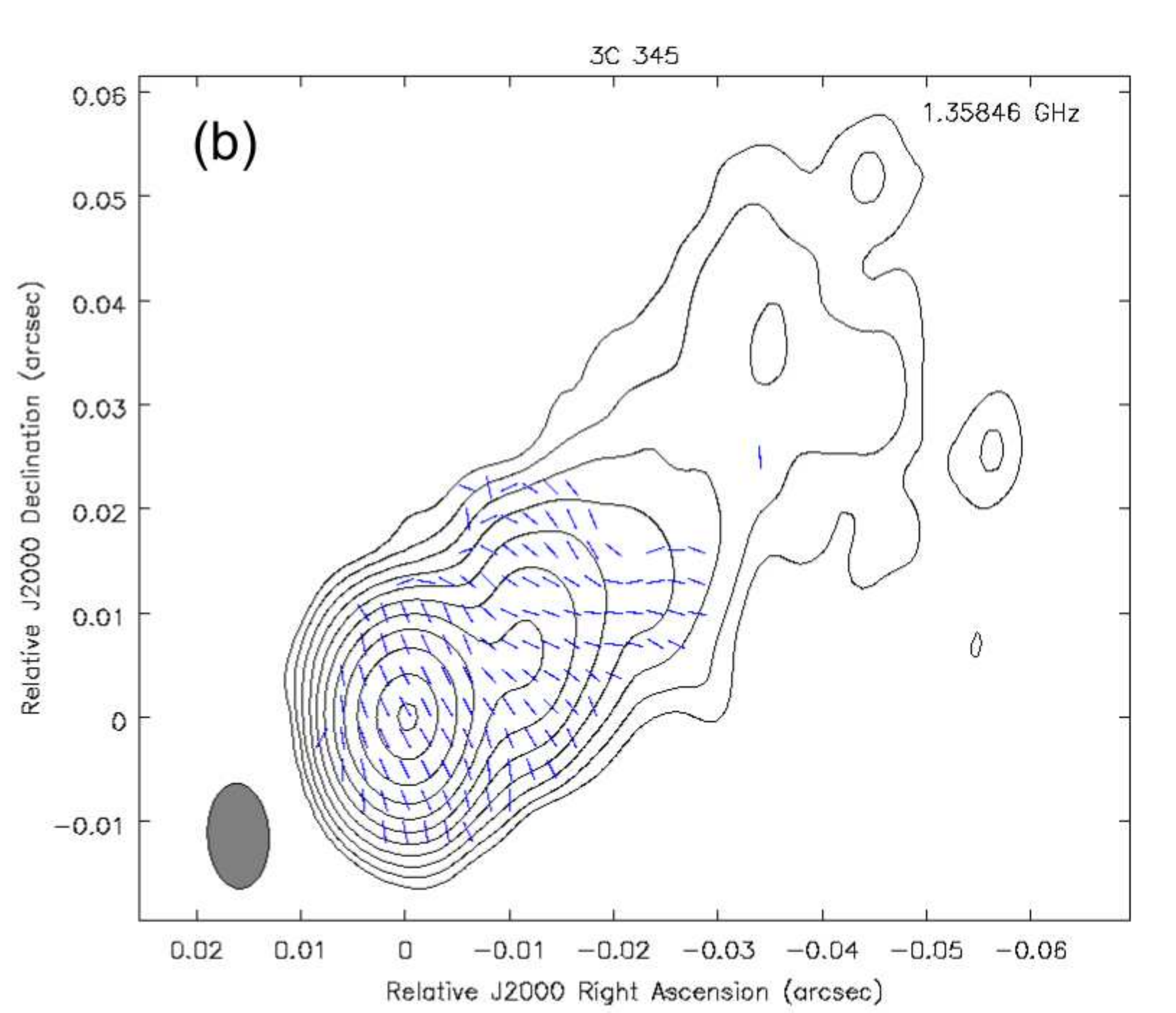}
	\end{minipage}
	\end{minipage}

	\begin{minipage}{\textwidth}
	\begin{minipage}{.45\textwidth}
		\centering
		\includegraphics[width=0.8\textwidth]{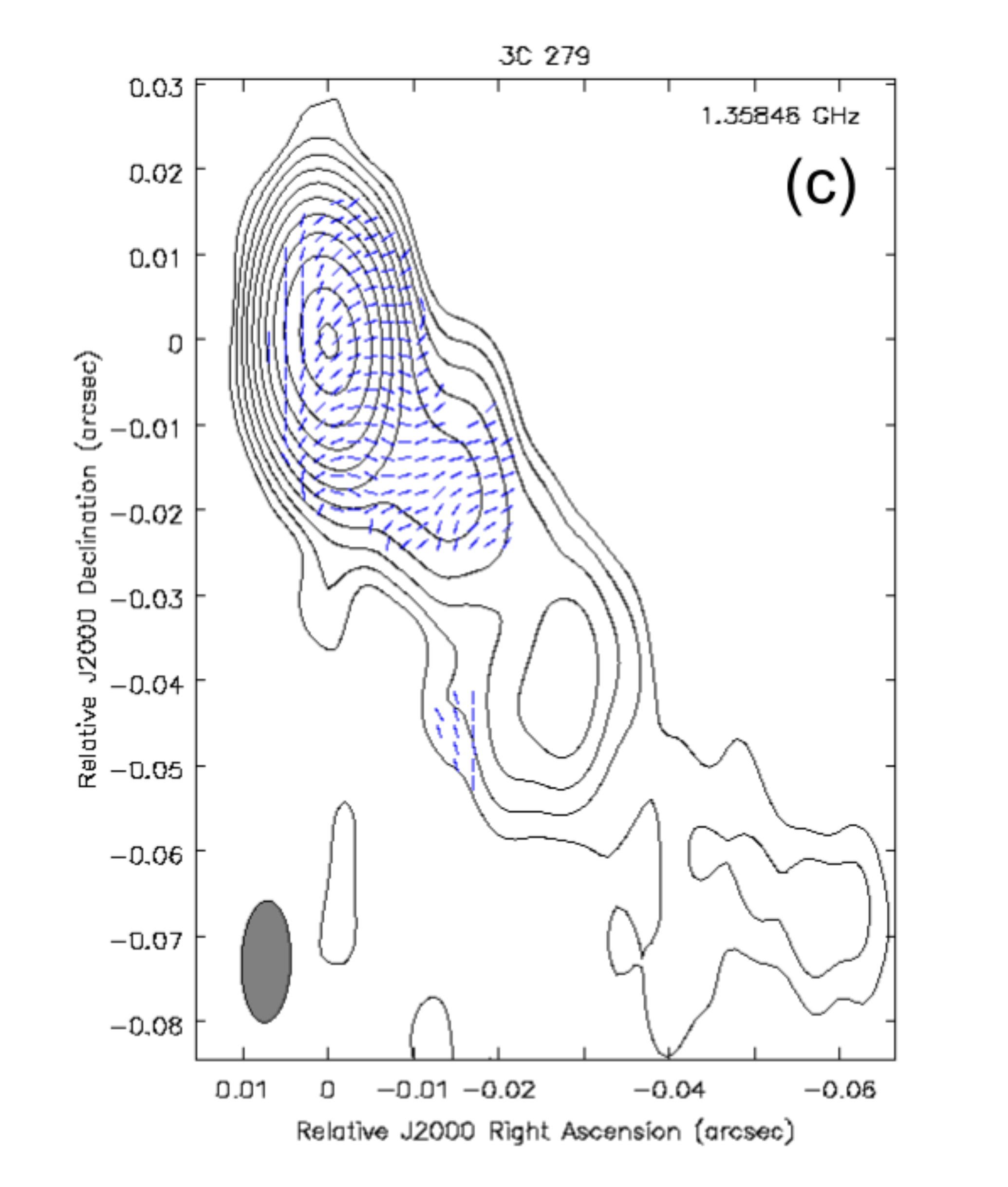}
	\end{minipage}
	\quad
	\begin{minipage}{.45\textwidth}
		\centering
		\includegraphics[width=0.8\textwidth]{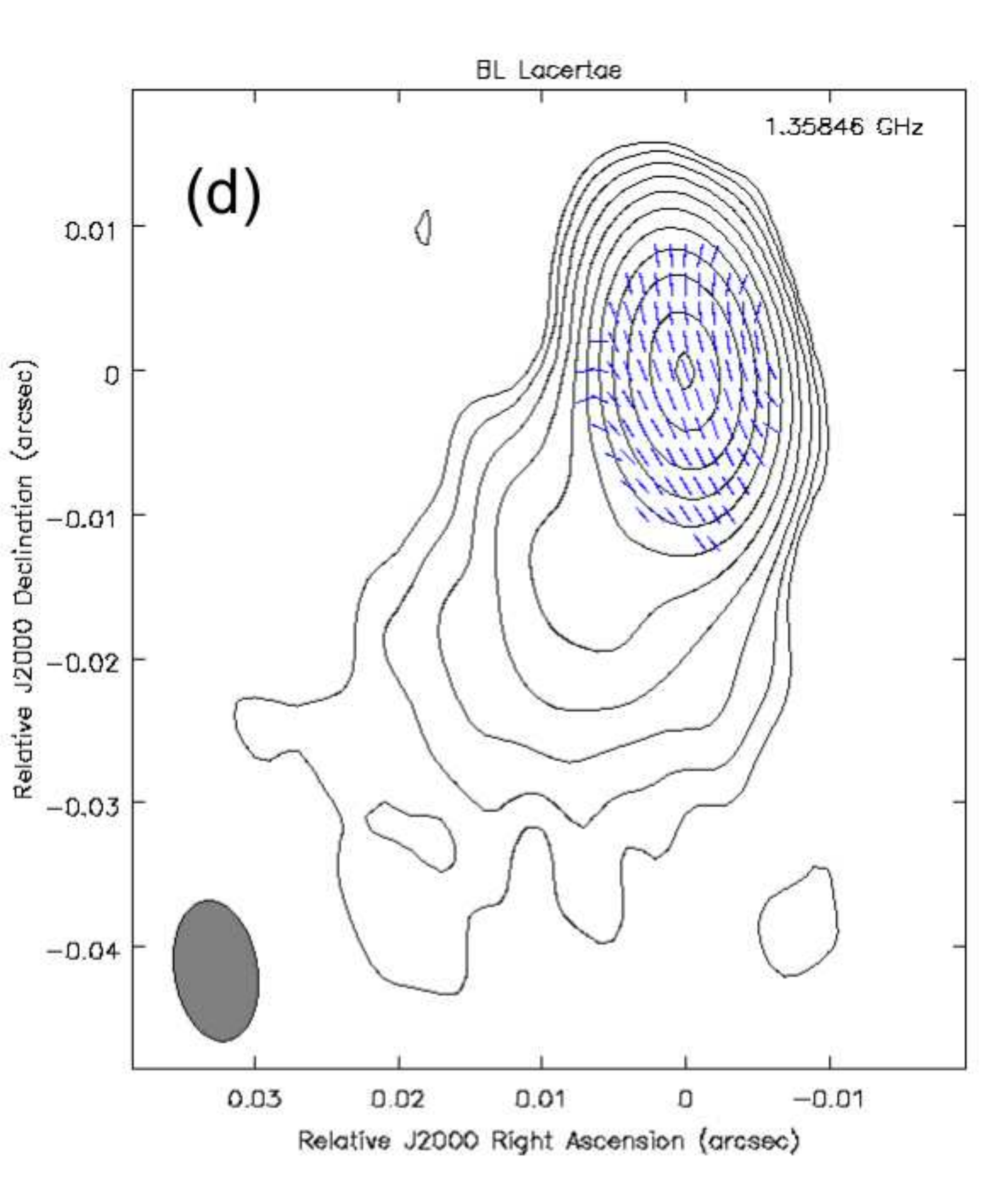}
	\end{minipage}
	\end{minipage}

	\begin{minipage}{\textwidth}
	\begin{minipage}{.45\textwidth}
		\centering
		\includegraphics[width=0.8\textwidth]{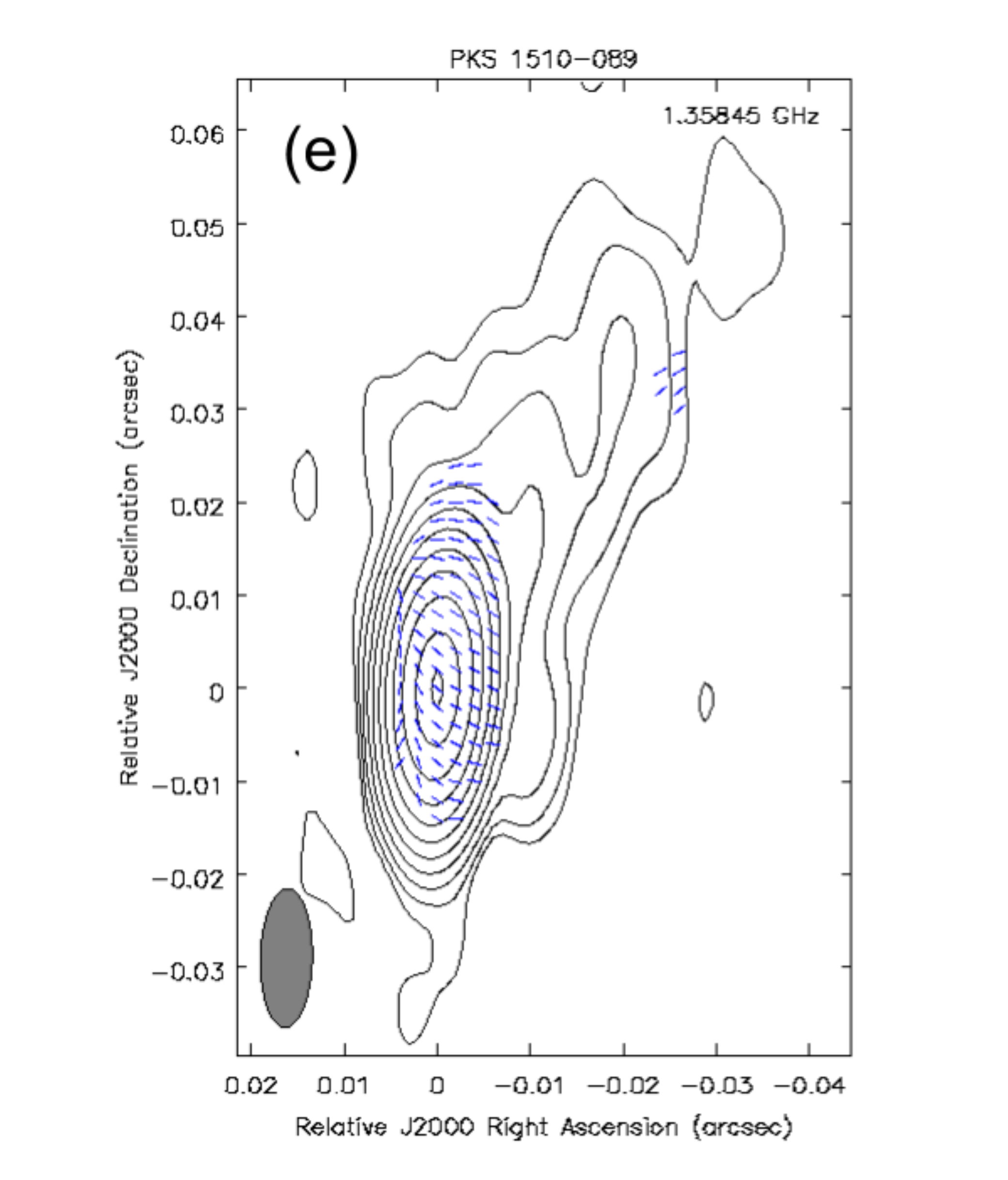}
	\end{minipage}
	\quad
	\begin{minipage}{.45\textwidth}
		\centering
		\includegraphics[width=0.8\textwidth]{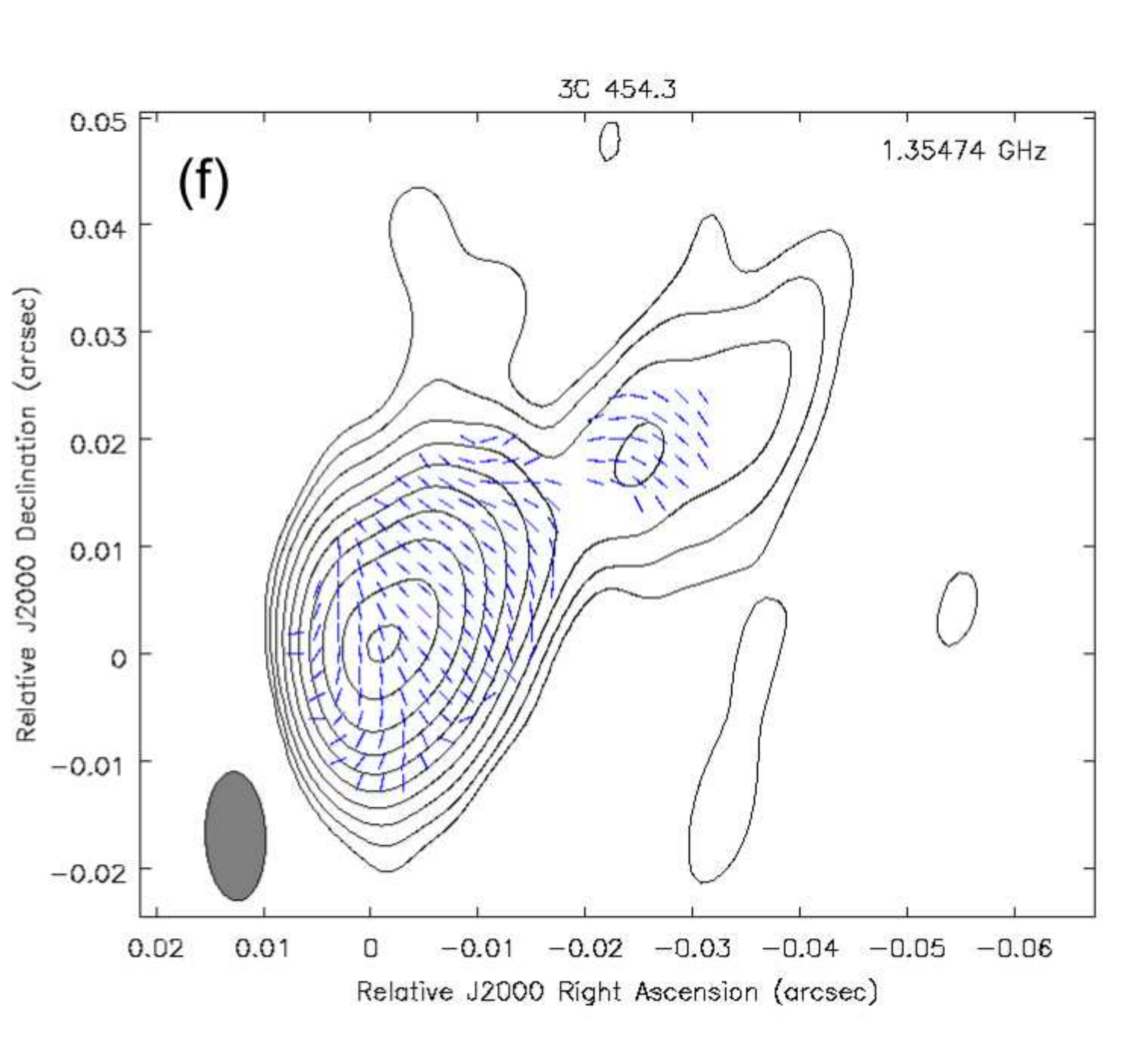}
	\end{minipage}
	\end{minipage}

\caption{The ticks depict the EVPAs corrected for the local and Galactic foreground Faraday rotation of OJ 287 (top-left), 3C 279 (middle-left), PKS 1510-089 (bottom-left), 3C 345 (top-right), BL Lac (middle-right) and 3C 454.3 (bottom-right) superimposed on the 1358 MHz total intensity contours. In all cases, the contour levels increase in steps of a factor of two. The gray ellipse on the lower left-hand corner of each map depicts the convolving beam. \label{EVPA_RMgal_RMlocal}
 }
\end{center}
\end{figure*}

Finally, in Fig. \ref{EVPA_RMgal_RMlocal}, we present the EVPAs of the six sources corrected for both the constant integrated and spatially variable local Faraday rotation superimposed on the 1358 MHz total intensity contours. The EVPA rotations introduced by the local Faraday rotation are small, and these two distributions are typically very similar, however, construction of the EVPA distribution corrected for the VLBA-scale local Faraday rotation provides a final check that the RMs that have been fitted using the VLBA data are not spurious in any region.

The statistical significances of the transverse gradients detected in our RM maps are shown in Table \ref{rmlist}. As is indicated above, we tested the robustness of these gradients by producing RM maps using circular beams with areas equal to those of the intrinsic elliptical beams (bottom-left panels of the RM figures). This is especially helpful in cases where the elliptical beams are appreciably elongated, and can have arbitrary orientations relative to the jet direction, which can occasionally give rise to spurious RM structures. Comparing the RM maps made with the intrinsic elliptical and circular beams, we can check if RM structures initially detected in the intrinsic-beam RM map remain visible and statistically significant in the circular-beam RM map, enabling the identification of spurious gradients.

As noted earlier, we used the approach of \citet{2012AJ....144..105H} when calculating the RMs uncertainties. We did not add the uncertainty in the EVPA calibration to the uncertainty in the EVPAs when determining the significances of possible RM gradients, since the EVPA calibration uncertainty affects all polarization angles equally for a given frequency, and, therefore cannot produce spurious RM gradients, as previously noted by \citet{2009MNRAS.400....2M} and \citet{2012AJ....144..105H}. When calculating the uncertainty of the difference between the RM values at the ends of a slice, we added the two RM uncertainties in quadrature; note that this is a more conservative approach than that used by \citet{2012AJ....144..105H}, helping to ensure that the significance of the RM gradients is not overestimated.

We identified monotonic, transverse Faraday rotation gradients in four of the six AGN, with statistical significances ranging from 2.5$\sigma$ to 4.4$\sigma$ when the RM maps are produced with the intrinsic elliptical beams. Three sources (PKS~1510-089, BL~Lac and 3C~454.3) have transverse Faraday rotation gradients with statistical significances higher than 3$\sigma$; a transverse RM gradient across the jet of 3C~345 has a significance of about 2.5$\sigma$, but is nevertheless significant, as it spans more than two beam widths \citep{2012AJ....144..105H}. We also found a tentative transverse RM gradient across the jet of OJ~287. Below, we present a brief summary of the results for each source.

\begin{table*}
 \scriptsize
  \caption{List of detected transverse RM gradients. \label{rmlist}
}
 \begin{threeparttable}
 \begin{center}

 \begin{tabular}{@{}lcccccc}
  \hline
  Source & Beam & Gradient & RM$_{1}$ & RM$_{2}$ & $|\Delta \mathrm{RM}|$ & Significance \\
  name & shape\tnote{a} & location & (rad m$^{-2}$) & (rad m$^{-2}$) & (rad m$^{-2}$) &  \\
  \hline
  OJ 287 & E & Core & 9.0$\pm$4.6 & -12.8$\pm$5.5 & 21.8$\pm$7.2 & 3.0$\sigma$  \\
  
  OJ 287 & C & Core & 7.3$\pm$4.2 & -10.7$\pm$8.8 & 18.0$\pm$9.8 & 1.8$\sigma$ \\ 
  
  PKS 1510-089 (S1) & E & Core & -5.3$\pm$4.2 & 47.8$\pm$11.3 & 53.1$\pm$12.0 & 4.4$\sigma$ \\
  
  PKS 1510-089 (S1) & C & Core & -12.0$\pm$5.5 & 18.4$\pm$5.7 & 30.4$\pm$7.9 & 3.8$\sigma$ \\ 

  PKS 1510-089 (S2) & E & Core & 15.3$\pm$11.3 & -22.4$\pm$9.7 & 37.7$\pm$14.9 & 2.5$\sigma$ \\
  
  PKS 1510-089 (S2) & C & Core & 22.4$\pm$10.6 & -19.0$\pm$13.6 & 41.4$\pm$17.2 & 2.4$\sigma$ \\
  
  3C 345 & E & Jet & -9.3$\pm$6.4 & 13.8$\pm$6.6 & 23.1$\pm$9.2 & 2.5$\sigma$ \\ 
  
  3C 345 & C & Jet & -11.6$\pm$8.8 & 18.3$\pm$10.0 & 29.9$\pm$13.3 & 2.2$\sigma$ \\
  
  BL Lac & E & Core & -0.02$\pm$0.40 & 9.3$\pm$2.3 & 9.3$\pm$2.3 & 4.1$\sigma$ \\

  BL Lac & C & Core & 0.8$\pm$0.5 & 7.3$\pm$1.8 & 6.5$\pm$1.9 & 3.4$\sigma$ \\
  
  3C 454.3 (S1) & E & Jet & 52.9$\pm$7.6 & 101.6$\pm$9.9 & 48.7$\pm$12.5 & 3.9$\sigma$ \\

  3C 454.3 (S1) & C & Jet & 43.1$\pm$9.4 & 119.6$\pm$9.7 & 76.5$\pm$13.5 & 5.7$\sigma$ \\

  3C 454.3 (S2) & E & Jet & 53.5$\pm$9.5 & 82.1$\pm$8.4 & 28.6$\pm$12.7 & 2.3$\sigma$ \\

  3C 454.3 (S2) & C & Jet & 43.5$\pm$8.8 & 80.5$\pm$7.4 & 37.0$\pm$11.5 & 3.2$\sigma$ \\

  \hline
 \end{tabular}
 \begin{tablenotes}
  \item[a] E denotes elliptical beam and C denotes circular beam.  
 \end{tablenotes}
 \end{center}
 \end{threeparttable}
\end{table*}

\begin{itemize}

\item \textbf{OJ~287}: Our maps and model fits show that the jet initially extends roughly towards the west and then appears to bend towards the south. The southern emission may be associated with plasma initially ejected in this direction, since the jet direction seen at the smaller scales sampled by the MOJAVE maps has PAs ranging from -90$^{\circ}$ to about -135$^{\circ}$. 

The EVPAs corrected for the integrated RM are somewhat offset from transverse to the local jet direction (Fig. \ref{Fig1_OJ287_3C279_PKS1510}a). The degree of fractional polarization (Figs. \ref{Fig1_OJ287_3C279_PKS1510}b and c) is less than about 4 per cent in the core and reaches tens of per cent in the jet. 

Our RM map (Fig. \ref{OJ287_RM}) shows a transverse Faraday rotation gradient in the core region, perpendicular to the local jet direction indicated by our model fitting, however, this gradient becomes non-monotonic near its ends. The statistical significance reaches about 3$\sigma$ if points just short of the end points are compared. The gradient structure remains visible when the RM map is constructed using a circular beam, but the statistical significance is reduced to about 1.8$\sigma$. The RM values display a sign change across the jet structure. The fractional polarization along the RM slice considered is fairly constant, but shows some tendency to increase towards the edges of the jet. Overall, we consider this a tentative transverse RM gradient.

The EVPAs corrected for both the integrated and local Faraday rotation (Fig. \ref{EVPA_RMgal_RMlocal}a) are overall similar to those in Fig. \ref{Fig1_OJ287_3C279_PKS1510}(a), consistent with a roughly longitudinal magnetic field.

\item \textbf{3C~279}: The jet extends towards the south-west at a PA of $\simeq$ -140$^{\circ}$, roughly in the same direction seen at the smaller scales on the MOJAVE maps at nearby epochs ($\simeq$ -130$^{\circ}$).

The EVPAs corrected for the integrated RM are predominantly transverse to the local jet direction (Fig. \ref{Fig1_OJ287_3C279_PKS1510}d). The polarization structure seen at the smaller scales sampled by the 2 cm observations is quite complex, with different regions in the jet having either predominantly orthogonal or longitudinal EVPAs. The degree of fractional polarization (Fig. \ref{Fig1_OJ287_3C279_PKS1510}e) is less than about 6 per cent in the core and reaches a few tens of per cent in the jet.

Our RM map (Fig. \ref{3C279_RM}) shows that the RM values are slightly enhanced in the core region. The RM map made with the intrinsic elliptical beam shows a possible transverse gradient in the core region; however, this is not apparent in the RM map made with the equal-area circular beam. We accordingly did not detect any statistically significant transverse RM gradient in this object.

The EVPAs corrected for both the integrated and local Faraday rotation (Fig. \ref{EVPA_RMgal_RMlocal}c) are roughly transverse to the jet direction in the jet, indicating a predominantly longitudinal magnetic field in this region. The polarization structure across the core region is complex, however, with fairly large changes in the orientation of the EVPAs between the eastern and western parts of the observed core that are not as prominent in the polarization map corrected only for the integrated Faraday rotation shown in Fig. \ref{Fig1_OJ287_3C279_PKS1510}(d). This suggests that some of the local Faraday rotation fits in the core may be unreliable, but we have not been able to find any evidence for this from the quality of the RM fits. We, likewise, found no evidence for significant optical depth effects in the core region that could affect the observed polarization angles. The origin of this behaviour of the core polarization angles in Fig. \ref{EVPA_RMgal_RMlocal}(c) is thus not clear, though it seems likely that some of the RM fits at the edges of the core are nevertheless unreliable. Although this introduces some uncertainty into our deductions concerning the direction of the magnetic field in the core region, the jet polarization certainly indicates a longitudinal magnetic field.

\item \textbf{PKS~1510-089}: The jet extends towards the north-west at a PA of $\simeq$ -30$^{\circ}$, in the same direction seen at the smaller scales mapped by the 2 cm MOJAVE observations.

The EVPAs corrected for the integrated RM are predominantly transverse to the local jet direction (Fig. \ref{Fig1_OJ287_3C279_PKS1510}f), in agreement with the overall trend of polarization structures seen in the MOJAVE maps. The degree of fractional polarization (Fig. \ref{Fig1_OJ287_3C279_PKS1510}g) is less than about 5 per cent in the core, and reaches tens of per cent in the jet.

Our RM map (Fig. \ref{PKS1510_RM}) shows monotonic, transverse gradients in the core region. The example slice shown in Fig. \ref{PKS1510_RM} displays an RM gradient with a statistical significance of about 4.4$\sigma$. The structure of the RM map remains the same in the version made with the circular beam; the transverse gradient across the core remains visible, with a statistical significance of 3.8$\sigma$. A sign change in the RM is observed across the jet. We note that Keck et al. (private communication) have tentatively found similar results for this source, with transverse RM gradients in the core region, using frequencies in the range 22 - 43 GHz.

Our RM map also displays a reversal in the direction of the transverse RM gradient further down in the jet of PKS~1510-089. This reversal is more evident in the RM map constructed using the circular beam; the corresponding statistical significances in both cases, elliptical and circular beam, are about 2.5$\sigma$. The fractional polarization slices show asymmetries across the jet, with higher polarizations indicated on the side of the jet where RMs with larger absolute values are observed, as would be expected if the transverse RM gradients are associated with helical magnetic fields.

Correction for both the integrated and local Faraday rotation (Fig. \ref{EVPA_RMgal_RMlocal}e) has rotated the EVPAs so that they are close to orthogonal to the jet direction. Our results thus indicate a predominantly longitudinal $\bm{B}$ field.

\item \textbf{3C~345}: The jet extends towards the north-west at a PA of $\simeq$ -55$^{\circ}$. At the smaller scales seen on the MOJAVE maps, the jet of 3C~345 initially extends towards the west and then it bends towards the north-west in roughly the same PA seen in our observations.

The EVPAs corrected for the integrated RM are predominantly transverse to the local jet direction (Fig. \ref{Fig2_3C345_BLLac_3C454.3}a) following the overall behaviour seen in the 2 cm observations. There is an interesting polarization structure about 35 mas from the core composed of orthogonal polarization near the jet edge and longitudinal polarization near its central axis that could be a `spine+sheath' composition. The fractional polarization of 3C~345 is roughly 2-5 per cent in the core (Fig. \ref{Fig2_3C345_BLLac_3C454.3}b), and reaches values of 15-30 per cent in the jet (Fig. \ref{Fig2_3C345_BLLac_3C454.3}c). The degree of polarization tends to increase towards the jet edges. This is expected if the jet is threaded by helical magnetic fields due to the higher ordering of the field lines in these regions.

Our RM map (Fig. \ref{3C345_RM}) shows a monotonic, transverse RM gradient in the jet at a distance of approximately 10 mas from the observed core, visible in both the intrinsic-beam (statistical significance 2.5$\sigma$) and circular-beam (statistical significance 2.2$\sigma$) RM maps. The RM values display a sign change across the jet. Although the statistical significance of this gradient is smaller than 3$\sigma$, the gradient spans more than two beamwidths, making it statistically significant, as is discussed in Section \ref{disc1}. The fractional polarization is asymmetric and varies from about 7 per cent to 15 per cent across the jet.

The EVPAs corrected for both the integrated and local Faraday rotation (Fig. \ref{EVPA_RMgal_RMlocal}b) are roughly transverse to the local jet direction, very similar to those in Fig. \ref{Fig2_3C345_BLLac_3C454.3}(a), indicating a predominantly longitudinal $\bm{B}$-field.

\item \textbf{BL~Lac}: The jet of BL Lac is initially directed slightly west of south, as can be seen on the 2 cm MOJAVE observations and in our model-fitting results, then bends towards the south-east. The EVPAs corrected for the integrated RM are predominantly longitudinal to the local inner jet direction (Fig. \ref{Fig2_3C345_BLLac_3C454.3}d), in agreement with the behaviour seen in the 2 cm polarization maps. The 1358 MHz fractional polarization is 2 - 8 per cent in the core and increases to $\simeq$25 per cent in the jet (Fig. \ref{Fig2_3C345_BLLac_3C454.3}e).   

Recalling that the innermost jet lies along a PA of $\simeq -174^{\circ}$, the RM map (Fig. \ref{BLLAC_RM}) shows a monotonic, transverse Faraday rotation gradient in the core region with a statistical significance of about 4.1$\sigma$. The gradient remains visible when the RM map is constructed using the circular beam, and the statistical significance remains above 3$\sigma$. The RM values display a sign change across the jet. Note that the slice shown in Fig. \ref{BLLAC_RM} was taken perpendicular to the direction of the innermost jet indicated by our model-fitting, $\simeq -174^{\circ}$. The variations in fractional polarization along the RM slice considered are relatively modest, with slightly higher degrees of polarization observed on the side of the core region with lower RM values.

The EVPAs corrected for both the integrated and local Faraday rotation (Fig. \ref{EVPA_RMgal_RMlocal}d) are aligned with the smaller-scale jet direction and suggest a predominantly transverse $\bm{B}$-field. This is in agreement with previous results published by \citet{2009MNRAS.393..429O} based on 2006 VLBA observations in the frequency range 4.6 to 43 GHz.

\item \textbf{3C~454.3}: The 1358 MHz jet of 3C~454.3 extends towards the north-west in a PA of $\simeq -55^{\circ}$. On the smaller scales probed by the 2 cm MOJAVE observations, the jet of 3C~454.3 extends towards the west before bending towards the north-west to lie in roughly the same PA shown in our map.

The EVPAs corrected for the integrated RM are predominantly transverse to the local jet direction (Fig. \ref{Fig2_3C345_BLLac_3C454.3}f). The degree of polarization ranges from 2 per cent to 8 per cent in the core region (Fig. \ref{Fig2_3C345_BLLac_3C454.3}g) and reaches tens of per cent in the jet (Fig. \ref{Fig2_3C345_BLLac_3C454.3}h).

Our RM map shows monotonic, transverse Faraday rotation gradients at projected distances of $\simeq 10-35$~mas from the core, as is shown in Fig. \ref{3C454.3_RM}. We show two slices, the first taken approximately 15~mas from the observed core, with a statistical significance of 3.9$\sigma$, and the second approximately 25~mas from the observed core, with a statistical significance of 2.3$\sigma$. Both gradients are also clearly visible in the RM map using a circular beam, with statistical significances of 5.7$\sigma$ and 3.2$\sigma$, respectively. The direction of both transverse RM gradients is the same, with higher RM values towards the southern side of the jet. The RM values display no sign change and the RM magnitude is appreciably enhanced in the core. Both fractional polarization slices show appreciable asymmetry, with higher degrees of polarization observed on the northern side of the jet, where the RM values are lower.

The EVPAs corrected for both the integrated and local Faraday rotation (Fig. \ref{EVPA_RMgal_RMlocal}f) are predominantly transverse to the local jet direction. Our results suggest an overall predominantly longitudinal $\bm{B}$-field in this source. This is in agreement with multi-frequency VLBI observations in the range 5 to 86 GHz presented by \citet{2013MNRAS.436.3341Z}, which have shown that the magnetic field in 3C~454.3 follows the total intensity contours even when they bend. 

\end{itemize}

\section{Discussion} \label{sec:dis}

\subsection{Reliability of the detection of transverse RM gradients in AGN jets} \label{disc1}

Much has been discussed in the literature about the reliability of the detection of transverse RM gradients in AGN. One possible issue was the extent of these gradients compared to the beam sizes of the observational experiments. However, the results of \citet{2012AJ....144..105H}, \citet{2013MNRAS.429.3551A}, \citet{2013MNRAS.431..695M} and \citet{2013EPJWC..6107005M} have shown that it is not meaningful to impose strict limits on the width spanned by an RM gradient in order for it to be reliably detected, and that the key criteria for its reliability are monotonicity and its statistical significance. 

The simulations carried out by \citet{2013MNRAS.431..695M} (for wavelengths 2-6 cm) and \citet{2013EPJWC..6107005M} (for wavelengths 18-22 cm) using simulated VLBA data for sources at different declinations with jets of various widths having RM gradients with various strengths convolved with beams with various sizes, showed that, with realistic noise and baseline coverage, transverse RM gradients can remain visible even for intrinsic jet widths of the order of 1/20 of a beamwidth, which have observed widths comparable to the beamwidth. \citet{2013EPJWC..6107005M} demonstrated that the occurrence of spurious 3$\sigma$ RM gradients, and of 2$\sigma$ RM gradients at least two beamwidths wide, in Monte Carlo simulations of VLBA RM maps based on the four wavelengths considered in this paper was appreciably less than 1 per cent of all cases, even for jets much narrower than the beam size.

Considering these results and the transverse Faraday rotation gradients presented in this work and listed in Table \ref{rmlist}, the gradients found across the jets of PKS~1510-089, BL Lac, 3C~454.3 and 3C~345 can be considered firm detections. The gradient across the jet of OJ~287 should be considered tentative.

\subsection{Transverse RM gradients and polarization structure as a diagnostic for the geometry of magnetic fields in AGN jets} \label{disc2}

Faraday rotation studies are an important tool for understanding the structure of magnetic fields in AGN jets. The RM depends on both the electron density in the region of Faraday rotation and on the line-of-sight component of the magnetic field. As already mentioned in Section \ref{sec:intro}, the occurrence of transverse RM gradients across AGN jets is expected in the presence of a helical or toroidal $\bm{B}$-field due to the systematically changing line-of-sight component of the magnetic field across the jet. Gradients in electron density could, in principle, also be responsible for transverse RM gradients. However, this can not explain the detection of monotonic, transverse RM gradients exhibiting a sign change.

We detected sign changes in the firmly detected RM gradients across the jet structures of PKS~1510-089, 3C~345 and BL~Lac, and also in the tentative RM gradient across the jet of OJ~287, thus indicating the presence of toroidal magnetic field components that could be associated with helical jet magnetic fields. In the case of BL~Lac, recent polarimetric results obtained from 15, 22 and 43 GHz VLBI observations presented by \citet{2016ApJ...817...96G} for this source, show gradients both in RMs and Faraday corrected EVPAs as a function of polar angle with respect to the centroid of the observed 15 GHz core, also suggesting that the core region in BL Lac is threaded by a helical magnetic field. In addition, a transverse RM gradient across the jet of BL~Lac based on 8--15~GHz VLBI observations has been reported by Gabuzda et al. (in prep). The collected results thus consistently point towards the idea that the jet of BL Lac carries a helical magnetic field.

We note, however, that the non-detection of a sign change in a transverse RM profile does not exclude the possibility that the gradient is caused by a helical magnetic field, since some combinations of helical pitch angles and viewing angles do not lead to changes in the observed RM signs across the jet. This could be the case in 3C~454.3, where we detected only positive values for the RMs across the jet. Moreover, evidence for a large scale helical magnetic field in 3C~454.3 has been reported by \citet{2013MNRAS.436.3341Z}, who found consistency between the observed jet asymmetries in the intensity, RM and degree of linear polarization profiles and the profiles expected in models with a helical magnetic field threading the AGN jets.

The reversal of the observed transverse RM gradient in the jet of PKS~1510-089 could be explained by magnetic-tower-type models \citep{1996MNRAS.279..389L}, or a nested helical $\bm{B}$-field structure composed of an `inner' and an `outer' helices. In this scenario, the two azimuthal components of the $\bm{B}$-field are oppositely directed as a consequence of the differential rotation of the accretion disc (see fig. 9 of \citealt{2009MNRAS.400....2M}), and the direction of the observed RM gradients depends on the dominance of either the `inner' or `outer' helix along the jet. A theoretical basis for this picture is described by \citet{2016A&A...591A..61C}.

The EVPA distributions corrected using only the integrated RM and corrected for both integrated and local Faraday rotation were very similar, as expected, giving indirect confirmation of the reliability of the local Faraday rotations derived. The only exception is the core region of 3C~279, where some of the RM fits at the edge of the core may be unreliable. The Faraday-rotation-corrected EVPA distributions indicate that the magnetic field is predominantly longitudinal in the jets of OJ~287, 3C~279, PKS~1510-089, 3C~345 and 3C~454.3, and predominantly transverse in BL~Lac. The dominance of either the longitudinal or transverse magnetic field component could be a manifestation of helical magnetic fields with either smaller or higher pitch angles, respectively \citep{2005MNRAS.356..859P}.

\subsection{Core-region transverse RM gradients} \label{disc3}

The compact `core' observed in VLBI images is often assumed to correspond to the region where the optical depth becomes equal to unity. However, due to lack of resolution, especially at lower frequencies, the observed core region is actually a blend of the theoretical optically thick base of the jet and optically thin regions in the inner jet, such that the polarization properties observed in the VLBI `core' are likely to be dominated by the emission of these optically thin regions, which have lower total intensities but higher degrees of linear polarization. As a consequence, in practice, the observed polarization angles in the core are usually orthogonal to the local magnetic field, as expected for predominantly optically thin regions.

Because the observed core polarization is usually dominated by the contributions of optically thin regions in the inner jet, this can also give rise to transverse RM gradients across the core region. This is especially true at low frequencies such as those we have considered in this work. \citet{2010ApJ...725..750B} carried out general relativistic magnetohydrodynamic simulations to build RM maps of parsec scale AGN jets with large scale toroidally dominated magnetic fields. Their results raised concern about the finite-beam effects on the observed RMs, especially for RMs detected in the core region. These simulations show that for unresolved transverse jet structures, non-monotonic behaviour in the transverse RM profiles may arise. On the other hand, any non-monotonic behaviour in the RM profiles can be smoothed by convolution with typical VLBA observing beams at the expense of detecting RMs whose magnitudes are generally smaller than the true values and hindering the detection of sign changes along the gradients [see the lower right panel of fig. 8 of \citet{2010ApJ...725..750B}]. Here, we should note that our goal is simply to detect the presence and direction of transverse RM gradients across AGN jets, and we do not aspire to determine intrinsic RM values. One possible manifestation that the observed core RM values are actually associated with predominantly optically thick regions would be deviations from linear behaviour in $\chi_{obs}$ versus $\lambda^2$ fits; it is noteworthy here that the $\chi_{obs}$ versus $\lambda^2$ fits shown in the RM figures do not show any clear evidence for significant deviations from linear dependences. The idea that the observed core polarizations are dominated by optically thin regions is also supported by the relatively high degrees of polarization in the core regions of the AGN considered here, which range from $\simeq 2-8$ per cent. 

Thus, we detected monotonic transverse RM gradients across the cores of PKS~1510-089 and BL~Lac with statistical significances exceeding 3$\sigma$. This can most straightforwardly be interpreted as being associated with helical jet $\bm{B}$-fields present in the jets of these AGN on scales slightly smaller than those probed in the observations analysed here. In the case of PKS~1510-089, 3C~345 and 3C~454.3, we have also detected transverse RM gradients across jet regions that are clearly dominated by optically thin emission.

\section{Conclusions} \label{sec:conc}

We have presented linear polarization, fractional polarization and Faraday rotation maps of six AGN constructed using VLBA data at four wavelengths in the range 18-22~cm. Our results show that the polarization structures implied by the Faraday-corrected EVPA distributions are consistent with those observed on smaller scales in the 2 cm MOJAVE maps. The degrees of polarization are 2 - 8 per cent in the cores of all the six sources, and reach tens of percent in their jets. 

We have detected monotonic, statistically significant transverse RM gradients in four out of six sources. These gradients are located in the observed core regions of PKS~1510-089 and BL~Lac, in the jet of 3C~345 and in an extended region across the jet of 3C~454.3. We also detected a tentative transverse RM gradient across the core region of OJ~287. These transverse RM gradients indicate the presence of toroidal magnetic fields, which may be one component of helical magnetic fields associated with these AGN jets.

The detection of sign changes in the RM distributions across the jets of OJ~287, PKS~1510-089, 3C~345 and BL~Lac, provides particularly strong evidence for the presence of helical fields, since electron density gradients could not cause this effect. We note, however, that the non-detection of a sign change in the RM profile of 3C~454.3 does not rule out the possibility that its transverse RM gradient reflects the presence of a helical jet magnetic field, particularly given the results of \citet{2013MNRAS.436.3341Z} that likewise suggests that the jet of 3C~454.3 carries a helical magnetic field.

Transverse profiles of the fractional polarization constructed in the regions where transverse RM gradients were detected show appreciable asymmetry.  It is interesting that the transverse fractional polarization profiles in PKS~1510-089, BL~Lac and 3C~454.3 display higher degrees of polarization on the side of the jet where the RM values have lower magnitudes; this is just the pattern expected if these jets carry helical magnetic fields, since the RMs are maximum where the line-of-sight magnetic field is maximum, while the degree of polarization is maximum when the magnetic field in the plane of the sky is maximum. This same pattern was observed in Mrk~501 \citep*{2013MNRAS.430.1504M}. In the case of PKS~1510-089, it is noteworthy that a reversal in the direction of the RM gradient is accompanied by a reversal in the direction of the fractional polarization profile. In 3C~345, the degree of polarization in the jet increases towards the jet edges, likewise consistent with a helical jet magnetic field. In general, the transverse intensity profiles are much more symmetric than is expected based on the corresponding fractional polarization profiles, if the emission arises in a region of helical magnetic field; the origin of this discrepancy between the symmetry of the intensity profiles and the asymmetry of the corresponding polarization profiles is not clear, but it is interesting that the same discrepancy was observed for Mrk~501, whose polarization profiles were fit very well by a helical field model \citep*{2013MNRAS.430.1504M}. A full analysis of the transverse polarization profiles in these AGNs goes beyond the framework of this study, and will be carried out in future work with these data.

Finally, we note that the six AGN considered here were chosen for this study for reasons that had nothing to do with the likelihood that their jets carry helical $\bm{B}$-fields; in other words, the likelihood that their RM maps would show transverse RM gradients was not known a priori. The detection of statistically significant transverse RM gradients in four of the six AGN sources studied is striking in this connection, and suggests that analysis of these 18-22~cm data for additional sources may yield many more detections of statistically significant transverse RM gradients on scales of tens of parsec. This, in turn, would suggest that a large fraction of AGN jets may possess helical $\bm{B}$-fields, or at least toroidal $\bm{B}$-field components, on scales out to of order 100 pc or more.

\section*{Acknowledgements}

JCM thanks CAPES (Coordena\c{c}\~{a}o de Aperfei\c{c}oamento de Pessoal de N\'{i}vel Superior) for the financial support under the grant BEX 3421-15-5, and CNPq (Conselho Nacional de Desenvolvimento Cient\'{i}fico e Tecnol\'{o}gico) under the grant 142041/2013-0. We thank Sebastian Knuettel for sharing his expertise during the course of this research.
This work has made use of data from the NRAO Very Long Baseline Array facilities.

\bibliographystyle{mn2e} 
\bibliography{mylibrary}

\end{document}